\documentclass{article}
\usepackage{arxiv}
\usepackage[utf8]{inputenc} 
\usepackage[T1]{fontenc}    
\usepackage{hyperref}       
\usepackage{url}            
\usepackage{booktabs}       
\usepackage{amsfonts}       
\usepackage{nicefrac}       
\usepackage{microtype}      
\usepackage{lipsum}
\usepackage{graphicx}
\usepackage{comment}
\usepackage{multirow}
\usepackage{subcaption}
\usepackage{amsmath}
\usepackage[psamsfonts]{amssymb}
\usepackage{amsxtra}
\usepackage{threeparttable}

\title{A Comprehensive Overview of Computational Nuclei Segmentation Methods in Digital Pathology}

\author{
  Vasileios Magoulianitis, Catherine A. Alexander and C.-C. Jay Kuo \\
  University of Southern California, Los Angeles, California, USA \\
  \texttt{magoulia@usc.edu, cc98905@usc.edu, cckuo@sipi.usc.edu}
}

\begin{document}
\maketitle

\begin{abstract}

In the cancer diagnosis pipeline, digital pathology plays an instrumental role in the identification, staging, and grading of malignant areas on biopsy tissue specimens. High resolution histology images are subject to high variance in appearance, sourcing either from the acquisition devices or the H\&E staining process. Nuclei segmentation is an important task, as it detects the nuclei cells over background tissue and gives rise to the topology, size, and count of nuclei which are determinant factors for cancer detection. Yet, it is a fairly time consuming task for pathologists, with reportedly high subjectivity. Computer Aided Diagnosis (CAD) tools empowered by modern Artificial Intelligence (AI) models enable the automation of nuclei segmentation. This can reduce the subjectivity in analysis and reading time. This paper provides an extensive review, beginning from earlier works use traditional image processing techniques and reaching up to modern approaches following the Deep Learning (DL) paradigm. Our review also focuses on the weak supervision aspect of the problem, motivated by the fact that annotated data is scarce. At the end,  the advantages of different models and types of supervision are thoroughly discussed. Furthermore, we try to extrapolate and envision how future research lines will potentially be, so as to minimize the need for labeled data while maintaining high performance. Future methods should emphasize efficient and explainable models with a transparent underlying process so that physicians can trust their output.

\end{abstract}

\section{Introduction}\label{sec:introduction}

\begin{figure*}[t]
\begin{center}
\includegraphics[width=1.0\linewidth]{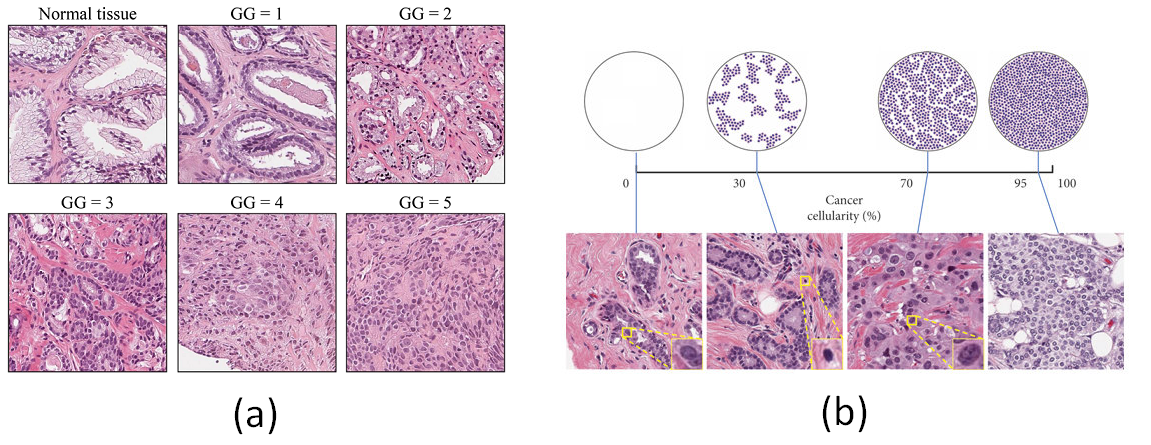}
\caption{Nuclei topology in forming glands is important for tumor grading. Nuclei segmentation task aims at detecting and highlighting nuclei over other areas, thus giving rise to visual patterns that are important for pathologists, such as assessing the cancer gleason grade in prostate tissue (GG) (a) or the cellularity percent (b)~(figures from \cite{salvi2023impact, peidirect}).} \label{fig:grading_cancer}
\end{center}
\end{figure*}

According to the Centers for Disease Control and Prevention (CDC), cancer is the second leading cause of mortality in the U.S. after cardiovascular diseases \cite{murphy2021mortality}. The toll accounts for more than 140 deaths per 100,000 population in the U.S. An instrumental step for cancer diagnosis, tumor grading, and staging evaluation is a biopsy. It is still the standard way for confirming cancer in patients. Tissue specimens are extracted from the suspicious areas, usually identified by radiologists, to reflect whether cancerous cells are present. Biopsy cores are processed onto slides and further stained using the popular Hematoxylin \& Eosin (H\&E) method to give rise to nuclei and cytoplasm. Before digital evolution, pathologists used to read the slides manually under microscopes, which was a time-consuming and laborious task, increasing the diagnostic expenditures while delaying the results turn in time\cite{gurcan2009histopathological}. Nuclei segmentation is a pivotal task towards cancer reading on histology images. The relative topology, size, and shape of nuclei can characterize cancer's development in a certain area, depending on the tissue type (see Fig.~\ref{fig:grading_cancer}). With the advent of whole slide image (WSI) scanners, it is possible to digitize the slides in a high resolution under a certain magnification level, thus enabling the pathologists to inspect the slides on the monitor using dedicated software \cite{guerrero2022software}. Yet, the very large size of each digitized image still requires much time from expert pathologists to be read. Furthermore, there is a reportedly high inter-reader variability because of different expertise levels \cite{elmore2016variability}. 

In the last decade, a lot of research has been conducted in developing accurate Computer-Aided Diagnosis (CAD) tools that can perform at the same level as pathologists and provide more objective decisions with no variability. These tools are meant to help pathologists in routinely performed tasks by taking on trivial cases. Several million  biopsies are performed in the U.S. alone, across several types of tissues. Given the high false positive rate of people sent to biopsy \cite{stolk2019false, ho2022cumulative}, pathologists tend to spend most of their time reviewing benign tissues. CAD tools can help to more efficiently screen trivial cases, which enables pathologists to focus on more ambiguous cases that need human's expertise. After recent AI advancements, it is possible now to develop CAD tools able to fully automate tasks in the diagnostic pipeline, hence easing the tissue reading process. Nuclei segmentation highlights the topology of nuclei, as well as their shape, so its output can help pathologists to review the slides much faster. Also, nuclei segmentation module can feed the input of another AI-based module that predicts a slide to be cancerous based on the nuclei segmentation output.

Following the last decade's research of AI and Deep Learning (DL), last year, the first ever FDA approval was given to the PAIGE tool \cite{kanan2020independent} meant for AI-assisted pathology reading. This brings modern pathology into a new era and paves the way for more approvals in the future for AI-powered tools in digital pathology. Yet, there are still commonly identified challenges \cite{lagree2021review} for the nuclei segmentation task that need to be addressed. Noise and artifacts during the staining process increase the intra-nuclei variance of appearance, which also varies across tissue from different organs. Another factor is the small amount of annotated data. Nuclei segmented images are hard and expensive to obtain since they require a considerable amount of time and expertise. Hence, publicly available nuclei segmentation datasets for research are scarce. Modern AI and DL-based solutions require large data for training, and thereby they face the challenge of generalizing into new images, especially in unseen organs \cite{zhou2019cia}.

A couple of surveys have been conducted about nuclei segmentation and histological image reading. Nasir {\em et al.}~\cite{nasir2022nuclei} have conducted an extensive review of several methods, comparing the nuclei- and gland-based segmentation methods. They provide various statistics and charts about methods and datasets accepted in journals, paper acceptance per year in the area, and analyze publicly available datasets. Various quantitative and qualitative analysis facts are derived, as well as comparisons between nuclei and gland segmentation problems. Hayakawa {\em et al.}~\cite{hayakawa2021computational} provide a brief survey on existing nuclei segmentation solutions, beginning from earlier approaches that used traditional methods and reaching up to recent state-of-the-art (SOTA) methods. That survey gives a quick overview of the whole field, focusing only on the nuclei segmentation problem, and is a nice guide to quickly navigate through the different categories of existing methods in the field. Irshad {\em et al.} ~\cite{irshad2013methods} carried out a broader survey for nuclei segmentation and detection, as well as general classification in digital pathology images. They provide a thorough technical review on earlier works before the advent of DL. In their work, different standard pre-processing methods are referred. Moreover, there are extensive explanations and formulas for earlier techniques used for nuclei segmentation, such as Gaussian Mixture Models (GMMs), clustering, active contours models and level sets, graph cuts and morphological operations. Other surveys \cite{lagree2021review} focus their review on DL-based methods for nuclei segmentation but target only breast tissue cancer which is a popular area for nuclei segmentation. Yet, breast nuclei segmentation methods are intertwined to some extent with the generic nuclei segmentation ones and thereby this paper provides a good overview of nuclei segmentation methods with Deep Learning and relevant datasets that comprise breast tissue. As nuclei segmentation may touch upon the general histology image classification problem, there are other surveys pertaining to the classification task that show how nuclei segmentation could couple with the WSI classification task \cite{zhou2020comprehensive}.

Our survey focuses on the nuclei segmentation problem, regardless of the type of tissue. It pertains to datasets with multiple organs, and we are looking into the generalization ability of methods in unseen organs as well. Despite recent surveys, this overview paper aims at providing a comprehensive technical review of existing literature in Section \ref{sec:method}, starting from the methods using traditional pipelines up to today's DL-based models with supervision, as well as self-supervised methods. After the extended review of proposed methods, our discussion focuses on how supervision may help improve nuclei segmentation to the degree that is needed. Motivated by the scarcity of annotated datasets, we try to delve deeper into weak supervision or self-supervision in order to answer two questions: (1) how much supervision do we need to solve the problem and (2) what are the areas of the problem that supervision helps to reach a higher performance. After comparing and identifying current issues and challenges in existing works in Section \ref{sec:evaluation} --using both quantitative and qualitative results-- we provide our thoughts about future directions (Section \ref{sec:future work}), abstract ideas on how supervision can help and how to identify the minimum amount of areas need supervision.

\section{Staining in Digital Pathology}\label{sec:staining}

For over a century, the dominant technique for staining histopathological images is using Hematoxylin \& Eosin (H\&E). Biopsy tissue specimens extracted from organ regions that are suspected to have developed tumor are stained to reveal the nuclei over other structures. In particular, Hematoxylin (H) gives rise to nuclei, reflecting on them by a dark-purple color, while Eosin (E) stains other structures in a light pink color, --not diagnostically relevant to cancer reading-- such as stroma and cytoplasm \cite{chan2014wonderful, fischer2008hematoxylin}. 

After staining the biopsy cores from glass slides, pathologists analyze the revealed nuclei cell structure microscopically to study the cellular morphology for cancer diagnosis. However, staining is a chemical procedure subject to high variations among different laboratories and organ tissues \cite{tosta2019computational}. Moreover, different microscope scanners are tuned under different parameters that cause more variation in the process. Hence, with regard to the automated nuclei segmentation process, noise can be induced at different stages before image acquisition. This is a challenge from existing systems that need to cope with the noise and defects that occur during the acquisition process.

\subsection{Staining Procedures And Color Variations}

The procedure for staining a histological specimen requires multiple steps, until the slide is ready for examination under the microscope. The concrete steps where color artifacts can occur are: (1) collection, (2) fixation, (3) dehydration and clearing, (4) paraffin embedding, (5) microtomy, (6) staining, and (7) mounting \cite{hiatt2007tratado}. 

\begin{figure*}[t]
\begin{center}
\includegraphics[width=0.5\linewidth]{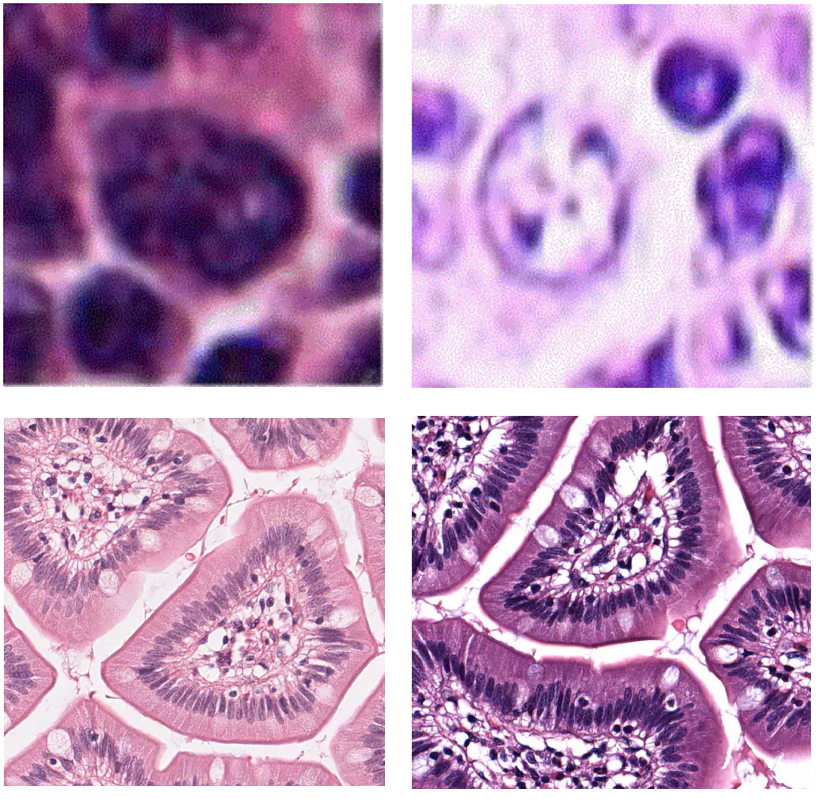}
\caption{Top row: different slice thickness results in very different nuclei color and appearance. Bottom row: Under-staining (left) and over staining (right) can change nuclei and background color \cite{michail2014detection, janowczyk2017stain}} \label{fig:color_variations}
\end{center}
\end{figure*}

Concretely, fixation time can vary the colorization results. Also, imperfect dehydration can leave behind water drops that obscure certain slide regions under the microscope. The thickness of slices after microtomy also affects a lot the nuclei appearance, as thinner slices may provide more detail to nuclei \cite{michail2014detection}. Finally, staining process entails chemical solutions and thus, different factors (i.e. staining time, solution pH etc.) can affect the tissue appearance \cite{janowczyk2017stain}. Image artifacts can be also caused during mounting the stained core onto the coverslip (e.g. bubbles or dust). Fig.~\ref{fig:color_variations} visualizes some common color artifacts during staining.

\begin{figure*}[t]
\begin{center}
\includegraphics[width=0.6\linewidth]{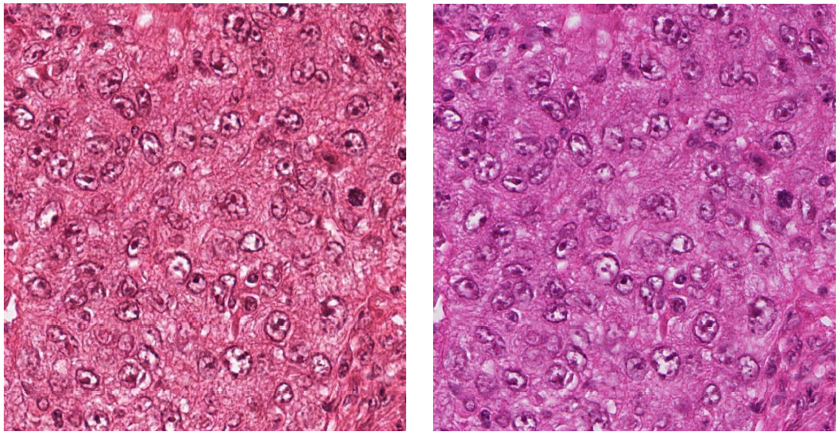}
\caption{Same tissue sample digitized under two different scanners. On the left, whole-slide image is acquired using Aperio XT, while the right one using Hamamatsu scanner. The large color and texture variation across different devices is a challenge for nuclei segmentation \cite{tosta2019computational}.} \label{fig:scanner_variations}
\end{center}
\end{figure*}

\subsection{WSI Scanners \& Image Digitization }

The next critical step after staining is the image acquisition at the microscopical level. The type of lens and magnification level parameter, camera chip type, as well as the illumination system within scanner, can largely affect the image output for the same stained tissue. There are a couple of scanners in the market for digitizing WSI. Hamamatsu, Aperio XT, Olympus, Philips, Huron, Leica, and others are some vendors selling digital scanners for pathological slides. In Fig.~\ref{fig:scanner_variations}, one can realize the color shift across different scanners. Moreover, the device parameterization is not standardized, and thus the digitized image may vary significantly across different laboratories using different scanner parameter adjustments. 

\subsection{WSI Variability Implications And Challenges}

Nuclei color and texture can be affected due to a number of imperfections. Artifacts and noise that may occur during the staining process, as well as the image digitization process (e.g., scanner's device systematic and random noise) can alter the color and nuclei texture. Unclear nuclei boundaries, overlapped nuclei, and variations in their color and texture are quite challenging for a generic nuclei segmentation pipeline. Mitosis also can look quite different on the digitized image due to staining and scanner variations \cite{bertram2019large, aubreville2021quantifying}.

Different pre-processing methods have been proposed in the past to normalize the color components \cite{roy2018study,vijh2021new} and mitigate the large color and nuclei appearance variations. Yet, today it still remains a challenge and the main reason modern AI nuclei segmentation models lack good generalization ability. The previously mentioned challenges due to the large variations during WSI acquisition are a subject for future research and the area of where different proposed methods contribute in handling certain aspects of the nuclei segmentation problem.

\section{Methods for Nuclei Segmentation}\label{sec:method}

\begin{figure*}[t]
\begin{center}
\includegraphics[width=1.0\linewidth]{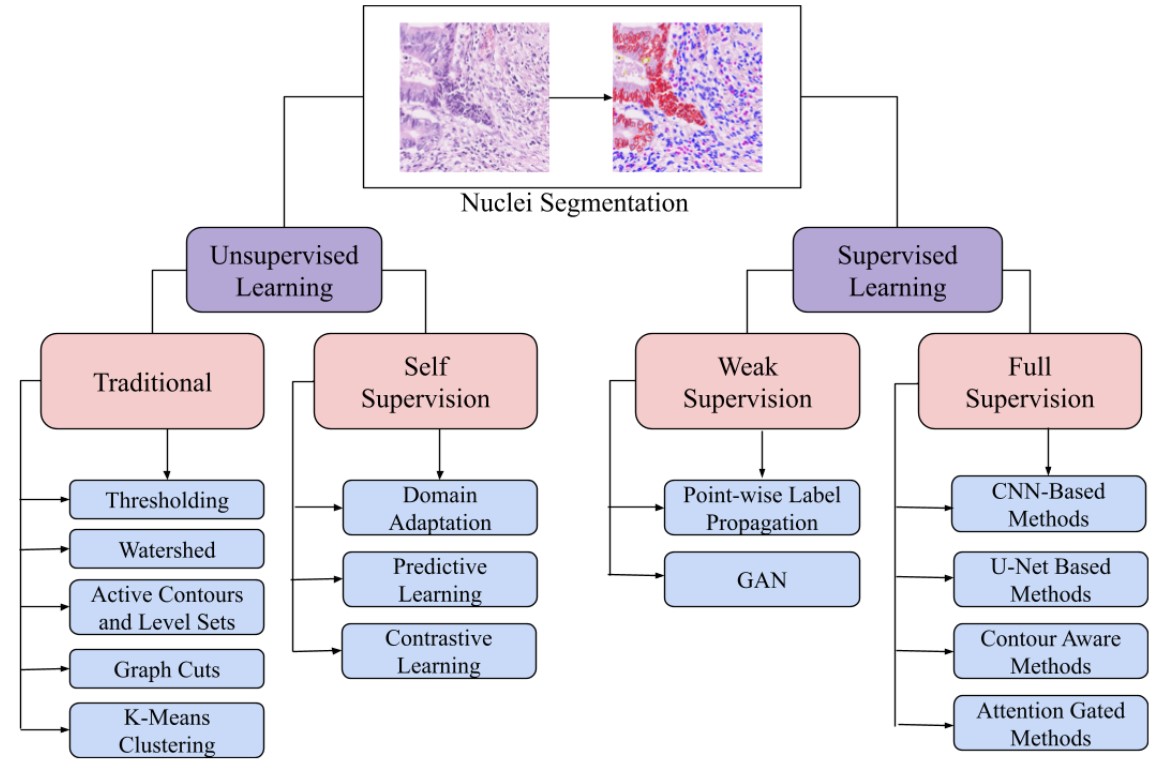}
\caption{Outline of the existing nuclei segmentation methods.} \label{fig:methods}
\end{center}
\end{figure*}

This section outlines the spectrum of nuclei segmentation methods as shown in Fig. \ref{fig:methods} and presents a comprehensive review of significant approaches. They are broadly classified into unsupervised and supervised learning methods. 

\subsection{Unsupervised}

Unsupervised methods perform segmentation without the help of any annotations. These methods are divided into two categories: traditional methods, which do not use any learning, and self-supervised learning-based approaches. 

\subsubsection{Traditional Methods}

Traditional nuclei segmentation methods were predominantly adopted before the deep learning era. They focused on applying different image processing techniques to obtain reasonable segmentation maps. A summary of the unsupervised traditional methods is listed in Table \ref{table:traditional_methods}.


\begin{table}
\small
\centering
\caption{A summary of traditional methods for nuclei segmentation.}\label{table:traditional_methods}

\begin{tabular}{p{0.05\textwidth}p{0.2\textwidth}p{0.2\textwidth}p{0.2\textwidth}p{0.2\textwidth}}

\toprule
\textbf {Ref.} &  \textbf {Dataset} &  \textbf {Methods} &  \textbf {Pre-Processing} &  \textbf {Post-Processing} \\ 
\toprule

\cite{gurcan2006image} & 20 slides of neuroblasts & Hysteresis Thresholding & Color space decomposition & Hole filling, smoothing, removal of false positives, watershed
    \\ 
\midrule
\cite{lu2012robust} & 30 cutaneous H\&E stained images &Local Region Adaptive Thresholding & Hybrid Morphological Reconstructions & Opening  
 \\ 
\midrule
\cite{phoulady2016nucleus} & Gold Standard Dataset & Hierarchical multilevel thresholding & Color deconvolution, opening & Dilation
 \\ 
\midrule
\cite{gautam2016automatic} & 20 leuokocyte images & Otsu’s threhsolding 
& Contrast stretching, histogram equalization  & Closing
 \\ 
\midrule
\cite{win2017automated} & 30 cytology pleural fluid images & Otsu’s threhsolding  & Median filtering, conversion into LAB color space & Opening
 \\ 
\midrule
\cite{magoulianitis2022unsupervised} & MoNuSeg & Adaptive Thresholding
& Data Driven Color Transform & Convex hull algorithm, nuclei area priors based thresholding, hole filling 
 \\ 
\midrule
 \cite{magoulianitis2022hunis}& MoNuSeg & Local Modified Adaptive Thresholding, Self supervised classification for uncertain pixels & H component extraction, Data Driven Color Transform & Convex hull algorithm, morphological operations
 \\ 
\midrule
\cite{veta2011marker} & 19 H\&E stained breast cancer images & Radial Symmetry Transform, Marker Controlled Watershed & Color deconvolution, morphological filtering & Size and solidity based refinement, ellipse approximation
 \\ 
\midrule 
\cite{vahadane2013towards}& 119 H\&E breast, gastrointestinal, and Feulgen prostate images & Seeded watershed based on image driven markers & Gaussian Smoothing, Morphological Operations & Morphological Operations 
\\ 
\midrule
\cite{shu2013segmenting} & 52 DAB stained colorectal cancer images & Region growing based seeded watershed & Global and local thresholding for foreground extraction & Intensity based auto thresholding, ellipse fitting
\\ 
\midrule
\cite{cui2016self} & Custom breast cancer H\&E dataset & Skeleton model based marker controlled watershed & Color deconvolution, Otsu’s thresholding, morphological operations
& Size based false positive removal, morphological operations
\\ 
\midrule
\cite{koyuncu2016iterative} & 34 fluorescence microscopy hepatocellular carcinoma images & Iterative marker controlled watershed  & Gradient map and distance transform of binary map & -
\\ 
\midrule
 \cite{rajyalakshmi2017supervised} & 120 H\&E breast cancer images & Circular Hough Transform based modified marker controlled watershed & Denoising, CLAHE, Morphological operations &-
\\ 
\midrule
\cite{hu2004automated}& H\&E esophageal images & Improved active contour with growing energy & Iterative dual thresholding, ultimate erosion & -
\\ 
\midrule
\cite{fatakdawala2010expectation} & 100 H\&E breast cancer images & Geodesic Active Contour & Expectation-Minimization algorithm
& Overlap resolution
\\ 
\midrule
\cite{faridi2016automatic} & 20 H\&E breast cancer images & DoG filtering and  thresholding followed by level set & Bilateral filtering, Gamma correction, morphological operations & -
\\ 
\midrule
\cite{beevi2016automatic}& MITOS dataset & Localized, Region Based Level Set  & Stain normalization, color deconvolution, filtering & -
\\ 
\midrule
\cite{rashmi2021multi} & KMC, BreakHis datasets (breast cancer images) & Modified Chan-Vese Model using multi channel color data & Color normalization, color channel selection & Morphological operations, area based false positive removal
\\ 
\midrule
\cite{7797086} & 45 synthetic cervical cytology images & Fuzzy C means clustering with spatial shape constraint & Complement-ing, histogram based binarization & False positive removal using area and shape priors, Closing\\
\bottomrule
\end{tabular}
\end{table}



\paragraph{Thresholding}

\begin{figure*}[htb]
\begin{center}
\includegraphics[width=1.0\linewidth]{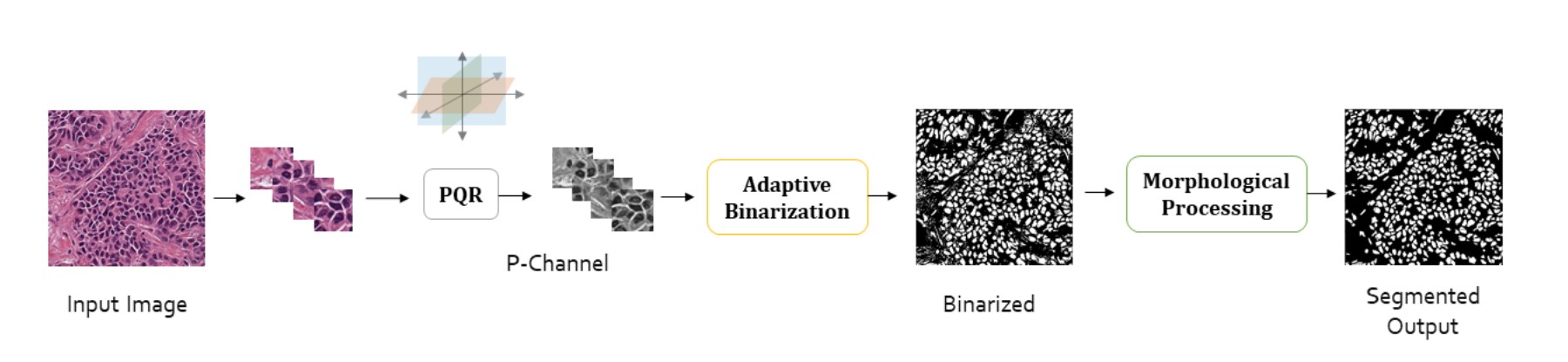}
\caption{ An example thresholding pipeline from \cite{magoulianitis2022unsupervised}.} \label{fig:thresholding}
\end{center}
\end{figure*}

Thresholding is one of the fundamental traditional segmentation algorithms. Different thresholding methods vary in how the threshold value for segmentation is computed. One widely adopted and automatic thresholding algorithm for bimodal images is Otsu's thresholding \cite{otsu1979threshold, cai2014new, phoulady2016nucleus, gautam2016automatic, win2017automated}. Here, the image histogram is separated into two clusters based on a threshold, decided either by minimizing the intra-class variance or maximizing the inter-class variance. Other approaches use global \cite{gurcan2006image} or local thresholding \cite{lu2012robust,magoulianitis2022unsupervised, magoulianitis2022hunis} in addition to morphological operations to refine the segmentation maps as in Fig. \ref{fig:thresholding}.

Gurcan {\em et al.}~\cite{gurcan2006image} proposed hysteresis thresholding based on morphological operations. They utilize the R component (from RGB) due to its high contrast and apply Top Hat transform to detect the high-intensity regions from the morphologically reconstructed complementary R component. Hysteresis thresholding is then used to remove tiny high-intensity regions near actual nuclei. It was observed that a few weakly stained nuclei were not detected through this method, thus reducing the true positive rate. Such global thresholding methods fail to account for the variations in staining intensities across the image and within the nuclei. To address this issue, local or regional thresholding was proposed. Lu {\em et al.}~\cite{lu2012robust} incorporated a two-module approach, with the first module performing Hybrid Morphological Reconstructions on the complementary image to reduce noise and intensity variations. The second module applied a local, regional adaptive threshold, classifying all pixels with intensity lower than the mean intensity of each block as nuclei. A refinement phase is followed, leveraging nuclei's elliptical shape and size distribution to correct under-segmentation issues due to local intensity variations. The opening operation was performed as the final step to remove ghost nuclei and smoothen the segmentation map. However, few nuclei with significant intensity variations are missed in this process. 

Cai {\em et al.}~\cite{cai2014new} propose an iterative thresholding method using Otsu's threshold and classify the pixels into three classes. The first iteration applies Otsu's algorithm to determine the threshold and the means of the two classes the threshold divides them into. These two means help classify the pixels into three different categories, with the nuclei being pixels with intensities greater than the largest mean and the background with pixel intensities lower than the smallest mean. The pixels with intensities between these two means are considered a separate class and subject to the same procedure, classifying the pixels into three classes again. This procedure is repeated on the third class of pixels between the two class means until a preset condition is reached when Otsu's threshold divides them into the nuclei and background. The nuclei pixels identified from each iteration are combined in a logical union, and a similar operation is performed on the background pixels to obtain the segmentation map. This approach is based on thresholding and helps recall some fine and weakly stained nuclei, which the original Otsu's algorithm may miss. 

Using the Beer-Lambert Law, Phoulady {\em et al.}~\cite{phoulady2016nucleus} first extract the Hematoxylin component from the H\&E stained image. They propose an iterative multilevel thresholding scheme, with each threshold determined using Otsu's method, separating the pixel intensities into several classes. The regions created in the initial stage are either shrunk or split into two or more smaller areas in the further steps. Morphological operations were then performed to remove any artifacts and improve the accuracy of the nuclei boundaries. Gautam {\em et al.}~\cite{gautam2016automatic} propose a similar scheme, where the H\&E stained image is first converted into grayscale, and a copy is made. One copy is histogram equalized, and the other copy is contrast stretched. Addition and subtraction are performed on these two copies resulting in minimum distortion in the nuclei. Otsu's thresholding is then applied to the entire preprocessed image, followed by the closing operation (combination of dilation and erosion) to fill in holes and remove false nuclei. Win {\em et al.}~\cite{win2017automated} apply the median filter to each component R, G, and B to remove noise in the image. They then convert the images into the LAB color space due to the dependency of the R, G, and B components on each other. Otsu's thresholding is performed on the grayscale adjusted and equalized image to binarize the image. The binarized image is then subject to the morphological opening operation to remove false nuclei. However, this method fails to segment overlapping or clustered nuclei. 

\begin{figure*}[htb]
\begin{center}
\includegraphics[width=1.0\linewidth]{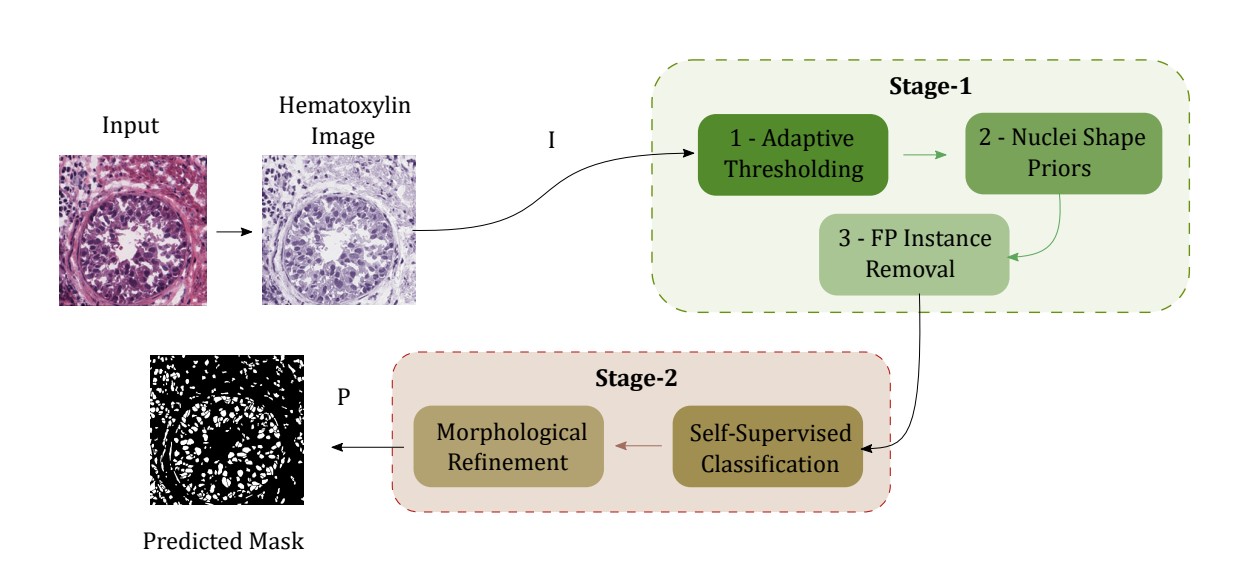}
\caption{ High Performance Unsupervised Nuclei Instance Segmentation (HUNIS) pipeline from \cite{magoulianitis2022hunis}.} \label{fig:HUNIS}
\end{center}
\end{figure*}

In the past year, Magoulianitis {\em et al.}~\cite{magoulianitis2022unsupervised} proposed a pipeline consisting of three modules: (1) Color transform, (2) Binarization, and (3) Morphological Processing called the CBM pipeline. They first split the image into 50x50-sized blocks comparable to the size of nuclei in the H\&E strained images. The strong correlation between the R, G, and B channels of an RGB image is exploited to reduce data dimensions. PCA is performed on the RGB image to minimize the three attributes to a single high energy attribute for further processing. A histogram of the normalized energy values from this attribute is plotted to visualize the distribution of values within each block. This histogram displays a bimodal distribution, the first peak representing the nuclei and the second peak representing the background. The valley between the two modalities may be used to determine the threshold to classify the pixels as nuclei or background. If this bimodal assumption does not hold, and there are more than two peaks in the histogram, or one peak is more prominent, the block is reduced to four smaller blocks or four blocks are merged to form one large block, respectively. Thresholding is applied individually to these new reduced or combined blocks. In the final stage, large connected nuclei are split using the convex hull algorithm, nuclei size priors are used to eliminate small erroneous nuclei, and false negatives in the image are removed using the hole filling filters. A follow-up work of CBM, namely HUNIS~\cite{magoulianitis2022hunis}, proposes a two stage approach, where stage-1 creates an initial segmentation output, and then stage-2 uses the output to train a pixel-wise classifier in a self-supervised manner (See Fig. \ref{fig:HUNIS}). In stage-1, PCA is performed on the Hematoxylin component extracted from the H\&E stained image to obtain a monochrome attribute to work with. The adaptive thresholding algorithm is slightly modified by considering two cases. When one peak in the histogram of the block is more prominent than the other, or there are more than two peaks in the image, the threshold for the block is adjusted adaptively based on the magnitude and direction of the slope of the line connecting the two peaks in the histogram for more precise segmentation. Unusually small nuclei are identified from a nuclei size distribution and eliminated. This is followed by using shape priors and morphological processing to split large nuclei. A false positive reduction module works on a larger tile to capture more nuclei instances. A global nuclei size threshold is used to group the instances of reasonable size as ground truth and the remaining as another set. Each element in the latter set is compared with the former set, and elements with a very low similarity are considered false positives and eliminated. The procedure is extended to a second stage, training a pixel-wise classifier on the pseudo labels obtained in the first stage. Utilizing the confidence scores from the classifier, pixels with high uncertainties which lie close to the boundary are reclassified into their correct classes. Final morphological processing steps are included to refine the segmentation maps further. 

\begin{figure*}[t]
\begin{center}
\includegraphics[width=0.55\linewidth]{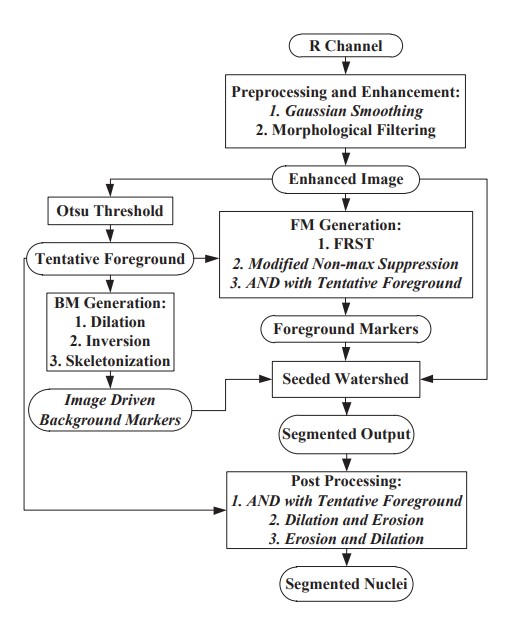}
\caption{ An example marker controlled watershed method from \cite{vahadane2013towards} in conjunction with thresholding.} \label{fig:watershed}
\end{center}
\end{figure*}

Some slight modifications in applying the thresholding help alleviate the errors due to staining variations. However, thresholding on its own needs help with segmenting overlapping or clustered nuclei. Several approaches combined thresholding with other methods to account for the clustered and overlapping nuclei, as in Fig. \ref{fig:watershed}.

\paragraph{Watershed}

The watershed algorithm \cite{roerdink2000watershed} employs topological information to segment an image into different regions of what are called catchment basins. The original version of the algorithm starts with finding the local minima in the image as the centers(seeds) of the catchment basins. Different colors are then flooded, beginning from the minima markers until they reach the boundaries of each catchment basin to form the watershed lines. The boundaries of each region distinguish one part from the other, resulting in a segmented image.  

When applied to nuclei segmentation, this process sometimes resulted in over-segmentation, meaning single objects were segmented into several regions, or under-segmentation, where multiple regions were combined into a single region. Marker-controlled approaches with novel marker selection methods were developed to overcome this difficulty. An example method using a marker-controlled watershed is illustrated in Fig. \ref{fig:watershed}.

Veta {\em et al.} proposed a marker-controlled watershed segmentation \cite{veta2011marker} using the Fast Radial Symmetry Transform (FRST). To remove irrelevant structures, color deconvolution (to extract the Hematoxylin component) and morphological operations are performed. FRST is used to detect the foreground and background markers based on the radial shape assumption for nuclei. Watershed segmentation is then applied using these regional markers. Regions of low solidity and size are removed to refine the segmentation further. Finally, they use the ellipse approximation to generate regular contours. This study observed that the FRST-based markers reduce the oversegmentation observed with regional minima markers. However, one drawback of this approach is the selection of background markers as everything around the nucleus of a specific area. This may not be true for all nuclei resulting in few errors. To overcome this drawback, Vahadane {\em et al.}~\cite{vahadane2013towards} suggested a background marker generation method. The image is first preprocessed using Gaussian smoothening and morphological processing to remove noise and enhance the foreground and background while preserving the edges. To obtain the background markers, the enhanced image is thresholded using Otsu's method, generating a tentative foreground, followed by inversion, dilation, and skeletonization. The nuclei markers are generated using FRST and refined using the tentative foreground. The nuclei and background markers thus obtained are employed in the watershed segmentation. Post processing through morphological processes reduces some false positives and splits connected nuclei. 

Shu {\em et al.}~\cite{shu2013segmenting} proposed a method combining thresholding and marker controlled watershed segmentation with a two step approach to seed detection. Since thresholding often misclassifies nuclei boundaries, they generate two masks, one with global thresholding and another with global and local thresholding. For the first step of seed detection, the latter is converted into a Euclidean Distance Map (EDM), and seeds for watershed are determined through Ultimate Eroded Points (UEP). In the second step, to account for the weakly stained nuclei, the remaining particles from this mask are considered seeds for a region growing process based on the global thresholded mask. This generates "necks," which helps watershed merge oversegmented regions and separate clustered nuclei. Seeds for watershed are obtained from the EDM, and segmentation is performed. Post processing consists of another round of local auto thresholding and ellipse fitting to refine the maps. The requirement of some empirically determined parameters restricts it to good performance on a single tissue type. 
To eliminate the requirement of empirical parameter determination, Cui {\em et al.} \cite{cui2016self} apply an ellipse detection algorithm to estimate nuclei size. Otsu's thresholding and morphological operations are performed on the Hematoxylin component before ellipse detection. From this binary image, connected components are identified. Based on the estimated nuclei sizes, these components are classified as noise, single nuclei, or multi nuclei regions. Skeletonization is performed on the multi-nuclei region to identify seeds for watershed segmentation. Once individual nuclei are identified, they are then post processed along with the components classified as single nuclei regions. However, this approach tends to mark some large nuclei as a multi nuclei region, causing oversegmentation. 

Seeds determined from regional minima approaches often include spurious markers in the form of noise which may degrade the performance of the segmentation algorithm. The H-minima transform is applied to suppress minima below a value 'h.' Selecting this value is critical, as low values may lead to oversegmentation, and higher values may cause under segmentation. Koyuncu {\em et al.}~\cite{koyuncu2016iterative} proposed an iterative H-minima based method for efficient selection of h. They first generate a gradient map and distance map from the image. The h value is varied with each iteration to generate markers from the gradient map, and regions with an area less than a specified threshold are eliminated. The markers obtained from each iteration are then combined, and the marker controlled watershed is used to grow regions identified from these markers on the distance map. To prevent oversegmentation, the region growing process is constrained to pixels that have not been previously classified as background or another nucleus. This process improves the segmentation of non-circular nuclei. 

Rajyalakshmi {\em et al.} \cite{rajyalakshmi2017supervised} propose a modified marker-controlled watershed. The H\&E stained image is preprocessed by applying Contrast Limited Adaptive Histogram Equalization (CLAHE) and morphological processes to remove noise. Local maximum points are identified with the help of Corner Detection Techniques, which are then used to detect nuclei circles using the Circular Hough Transform. Otsu's thresholding is applied to remove false positives. To eliminate certain dark and bright regions, a map of variable size structured elements is constructed and applied to the thresholded image. Marker-controlled watershed is applied to the resulting image to extract the boundaries of overlapped nuclei. 

\paragraph{Active Contours \& Level Sets}\label{paragraph:AC & LS}

Active contours, also known as snakes, start with an initial contour of points and evolve to fit the points on the object by energy minimization. The initial contour is often generated from a representation of parameters or a formulation. The contours evolve iteratively until a potential minimum energy boundary is obtained.

However, this method is highly sensitive to initial contour placement. One of the early implementations of the snakes algorithm by Hu {\em et al.}~\cite{hu2004automated} detects the nuclei centers using a dual threshold algorithm followed by ultimate erosion. To overcome the issue of initial contour placement, they propose an improved snake energy minimization function by adding a region similarity based growing energy function. This algorithm also restricts the movement of the contour along radial directions, which reduces the computational time and broadens the boundary attraction range. 

\begin{figure*}[t]
\begin{center}
\includegraphics[width=0.75\linewidth]{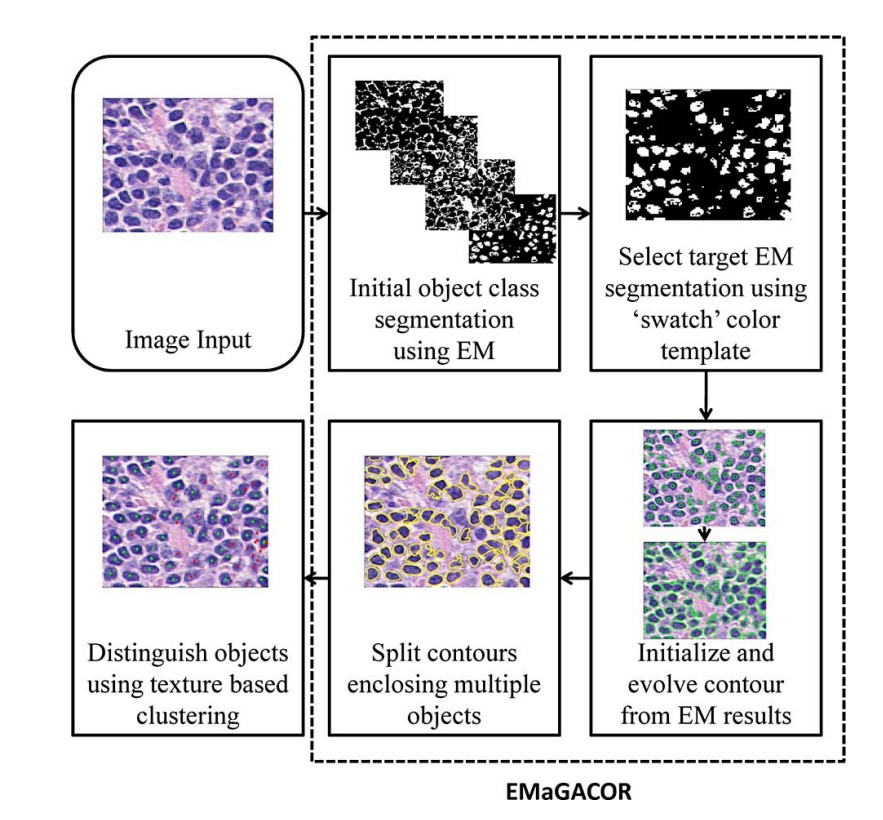}
\caption{An example active contour based pipeline from \cite{fatakdawala2010expectation}.} \label{fig:active_contour}
\end{center}
\end{figure*}

Active contours are also limited in their ability to segment overlapping objects. Fatakdawala {\em et al.}~\cite{fatakdawala2010expectation} employ an expectation-minimization (EM) algorithm with four classes to initialize a geodesic active contour as depicted in Fig. \ref{fig:active_contour}. The EM step generates an initial segmentation map, which helps reduce the impact of dataset variability. A swatch color template selects the target initial segmentation map representing the nuclei. A Magnetostatic Active Contour model is applied to the chosen map using a bidirectional force to evolve the nuclei boundaries. A final step to resolve overlap between nuclei contours is implemented by identifying points of high concavity in multi nuclei regions, followed by an edge path algorithm to split the contours using the edge information and a size heuristic. Nonetheless, good performance from this approach is subject to well-stained and low noise samples, as the EM algorithm depends on the R, G, and B values. 

Level sets are an alternative mathematical implementation of Active contours, in which the boundary is viewed as the level set function $ \phi = 0 $ or the zero-level-set function. Starting from an initial contour represented by a level set function, this contour is evolved to fit the object's boundary by constructing a zero level set function. 

Faridi {\em et al.}~\cite{faridi2016automatic} proposed a level set based technique to detect and segment cancerous nuclei. A bilateral filter is applied to the image to smoothen the image while preserving the edges. The green channel indicates a higher probability of cancerous nuclei, hence chosen for further processing. Gamma correction is applied to this channel and thresholded to obtain a binary image. The difference of Gaussian (DoG) filter is applied to the morphologically processed image from the previous step and thresholded to obtain nuclei regions. Initial contours for the level set algorithm are generated by dilating the detected nuclei centers (regions). Nuclei with smooth contours are obtained as an output of the level set algorithm. False positives may still occur due to staining variations, and not all critical cancerous nuclei may be detected. 

Beevi {\em et al.}~\cite{beevi2016automatic} implemented an approach combining the Krill Herd Algorithm (KHA) based multilevel thresholding and localized Level Set. Upon stain normalization, the R component of the image is chosen for further processing due to the high contrast between nuclei and background. A Weiner filter is applied to the image to enhance the weakly stained nuclei and edges. Initial contours are obtained by employing the KHA optimized multilevel thresholding, where out of the three thresholds obtained, the lower threshold values are detected as nuclei. Localized level set algorithm works on the principle of maximizing the mean intensity difference between the foreground and background along the contour. In addition, energy minimization is done by region based techniques and local information accounting for the intensity variations. KHA exhibited fast convergence, eliminated the risk of oversegmentation, and improved the segmentation of overlapping and touching nuclei. 

The Chan-Vese model~\cite{chan2001active} for image segmentation based on the Level Set implementation has been a widely adopted method. Instead of the regular image gradient based stopping criterion in regular active contours, this model applies a stopping criterion based on the Mumford-Shah segmentation algorithm~\cite{mumford1989optimal}, thus detecting even irregular boundaries. 

Rashmi {\em et al.}~\cite{rashmi2021multi}, apply a multichannel Chan-Vase model to perform nuclei segmentation. The Green Channel (from RGB color space) and inverted S Channel (from the HSI color space) are obtained from the color normalized H\&E stained images due to their ability to distinguish weakly stained nuclei. The energy function in this Chan-Vese implementation utilizes both channels to get the zero level set contour. The output of this step is postprocessed by filling the holes using the closing operation and thresholded to remove false positives. The complementary information supplied by the Green and Saturation Channels contributes to improving the segmentation performance. 

\paragraph{Graph Cuts}

In graph-based approaches, each pixel is considered a node in a graph, and each edge is weighted based on the degree of similarity between its connecting nodes. A cut in the graph partitions the graph into two disjoint sets of nodes. The best cut will have minimum cost or energy. 

\begin{figure*}[t]
\begin{center}
\includegraphics[width=0.95\linewidth]{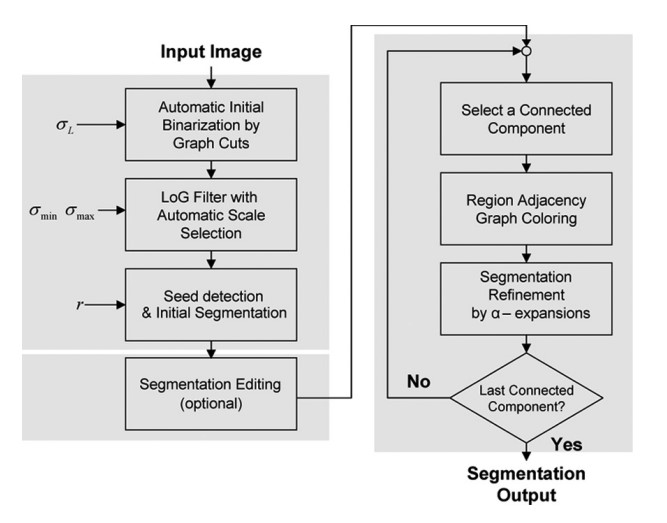}
\caption{ An example graph cuts based pipeline from \cite{al2009improved}.} \label{fig:graph_cuts}
\end{center}
\end{figure*}

Al-Kofahi {\em et al.}~\cite{al2009improved} proposed graph based methods for initial binarization and further refinement. The normalized image histogram is computed and a minimum error thresholding based on Poisson distribution is performed before applying the fast maxflow Graph cut algorithm to obtain initial segmentation. To detect seeds, the scale normalized Laplacian of Gaussian (LoG) filter response at multiple scales is computed. To overcome undersegmentation, maximum scale values are constrained using the Euclidean Distance map. They use these seeds as nuclei centers in a local maximum clustering algorithm, and foreground pixels are assigned to these centers forming clusters. A region adjacency graph coloring method to divide large clusters in the initial segmentation into smaller groups of nuclei precedes the $\alpha$-expansion (algorithm to obtain multi-way cuts in a graph) to delineate nuclei boundaries in clusters. This pipeline is illustrated in Fig. \ref{fig:graph_cuts}.

Dan{\v{e}}k {\em et al.}~\cite{danvek2009segmentation} proposed a two-stage graph cut model where the first stage distinguishes the foreground and background, and the second stage segments touching nuclei. Bilevel histogram analysis is employed to generate initial weights to the edges of the graph. The centers of the two peaks depicting the nuclei and the background are considered as thresholds. Weights of links of background voxels less than the threshold and foreground greater than the threshold are given the value $ \infty $. The rest of the voxels are not given any weight. The background segmentation is obtained by performing the two terminal graph cut algorithm. In the second stage, centers are identified from the nuclei clusters by calculating a distance transform inside the cluster and using the maxima transform to find the peaks. The graph weights in this stage are determined by the Euclidean distance from the nuclei centers to the current voxel. Since the standard maxflow algorithms may not be used with multiple nuclei, an iterative algorithm to find the best cut for label pairs is implemented. Strong gradients in the nuclei centers are ignored by including nuclei shape a priori information while performing graph cuts. 

Zhang {\em et al.}~\cite{zhang2014segmentation} use an adaptive and local graph cuts approach. Preprocessing involves converting the image from RGB to HSV space and extracting the V component, enhancing it via linear stretching and removing noise using a median filter. They employ an adaptive thresholding algorithm suggested by Sauvola {\em et al.}~\cite{sauvola2000adaptive} using textural and intensity information to detect approximate nucleus regions. Each of these regions is refined using a Poisson distribution based localized Graph Cuts with the help of boundary and regional information. This approach improves the performance in case of non uniform chromatin distribution and low contrast difference between nuclei and background. The segments with maximum overlap with the region obtained in the adaptive thresholding are retained, and an empirically determined condition on roundness reduces computational time by determining the need for further refinement. 

\paragraph{K-Means Clustering}

For K-Means Clustering Based approaches, a value K is selected as the number of clusters, and cluster centers are chosen randomly or heuristically. Each pixel is assigned a cluster label based on the minimum distance between the pixel and the cluster centers. New cluster centers are computed using the new labels. This process is repeated until it converges or no changes occur. 

Sarrafzadeh {\em et al.}~\cite{sarrafzadeh2015nucleus} proposed a K-means clustering algorithm integrated with a region growing mechanism to segment nuclei. They apply a median filter to each of the image's R, G, and B components to remove noise and preserve the edges. The filtered image is converted to the LAB color space to decouple the intensity and color bands. K-means clustering is applied to the a\* and b\* color spaces, creating three clusters based on Euclidean distance. The cluster with the minimum mean of RGB bands is detected as the nuclei. The opening and hole filling operations follow to smoothen boundaries, eliminate false segments, and complete segmented regions. To separate touching nuclei, connected components are identified as those regions with an area greater than the area of cells. With the green component being the most suitable for edge detection, the Sobel filter is applied to this component, and the detected edges are superimposed on the binary map obtained from the previous step. Seed points for region growing are determined by computing the center of mass after applying erosion on the edge included map. From these seeds, the regions are expanded or grown until there are two or more edge points in an 8-connected neighborhood of the central pixel. This results in splitting connected nuclei, giving distinct nuclei in the final map. 

\begin{figure*}[htb]
\begin{center}
\includegraphics[width=1.10\linewidth]{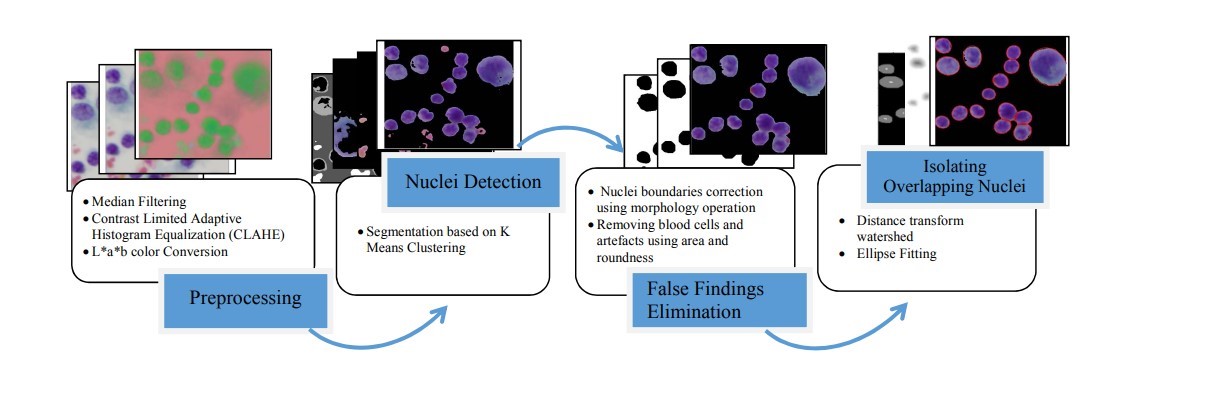}
\caption{ An example K-Means Clustering based pipeline from \cite{win2017k}.} \label{fig:k-means}
\end{center}
\end{figure*}

Win {\em et al.~\cite{win2017k}} proposed an approach on the same lines as \cite{sarrafzadeh2015nucleus}. To preprocess the images, in addition to median filtering, CLAHE is applied to enhance the contrast. They use the same K-means procedure with k as three, followed by morphological processing to improve the nuclei boundaries. Fig.~\ref{fig:k-means} illustrates the flow chart of the adopted method. These two methods differ in the procedure adopted to split overlapping nuclei. Distance transform is applied, where the value of each pixel is replaced by its distance from the closest background pixel. The seeds are assumed to be the darkest parts of each object, and the watershed is performed using these seeds. Finally, ellipse fitting is done to generate smooth contours.  

Chang {\em et al.}~\cite{chang2016quantitative} proposed K-means clustering using morphological features. They extracted features from H\&E stained images using the Gabor filters of different orientations and frequencies. The impulse responses from Gabor filters and other features, like intensities, are stacked to form an n-dimensional feature space. Each pixel in the image is mapped to a point in this feature space. The pixels are then enhanced using chosen features, and pixels with similar features are clustered using the K-means clustering algorithm. Cytological profile, including features like nuclei shape, size, intensities, etc., may be used to eliminate false positives and improve segmentation. 

Fuzzy C-means(FCM) clustering is an extended, robust clustering-based segmentation algorithm among all the fuzzy clustering methods. It offers the flexibility of allowing partial memberships in clusters. The standard algorithm utilizes only intensity information and hence is susceptible to artifacts. Including spatial information, however, tends to result in poor segmentation. Saha {\em et al.}~\cite{7797086} proposed a spatial shape constrained FCM to segment nuclei\. The input image is first complemented and binarized by subtracting the background using the second peak from the image histogram as the threshold. A fuzzy partition matrix is initialized before applying the spatial shape constraints. The circular shape function (CSF) is defined from seeds identified using an adjacency graph. In each connected component, nodes with maximum intensity are considered seed points. CSF is calculated for each pixel in the image as a function of its spatial coordinates with respect to other seed points. This value is then used to modify the pixel's intensity, thus moderating which target cluster the pixel belongs to. This CSF based FCM procedure is repeated until convergence. CSF thus helps in differentiating pixels in spatially different locations with similar intensities. Features like area, eccentricity, and circularity were used to remove irrelevant areas from the output, followed by morphological processing to smoothen boundaries. 

\subsubsection{Self-Supervised}

Although results from deep learning based methods were favorable, the requirement of large amounts of data and their annotation efforts are a matter of concern. This led to the development of self-supervised learning based techniques. This class of methods is categorized under unsupervised learning, considering these methods require no labels. All approaches in this section have a two-stage pipeline consisting of a pretext task(pre-training) and a downstream task(fine-tuning). They are further classified into Domain Adaptation, Predictive Learning, and Contrastive Learning. Table \ref{table:self_supervised_methods} summarizes the self-supervised methods. 


\begin{table}
\small
\centering
\caption{A summary of self-supervised nuclei segmentation methods.} \label{table:self_supervised_methods}

\begin{tabular}{p{0.02\textwidth}p{0.2\textwidth}p{0.2\textwidth}p{0.2\textwidth}p{0.2\textwidth}}
\toprule
    \textbf {Ref.} &  \textbf {Dataset} &  \textbf {Methods} &  \textbf {Pre-Processing} &  \textbf {Post-Processing} \\ 
\toprule

\cite{hsu2021darcnn} & COCO, BBBC, Kumar, TNBC & Mask RCNN based domain adaptation using pseudo labeling & DARCNN pretrained with source dataset & -
\\

\midrule
\cite{zheng2018fast} &400 single WBC images split into two datasets of 300 and 100 images &Unsupervised initial segmentation with K-means followed by supervised refinement using SVM classifier &Conversion to HSI color space &-
\\ 
\midrule
\cite{sahasrabudhe2020self} &MoNuSeg &Attention network for scale identification with segmentation maps as auxiliary outputs &Tile extraction, stain normalization &Opening, closing, distance transform 
\\ 
\midrule
\cite{ali2022multi} & MoNuSeg, TNBC & Multi scale representation based Self supervised Learning using U-Net & Cropping, resizing, ResNet-18 pretrained with zoomed in and zoomed out tiles & -
\\ 
\midrule
\cite{punn2022bt} &Kaggle DSB18, BUSIS, ISIC18, BraTS18 &Redundancy reduction based Barlow Twins U-Net &U-Net encoder pretrained with Barlow Twins approach (Siamese Net) &-
\\ 
\midrule
\cite{liu2020unsupervised} &BBBC039V1, Kumar, TNBC &Cycle Consistency Panoptic Domain Adaptive Mask RCNN &Normalization, random sample cropping, removal of samples with less than 3 objects, complementing &-
\\ 
\midrule
\cite{xie2020instance} & MoNuSeg & Contrastive Learning using Scalewise Triplet Loss and Count Ranking to pretrain U-Net encoder
& Anchor, positive and negative tile sampling & -
\\ 
\midrule
\cite{boserup2022efficient} &MoNuSeg, CoNSeP &Positive and Negative Patch based Contrastive Learning using FCN &- &-
\\ 

\bottomrule

\end{tabular}

\end{table}

\paragraph{Domain Adaptation}

\begin{figure*}[htb]
\begin{center}
\includegraphics[width=0.75\linewidth]{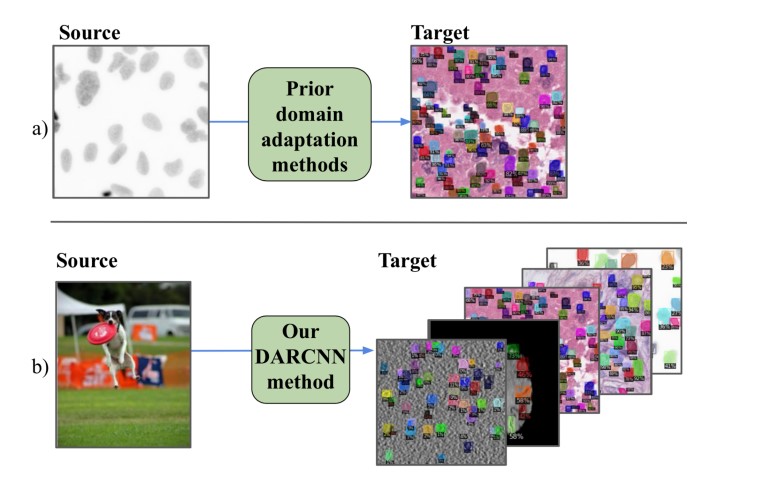}
\caption{ An example image from \cite{hsu2021darcnn} illustrating the idea of domain adaptation. The source data can vary from a common image set to labelled biomedical images. } \label{fig:domain_adapt}
\end{center}
\end{figure*}

Publicly available fully annotated biomedical datasets are only a few, but there a plenty of completely labeled general datasets. As illustrated in Fig. \ref{fig:domain_adapt}, domain adaptation takes advantage of these large volumes of labeled data, called the source, in the pre-training stage to learn features. The relevant biomedical datasets, called the target, are then used in the second stage to fine tune the performance of the model.

Hsu {\em et al.}~\cite{hsu2021darcnn} proposed a Domain Adaptive Region-based CNN (DARCNN) that learns object definition from a large annotated vision dataset COCO and adapts it to biomedical datasets. The pipeline consists of two stages of feature level adaptation and pseudo labelling at the image level. DARCNN utilizes source dataset weights for pretraining, and trains on combined batches of the source and target datasets. The two step framework of the Mask RCNN along with a domain separation module form the main structure of the DARCNN. The large domain shift between the general vision dataset to the biomedical dataset is handled by the domain separation module, which learns domain specific and domain invariant features that are input to the mask segmentation and regional proposal networks respectively. The domain specific features contain unconstrained embedding space in addition to information about the discriminability of the source and target domains. On the other hand, the domain invariant features contain information on objectness in the inputs from both domains. The loss function used in DARCNN consists of four losses: $ L_{sim} $ representing domain invariant features, $ L_{diff}$ representing domain specific features, $ L_{source}$ representing the Mask RCNN losses to train the source dataset, and $L_{target}$ representing the self supervised consistency loss. This approach gives space for background variation within the dataset by assuming an independent background consistency in each image. This is done with the help of the region proposal network, which minimizes the variations in the representation of the background. $L_{target}$, also known as the self supervised representation consistency loss, is responsible for this minimization. The output of the first stage gives a coarse segmentation map, which needs to be further refined to obtain image level supervised results. Image level supervision is achieved in the second stage of the DARCNN, which trains only the target branch on the pseudo labels with high confidence generated from the first stage. These pseudolabels are strengthened with the help of augmentations, accounting for variations in illumination and image quality. They also show the generalizability of DARCNN by adapting the model to three diverse biomedical image datasets. 

In \cite{liu2020unsupervised}, Liu {\em et al.} proposed a Cycle Consistent Panoptic Domain Adaptive Mask RCNN (CyC-PDAM). They choose fluorescence microscopy images as their source domain and synthesize H\&E stained images using CycleGAN. The fluorescence microscopy images are preprocessed to generate square patches of size 256x256. With CycleGAN on its own, the synthesized images appear to have some undesirable nuclei, that, in further tasks, tend to be marked as background. An auxiliary task of nuclei inpainting is presented to remove these unlabeled nuclei. From the synthesized image and its mask, an auxiliary mask with all the unlabelled nuclei is generated. Using this, a fast marching based nuclei inpainting is applied to replace these nuclei with unlabelled background pixels, thus eliminating all the undesirable nuclei. Mask RCNN, used as the baseline model, is built using ResNet-101 and a feature pyramid network(FPN). This Mask RCNN has domain bias in the semantic features, as it focuses mainly on local features and lacks a global view. To introduce this panoptic view, a semantic branch including a domain discriminator, is appended to the FPN. This branch in addition to the instance level segmentation branch help in reducing cross domain discrepancies and produces domain invariant features. Finally, to decrease the bias towards the source domain, a reweighted task specific loss is introduced. This network performs better than its fully supervised equivalent on unseen datasets, proving the efficacy of the domain invariant features that prevent the network from being influenced by dataset bias. 

\paragraph{Predictive Learning}

\begin{figure*}[t]
\begin{center}
\includegraphics[width=1.0\linewidth]{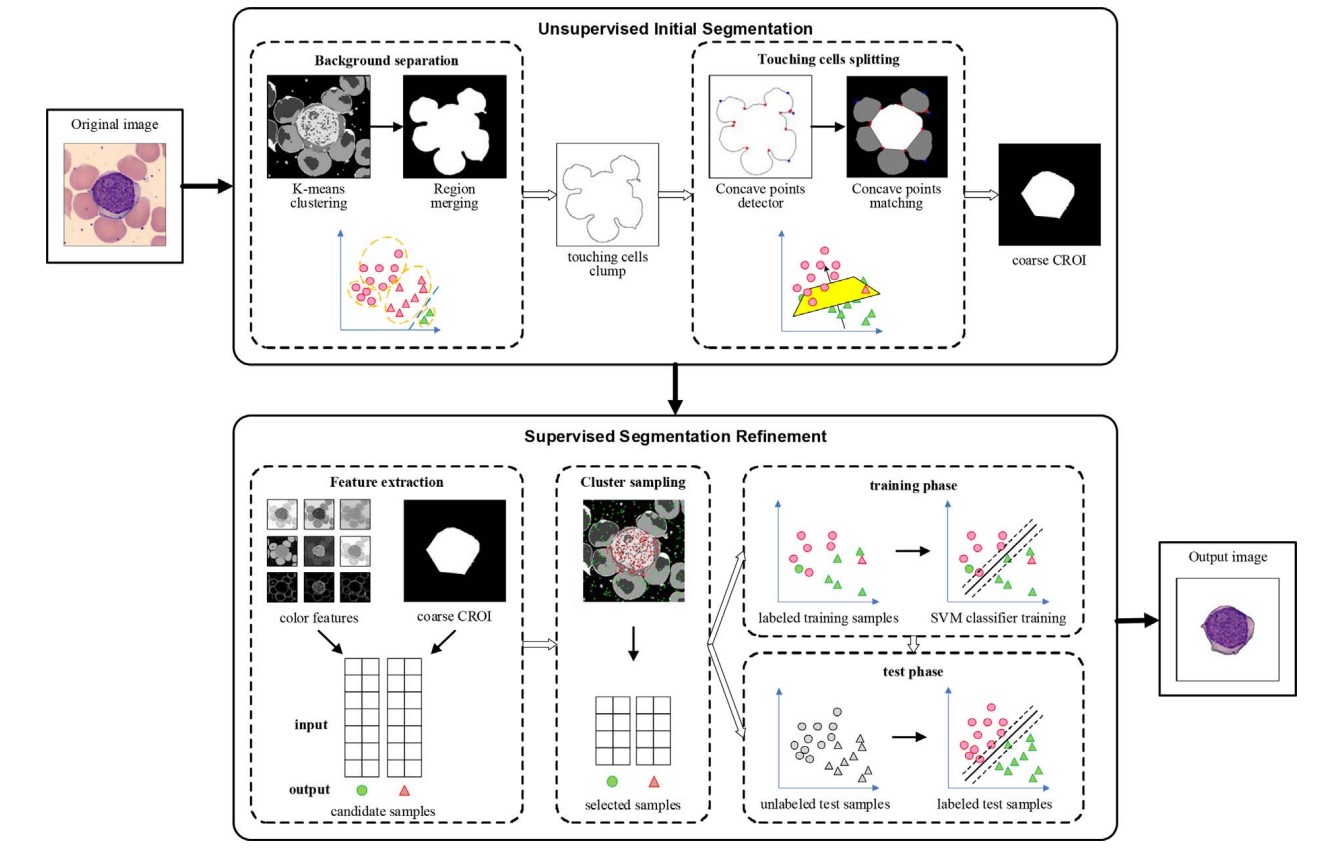}
\caption{ An example image from \cite{zheng2018fast} showing the use of pseudolabels generated from the first stage to train a classifier in the second stage. } \label{fig:predictive_learning}
\end{center}
\end{figure*}

This method uses a framework to generate pseudo labels for the training set that are further refined in the next stage. Such an example is shown in Fig.~\ref{fig:predictive_learning}. If a deep learning based model is used, the weights obtained from the pretraining stage are transferred to the second stage to finetune the segmentation model. 

Zheng {\em et al.}~\cite{zheng2018fast} proposed a self supervised method with an unsupervised initial segmentation to generate pseudo labels, which are later used in a supervised refinement phase. The first module in the pipeline performs background separation in the HSI color space with the help of K-means clustering. Several oversegmented regions in the border of the image are removed. Regions with colors similar to the removed regions are also eliminated. The remaining image is merged into the foreground leading to touching or overlapping clumps. In order to split these connected nuclei components, concave points are identified on the contour of the clumps. The clumps are then iteratively split by connecting pairs of concave points on the contour. The second module is responsible for fine tuning the results from the first module, by a supervised classification approach. Features like RGB colors, topological structure, and HSV based weak edge enhancement operator values are extracted for each pixel. To speed up the classifier training, a cluster sampling technique, selecting representative pixels from the oversegmented regions in the background separation step, is applied. The final step is to train an SVM classifier using the selected representative points. This trained classifier can then be used to classify the pixels of an image as the required region of interest or the background. 

Sahasrabudhe {\em et al.}~\cite{sahasrabudhe2020self} proposed a method based on the assumption that the magnification level of an image can be determined by the texture and size of nuclei. This approach shows that identifying the magnification or scale of the image acts as a self supervised signal for nuclei location. The required segmentation maps are obtained as an auxiliary output in this scale classification network. They use the concept that if a tile of nuclei can determine the magnification level, its element wise multiplication with an attention map representing the corresponding nuclei will also be able to determine the magnification level. A confidence map is generated using a fully convolutional feature extractor, which is then activated by a sigmoid function to generate the attention map. A sparsity regularizer is applied to this map to focus attention on the input patch. Now, the elementwise multiplication of this attention map and the original image tile is input to a scale classification network built out of ResNet-34, that gives the scale classification probability. The entire network is trained end to end, with the auxiliary output generating the nuclei segmentation map. A smoothness regularizer is applied to the attention maps to remove high frequency noise. In addition, to impose semantic consistency on the feature extractor, transformation equivariance is described by applying transformations like rotation, transpose, and flips. To obtain the final segmentation output, the attention maps are subject to opening and closing operations. Distance transform is computed from this image, and local maxima are identified to locate seeds for the marker controlled watershed that followed. From their observations, this model generalizes well on unseen organs. 

Based on a similar scale based approach, Ali {\em et al.}~\cite{ali2022multi} proposed a multi-scale self-supervised model. Small patches from the whole slide images are extracted, and each patch is again cropped and resized. These patches have images that are zoomed in or zoomed out. The initial self supervised stage uses a ResNet-18 model, trained to classify these pairs of patches as zoomed in or zoomed out. The second stage employs a U-Net architecture with a ResNet-18 encoder and a Feature Pyramid Network decoder. The U-Net model is trained with the weights transferred from the first stage to perform the actual segmentation task. This model was fine tuned using the Adam optimizer with the Cross Entropy Loss. The results of this approach support the effectiveness of transferred weights from the same domain, as opposed to domain adaptation from general datasets like ImageNet. 

Punn {\em et al.}~\cite{punn2022bt} proposed a self-supervised framework, known as the BT-UNet, employing the redundancy reduction based Barlow Twins approach to pretrain the encoder in the U-Net. Two distorted images are generated from an image by introducing distortions like cropping or rotation. The first stage comprises pretraining the U-Net encoder with the help of the Barlow Twins strategy, followed by a projection network to obtain encoded feature representations. The Barlow Twins approach uses a twin encoder and projector based Siamese net sharing similar weights and parameters. The feature maps generated by the encoder network lead to feature representations by passing through blocks of global average pooling, fully connected (FC) layers, ReLU activation, batch normalization, and a final FC layer. A cross correlation matrix is computed from these representations. The model is then refined to make the cross correlation matrix similar to an identity matrix with a Barlow Twins Loss function. In the second stage, the weights learned by the twin encoders are transferred to the U-Net model initializing the encoder, while the decoder is initialized with default weights. A limited number of annotated samples are utilized to train this U-Net segmentation model, which uses the average of the binary cross entropy loss and dice coefficient loss as the loss function. 

\paragraph{Contrastive Learning}

Contrastive learning creates positive patches and negative patches from an image, and a model learns attributes by contrasting the patches against each other. This helps the model find the similarities and dissimilarities among the image patches. 

\begin{figure*}[t]
\begin{center}
\includegraphics[width=0.7\linewidth]{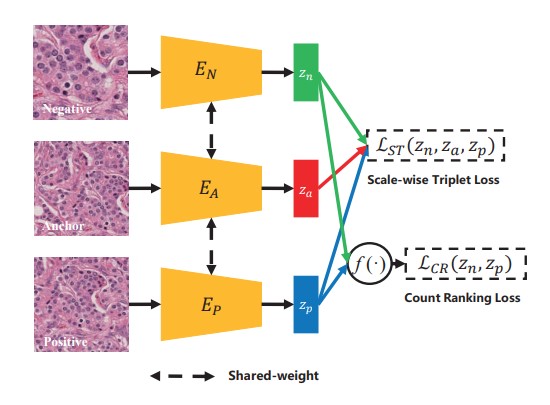}
\caption{ A contrastive learning example from \cite{xie2020instance} using three different patches generated from an image. } \label{fig:contrastive_learning}
\end{center}
\end{figure*}

Xie {\em et al.}~\cite{xie2020instance} proposed an instance-aware self supervised method involving scalewise triplet learning and count ranking. This implicitly helps the network learn the nuclei size and quantity information from the raw data. In triplet learning, three samples are generated from the original input image. A random sample of a specific dimension is cropped from the input image, called the anchor. Another sample of the same size is cropped from the same image, called the positive image. These two samples will have identical nuclei sizes as they are same-sized samples. To include nuclei size information, a negative patch, a sub-patch from the positive patch, is sampled and resized to the size of the anchor and positive images. This sub patch is randomly sampled from a set of three sizes to introduce diversity. The anchor, positive and negative samples form a triplet, used in this self supervised proxy task shown in Fig.~\ref{fig:contrastive_learning}. While these patches implicitly account for the nuclei size, they also account for nuclei quantity, as negative patches will always have a smaller number of nuclei than the positive and anchor patches. This introduces another metric, the pairwise count ranking for self supervised learning. The proxy task comprises three encoders with shared weights trained on the count ranking loss and scalewise triplet loss. These encoders aim to embed the features into a 128 dimensional feature space. Triplet learning focuses on narrowing the distance between samples with similarly sized nuclei and enhances the dissimilarity between the samples with differently sized nuclei. Count ranking enables the network to identify large crowds of nuclei. Fine tuning for the segmentation task is done using a U-Net with a three way classification, including the nuclei, nuclei boundary, and background. ResNet-101 is the backbone of the encoder and is pretrained with the proxy task. These weights are transferred to the U-Net encoder, while the decoder weights are randomly initialized. Joint training on the two proxy tasks appears to substantially improve the segmentation performance, compared to employing just one of the two. 

Boserup {\em et al.}~\cite{boserup2022efficient} proposed a patch based contrastive learning based network. A confidence network is used to predict a set of confidence maps for each image, representing the confidence level that each pixel belongs to a particular class 'k'. The high confidence level of an image representing class 'k' implies that the image contains objects of a particular class. Such a confidence map, implemented using a fully convolutional neural network, is trained to distinguish between objects of different classes by contrastive learning. The selection of positive and negative patches is highly critical as they determine the performance of the confidence network. Positive patches are those which are believed to have objects of a specific class, and negative patches are those which do not have the object of that class. To obtain positive and negative patches, an entropy based patch sampler is put to use. The average patch entropy is defined as a function of the Bernoulli Random Variable of the confidence value of a pixel belonging to a particular class. Ideal choices of these patches would correspond to higher certainty from the confidence network. From a set of patches sampled from an unnormalized Bernoulli distribution for each class, positive and negative samples are partitioned based on their confidence scores. The similarity between patches is calculated from the pixelwise product of an image and its confidence map. Mean squared error and mean cross entropy are the pixel based similarity measures employed for this purpose. This scaling with the confidence map connects the gradients between the confidence network and the sampling process, which aids in backpropagation for end-to-end training of the model. This approach uses a combined loss, including the inter-class and intra-class contrastive losses, to ensure distinct features are identified for each class, in addition to maximizing and minimizing the similarity among positive and negative patches, respectively. The required segmentation maps are obtained from the confidence maps of the network after convergence. 

\subsection{Supervised}

Supervised learning methods require labels to train the model. Depending on the level of supervision required, the approaches are classified into Full Supervision and Weak Supervision. 

\subsubsection{Full Supervision}

Full Supervision refers to deep learning models that require 100\% of the training set to achieve a good performance. A summary of the fully supervised methods is shown in Table \ref{table:full_supervision}. 



\begin{table}
\small
\centering
\caption{A summary of fully supervised nuclei segmentation methods.}\label{table:full_supervision}

\begin{tabular}{p{0.02\textwidth}p{0.2\textwidth}p{0.2\textwidth}p{0.2\textwidth}p{0.2\textwidth}}
\toprule
    \textbf {Ref.} &  \textbf {Dataset} &  \textbf {Methods} &  \textbf {Pre-Processing} &  \textbf {Post-Processing} \\ 
\toprule
\cite{liang2022region} & DSB2018, MoNuSeg & Region based Mask RCNN with Guided Anchor RPN  & Resizing & Soft Non Maximum Suppression
\\
\midrule
\cite{roy2023nuclei} & DSB2018 & Mask RCNN with ResNet-101 backbone & Pretrained with weights of COCO dataset  & Clump identification followed by marker controlled watershed
     \\ 
\midrule
\cite{graham2019hover}& CoNSeP, Kumar, TNBC, CPM-15, CPM-17, CRCHisto & HoVerNet : Three branch U-Net with horizontal and vertical distance maps to separate nuclei clusters  & Patch extraction & Gradient based marker controlled watershed from distance map
\\ 
\midrule
\cite{chanchal2021high} & Kidney dataset, TNBC, MoNuSeg & High resolution wide and deep transferred ASPPU-Net & Patch extraction, Data augmentation & -
\\ 
\midrule
\cite{kiran2022denseres}& MoNuSeg, TNBC, CryoNuSeg, BBBC039V1 & Dense ResU-Net with residual connections of atrous blocks & Color Normalization, patch extraction, Data Augmentation & -
\\ 
\midrule
\cite{saednia2022cascaded} & Post-NAT-BRCA, MoNuSeg & Cascaded U-Net framework (U-Net with weighted pixel loss  & Zero padding, patch extraction, weighted mask generation & Erosion, Dilation, Reconstruction
\\ 
\midrule
\cite{hancer2023imbalance} & MoNuSeg & Enhanced lightweight U-Net with generalized Dice loss & Stain normalization, resizing, patch extraction, data augmentation & Opening
\\ 
\midrule
\cite{chen2023cpp} & DSB2018, BBBC006v1, BBBC039, PanNuke & CPP-Net with Context Enhancement, Confidence Based Weighting and Shape Aware Loss & - & Semantic segmentation decoder, NMS, Reassignment of pixels to correct categories 
\\
\midrule
\cite{kumar2017dataset}  & Kumar dataset  & Ternary CNN with boundary class & Color normalization, (boundary annotation, pixel mapping in training stage), patch extraction & Seed detection by thresholding followed by region growing using boundary class
\\ 
\midrule
\cite{zhou2019cia} & MoNuSeg & Contour Aware Information Aggregation Network & Stain normalization, data augmentation & Nuclei and contour outputs subtracted, connected component identification
\\ 
\midrule
\cite{chen2020boundary} & Kumar, CPM-17 & Boundary assisted Region Proposal Network & Stain normalization, data augmentation & -
\\ 
\midrule
\cite{qin2022reu}& HUSTS, MoNuSeg, CoNSeP, CPM-17 & Region Enhanced multitask U-Net with auxiliary tasks of rough segmentation and contour extraction & Patch extraction, data augmentation & Marker controlled watershed 
\\ 
\midrule
\cite{lal2021nucleisegnet}& KMC Liver dataset, Kumar dataset & NeucliSegNet (U-Net based) with residual blocks, bottleneck block, attention decoder block & Resizing, patch extraction. No pretraining  & -
\\
\midrule
\cite{yang2022gcp} & ClusteredCell, MoNuSeg, CoNSeP, CPM-17 & Gating Context Aware Pooling integrated modified U-Net & Resizing, patch extraction, ImageNet pretrained ResNet-34 & -
\\ 
\midrule
\cite{thi2022convolutional} & DSB2018, MoNuSeg & U-Net based Convolutional Blur Attention Network & Patch extraction, training data generation, Biorthogonal wavelet denoising & -
\\ 
\bottomrule

\end{tabular}
\end{table}

\paragraph{CNN Based Methods}

\begin{figure*}[t]
\begin{center}
\includegraphics[width=1.0\linewidth]{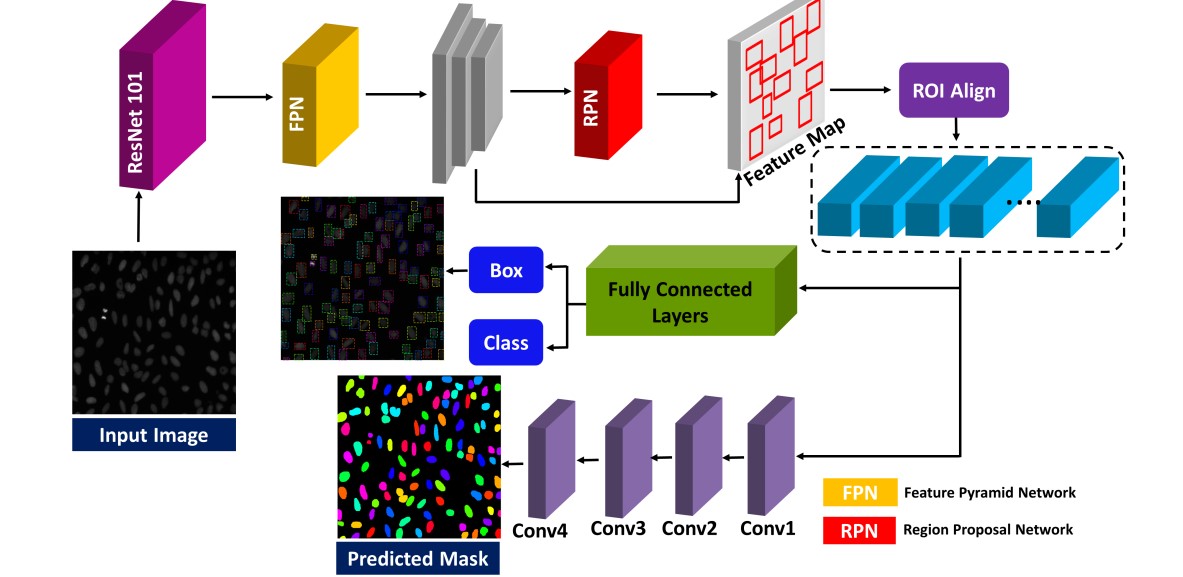}
\caption{ An example Mask RCNN based nuclei segmentation framework from \cite{roy2023nuclei} called NucleiNet. } \label{fig:cnn}
\end{center}
\end{figure*}

This section discusses some landmark CNN based works. In addition to the conventional CNN, the advent of Region based CNNs lead to the Mask RCNN, designed to predict object masks in addition to bounding boxes. Such a Mask RCNN based framework is shown in Fig.~\ref{fig:cnn}. 

One of the initial CNN based nuclei segmentation methods was proposed by Xing {\em et al.}~\cite{xing2015automatic}. They propose a supervised deep CNN to generate a probability map that assigns each pixel a probability of how close it is to the nucleus center. The CNN is trained with images in the YUV space. Each image is manually annotated for nuclei centers, and patches with the centers at a radius of 4 pixels from these centers are considered to be positive, and others as negative. Rotation invariance is achieved by rotating the positive patches before training, thereby augmenting the training data. The CNN uses softmax with two neurons as its final layer to generate the probability of each patch being positive or negative. Patches with a positive probability of less than 0.5 are eliminated from further processing. Additionally, a region size threshold is employed to eliminate patches with small areas that could indicate noise. From the probability maps, the distance transform and H-minima transform are applied to generate minima as markers for an iterative region growing algorithm. A smoothing operation is performed to preserve the shapes of nuclei for the next step. These initial shapes are used in a selection based dictionary learning to generate a shape repository representing a subset of the nuclei. This method works by minimizing the ISE (integrated square error) based on a sparse constraint. To account for the wide variability in shapes, the different shapes are clustered into groups using K-means, and a shape-prior model is learned for each of these groups. They then perform an alternative shape deformation and shape inferencing algorithms to perform the segmentation. The shape deformation is implemented using a Chan-Vese model \cite{chan2001active} incorporated with an edge detector and a repulsive term to introduce robustness and split touching nuclei. The shape prior model is used to perform shape inferencing, thus allowing the contours to iteratively evolve towards the nuclei boundaries. This approach ensures that the contours don't split or merge due to any heterogeneities in intensities as opposed to the level sets in Sec. \ref{paragraph:AC & LS}. 

Liang {\em et al.}~\cite{liang2022region} proposed a region based CNN, employing a guided anchored region proposal network (RPN). This network uses a Mask RCNN and FPN as its baseline. Rather than applying the conventional, dense, predefined anchors, guided anchoring is used in the dynamic prediction of anchors with different shapes and sizes. The GA-RPN module consists of two branches responsible for location and shape prediction, respectively. This module generates multi level anchors, collecting anchors from multiple feature maps generated at different levels by the FPN. An Intersection of Union (IoU) branch is designed to regress the IoU between the ground truth and the predicted bounding box. Generally, Mask RCNN uses non maximum supression (NMS) to order the boxes by classification score. However, this approach may eliminate certain boxes of low classification scores with high quality while preserving some false positives. The IoU module overcomes this by introducing the IoU regression score. A new metric called the Fusioned Box Score (FBS), the geometric mean of the classification score and the IoU score, is used to classify the boxes into their correct classes. NMS with a low threshold may cause an increase in the miss-detection rate as it can classify clustered nuclei as one object. To overcome this limitation, they propose a soft NMS that decays the FBS of boxes that have a significant overlap with the box with the highest FBS. This would penalize boxes close to one with the largest FBS more than the boxes farther away. The results from this method show fewer undetected nuclei compared to other SOTA nuclei segmentation methods. 

Roy {\em et al.}~\cite{roy2023nuclei} proposed the Nuclei-Net, a Mask RCNN based multistage network. The first stage employs a Mask RCNN with a ResNet-101 backbone, transfer learned from the COCO dataset. An RPN follows the backbone network, using the sliding window method to scan the feature maps from the previous step to generate anchor boxes of reasonable size and aspect ratios. On extracting the proposals after the application of ROI Align, an FCN is applied to calculate the offsets in each box. These offsets are used to refine the originally obtained proposals, which are then classified as nuclei or background using a classification head. Four consecutive convolution layers are applied to each feature map from a bounding box to generate the binary feature maps. The model is trained using two losses: a classification loss based on the cross entropy loss and a mask loss based on binary cross entropy loss. Most of the regions in this coarsely refined segmentation map have well defined boundaries except few complex clumps. Such regions are identified using an area based threshold to perform further refinement. In the second stage, marker controlled watershed is performed to retrieve individual nuclei from the clumps. They propose to generate markers from these clumps by first identifying boundary points in each clump and initializing them as potential markers. An iterative algorithm is used to reach the markers from the boundary points by removing points with a distance less than a heuristically determined threshold from a specific boundary point. This stage splits any connected nuclei clumps from the previous stage to obtain individual nuclei. 

\paragraph{U-Net like Methods}

U-Nets were initially designed for biomedical image segmentation \cite{ronneberger2015u}. They are similar to encoder-decoder architectures, except that there are skip connections between the encoder and decoder, allowing coupling between the two. This allows the transfer of some low level features from the encoder to the high level stages of the decoder, enriching the segmentation result. 

\begin{figure*}[t]
\begin{center}
\includegraphics[width=1.0\linewidth]{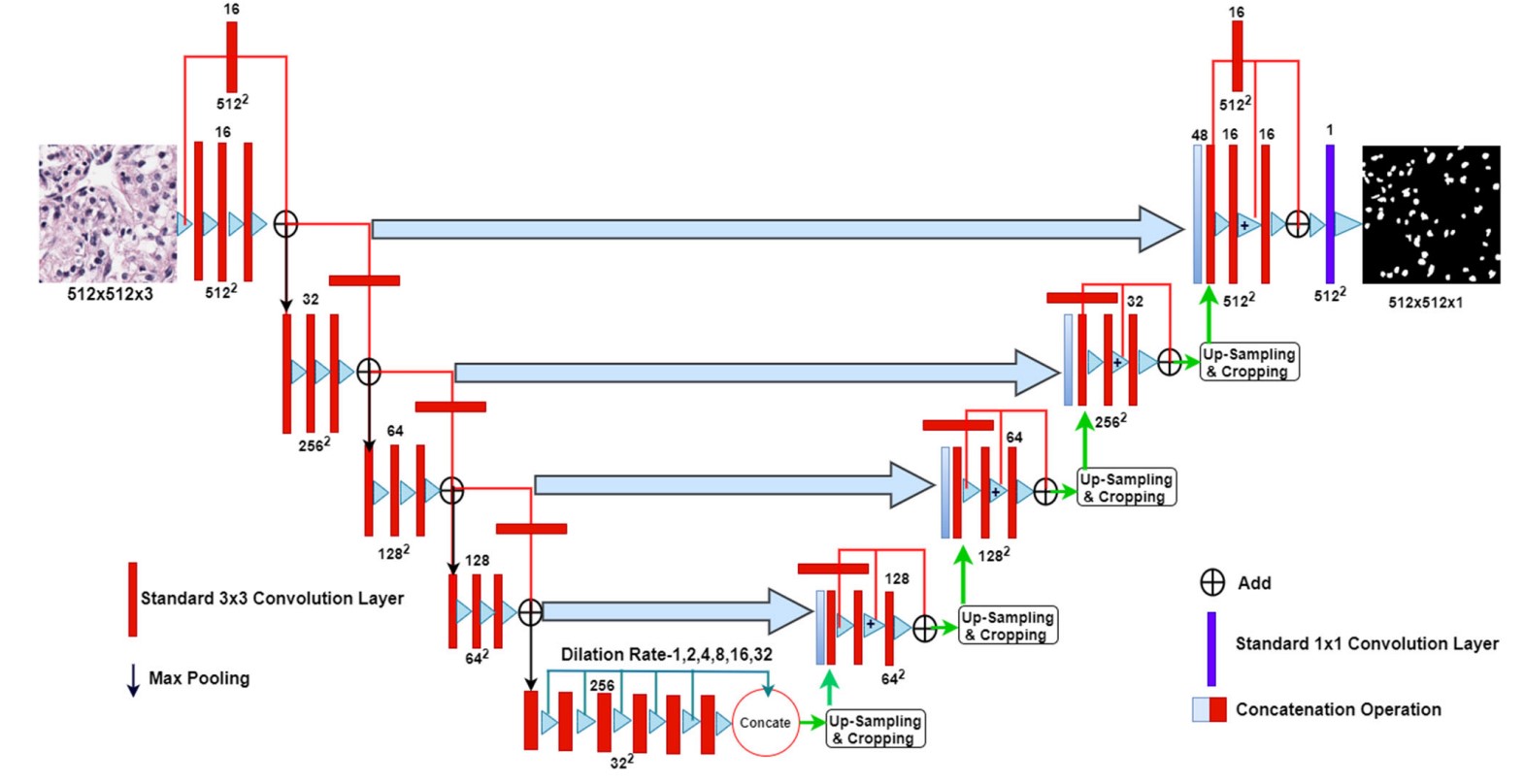}
\caption{ An example U-Net architecture from \cite{chanchal2021high} with an atrous spatial pyramid pooling bottleneck block. } \label{fig:unet}
\end{center}
\end{figure*}

Graham {\em et al.}~\cite{graham2019hover} proposed a novel encoder-decoder framework called HoVerNet, with two decoder branches for nuclei segmentation leveraging features encoded by the horizontal and vertical distances from the centers of mass. This network employs the preactivated ResNet-50 to extract a powerful feature set. To ensure minimum loss of information in the initial stages, the downsampling factor is reduced to 8 from 32. Following the feature extraction are the two nearest neighbor upsampling branches, the nuclear pixel and HoVer branches. The nuclear pixel branch determines whether a pixel represents the nuclei or the background. The HoVer branch is responsible for determining the horizontal and vertical distances from the centers of mass of nuclei, and hence splits touching nuclei. These branches consist of upsampling units followed by multiple stacked dense units. Convolution layers in between the upsampling stages help in improving predictions at the boundaries. The loss function concerned with each branch is a combination of two individual losses. For the nuclear pixel branch, the cross entropy loss and dice loss are added. The HoVer branch loss comprises of the mean squared error between ground truth and the horizontal and vertical distances. In addition, it includes the mean squared error of the gradients from the horizontal and vertical maps and their respective ground truths. From the horizontal and vertical distance maps, a significant difference was observed between the pixels of different instances. Computing gradients can shed light on where the nuclei boundaries exist. Finally, a marker controlled watershed, with the help of the calculated gradient information, can help split the touching or overlapping nuclei. Though trained on images from a single tissue, this network exhibits generalizability when tested on diverse tissue samples. 

Chanchal {\em et al.}~\cite{chanchal2021high} proposed a high resolution deep and wide transferred ASPPU-Net, consisting of an atrous spatial pyramid pooling (ASPP) bottleneck module amidst an encoder-decoder architecture (see Fig.~\ref{fig:unet}). The high resolution encoder has four levels of convolution layers followed by max pooling layers. A residual connection from each layer to its corresponding layer in the main network minimizes losses occurring during pooling. The powerful decoder concatenates features at a similar level to extract residual information. The performance of the network is improved by introducing the ASPP bottleneck with a multiple dilation rate CNN. Dilation rate helps visualize larger areas, and applying multiple dilation rates in one layer extracts multilevel features. The addition of the ASPP bottleneck aids in extracting more relevant features. This model achieves excellent performance by not producing any false positives and extracting maximum information. Nevertheless, overlapping nuclei and blurry boundaries still pose a challenge. 

In \cite{kiran2022denseres}, Kiran {\em et al.} proposed the DenseResU-Net that employs dense units in the higher layers of the encoder focusing on relevant features from the previous layers. The input H\&E stained images are preprocessed using color deconvolution. This helps reduce the detection of false positives and gives a better picture of nuclei and boundaries for further segmentation. Cropping, flipping, rotation, and other basic augmentation operations are performed to enhance the generalizability of the model on unseen organs. Distance mapping is applied to detect nuclei and binary thresholding at the value of 0.4 is used to obtain contour information. DenseResU-Net comprises a five stage architecture, with five dense blocks in the final layer of the contracting path, which helps in preventing the model from learning redundant features. These dense blocks give rise to high computational efficiency. Since skip connections between the encoder and decoder cause semantic gaps, residual connections using the atrous block and non-linear operations are implemented similar to \cite{chanchal2021high}, extracting spatial features. The decoder retrieves the segmented output by reconstructing the feature maps through upsampling. This model shows excellent performance on images from different organs, proving its robust nature and generalizing ability.

Saednia {\em et al.}~\cite{saednia2022cascaded} proposed a cascaded U-Net framework, with a weighted U-Net followed by a vanilla U-Net with the VGG-16 baseline trained using the soft Dice loss. The model was pretrained using the public Post-NAT-BRCA dataset, prior to training on the MoNuSeg dataset (elaborated in Sec. \ref{subsec:datasets}). Weighted masks are generated from each image using the binary mask such that pixels between adjacent nuclei are given larger weights. These weights ensure the model learns the separations between nuclei in such regions on application of the weighted loss function. The weight maps are employed while calculating the loss function to penalize the loss function at boundary regions between touching or overlapping nuclei. A weighted cross entropy loss is used to train the weighted U-Net model. This model was trained with the input images, their binary and weighted masks to generate an output probability map. These probability masks and binary masks are input to the vanilla U-Net. The second stage is implemented with a VGG-16 backbone to reduce the number of parameters, thus promoting generalizability. The soft Dice loss function is selected to penalize the network for predicting nuclei with a low confidence level. The final segmentation maps from the cascaded network were post processed using morphological operations to remove small noisy structures. The second stage accounts for some parts of nuclei that were missed in the first stage, especially small nuclei and centers of large nuclei. This cascaded model performs on par with most deep learning based segmentation models. Accurate boundary detection still remains a challenge. 

Hancer {\em et al.}~\cite{hancer2023imbalance} proposed an imbalance aware method for nuclei segmentation using a lightweight enhanced U-Net model. They apply Macenko's stain normalization technique by obtaining optical density vectors from the RGB image and performing single value decomposition to get accurate stain vectors. To resize the high resolution images without losing any pixels, nearest neighbor interpolation technique is implemented. The next step is data augmentation, where techniques like rotation, reflection, and translation are used to increase the number of training samples. Class imbalance may impact the model training, leading to a biased model. Loss functions have been the common solution to such challenges and in this work, they incorporate the generalized Dice loss to account for the class imbalance. This loss involves per-class weight as the inverse square of the class volume. The generalized Dice loss is used to train a lightweight U-Net model, with a depth of three layers as opposed to the original four layer U-Net. In addition, the final layer of the network uses a Dice pixel classification layer that assigns a categorical label to each pixel. Finally, morphological operations are performed to refine the segmentation maps. 

Chen {\em et al.}~\cite{chen2023cpp} proposed the Context-Aware Polygon Proposal Network (CPP-Net) with a U-Net backbone. They make use of polygons to represent nuclei, which can help the task of differentiating between nuclei that touch or overlap. Following the U-Net are three unit sized convolutional layers to predict the distance maps, confidence maps, and the centroid probability map. The next step is the Context Enhancement Module that samples a set of points from an initial point toward its predicted boundary. The multiple predicted distances are then merged to update the distance between the pixel and its boundary. To perform this merging adaptively, they use a Confidence Based Weighting Scheme with the help of confidence maps. In addition to the BCE loss for centroid probability and weighted L1 loss for the distance regression, CPP-Net includes a Shape Aware Perceptual loss that penalizes the difference in shape between predicted and ground truth instances. As a part of the fine grained post processing, a semantic segmentation decoder attached to CPP-Net's encoder identifies the foreground pixels. NMS converts each polygon to a mask and certain pixels are reassigned to their correct categories. This step helps refine boundaries, thus improving the quality of segmentation. One limitation of this approach is that irregularly shaped nuclei may not be efficiently represented by a polygon, and CPP-Net may fail in such cases. 

\paragraph{Contour Aware}

\begin{figure*}[t]
\begin{center}
\includegraphics[width=1.0\linewidth]{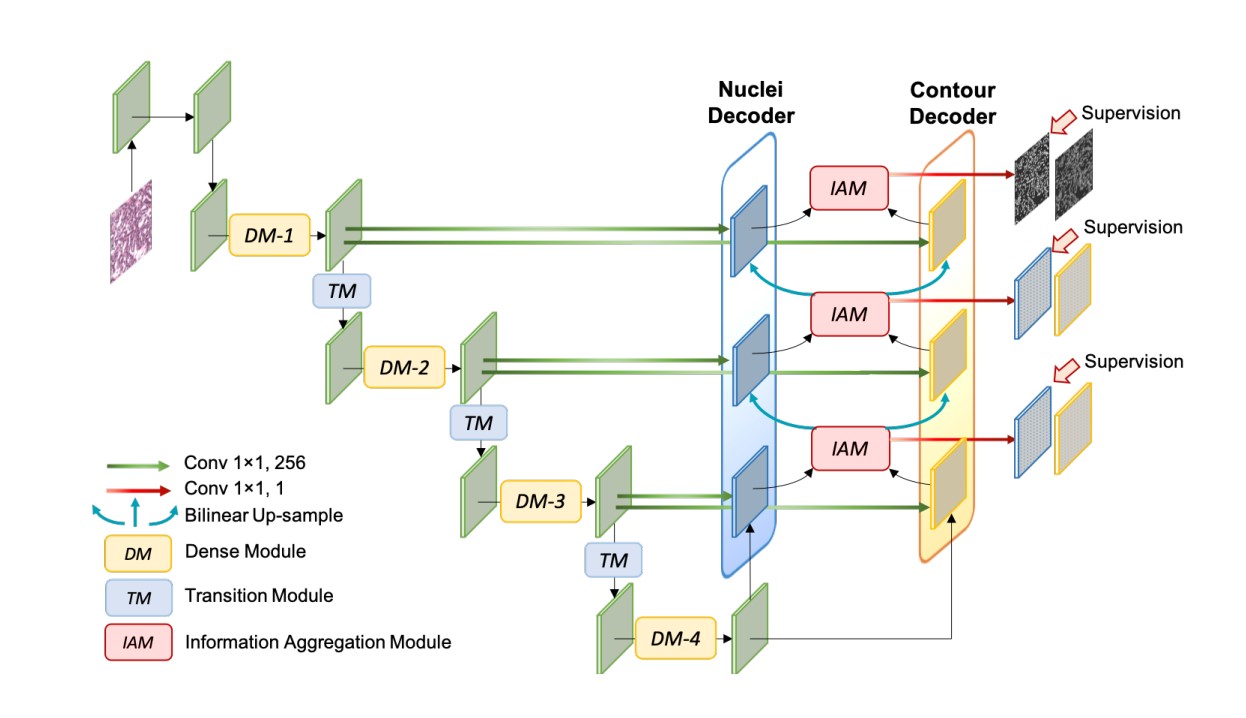}
\caption{ An example of the contour aware CIA-Net from \cite{zhou2019cia} that uses a U-Net architecture with two decoding paths dedicated to nuclei and contour decoding respectively. } \label{fig:contour}
\end{center}
\end{figure*}

From error analysis, it was observed that a majority of the misclassifications were from regions around the boundary of nuclei. This was due to the presence of overlapping, touching nuclei in dense clusters. With the idea of giving importance to contours, several approaches introduced a way to focus especially on the nuclei boundaries as shown in Fig.~\ref{fig:contour}.

Kumar {\em et al.}~\cite{kumar2017dataset} proposed a CNN that produced a ternary map, as opposed to the binary map distinguishing between nuclei and background. This method classifies each pixel into three categories, namely, nuclei, nuclei boundary, and background. The three-way classification is visualized to identify nuclear boundaries even among dense clusters and chromatin sparse nuclei. The optical density vector is first obtained from the H\&E stained image by applying Beer Lambert Transform. Sparse non-negative matrix factorization or SNMF is applied to this vector to generate two sets of matrices, one contains stain density maps, and the other contains optical density components of each prototype i.e., H \& E prototypes. To obtain the color normalized image, the stain density map is multiplied by the basis matrix for its corresponding color prototype and inverse Beer Lambert Transform is performed. The normalization procedure helps in reducing the contrast among nuclei in different images while preserving the nuclei to background contrast. The constructed CNN architecture consisted of three convolution layers, with max pooling layers between them and activated by ReLU activation. These layers were followed by two FC layers and an output layer with three nodes and softmax activation. To obtain the final segmentation map, the nuclei body probability map was thresholded at 0.5. This operation provided seeds for a region growing mechanism. As the seeds are grown, the average boundary class probability increases, and the average nuclei probability decreases. The regions are grown to obtain the segmentation map until the boundary class probability reaches a local maxima as long as it doesn't interfere with other nuclei or boundaries. This gives rise to an anisotropic region growing method, with the regions growing at different rates and directions, leading to non-circular shapes. Including boundary supervision is shown to improve accurate detections and has a slight edge with segmenting chromatin sparse nuclei. 

Oda {\em et al.}~\cite{oda2018besnet} proposed a U-net based network with two decoding paths, and special emphasis on the boundaries called Boundary Enhanced Segmentation Net or BESNet. The first decoding path focuses on boundary prediction and is trained on boundary labels. The second path (main decoding path - MDP) utilizes the responses from the boundary decoding path to weigh the training loss for segmentation adaptively. Specifically, information on the difficulty of determining boundaries is combined with the MDP. The input image is fed into the encoder path to generate feature maps. Feature maps from the boundary and main decoding paths are concatenated. The boundary decoding path is trained using the cross entropy loss. The output in this path will be high at boundary regions but deteriorate at unclear regions, meaning that such regions have a higher training difficulty. These regions are given more importance in the MDP by the adaptively weighted Boundary Enhanced Cross Entropy (BECE) Loss. The additional decoding path, however, adds computational burden on the system compared to other U-Net based networks. Though this network uses boundary information to gain insight into the entire nuclei body, it doesn't leverage the nuclei information to learn about the boundary. Since contours have a greater intra-variability, networks may benefit from mutual information from the nuclei and contours, thus improving the prediction performance. 

To leverage the advantages of mutual dependencies between nuclei and their boundaries, Zhou {\em et al.}~\cite{zhou2019cia} proposed the fully convolution Contour Aware Information Aggregation Net (CIA Net). The U-Net based design has a densely connected encoder using the FPN for feature extraction. The encoder is built using four dense modules stacked hierarchically, with transition modules following each dense module. To take advantage of multi scale features as in an FPN, CIANet proposes lateral connections at each level between the encoder and decoders. Local and textural information from the initial layers is summed with the more robust semantic features from the upsampled layers in the decoder. This network uses two decoders, one for nuclei and the other for contours, with multilevel information aggregation modules(IAMs) between them. The IAM helps in bidirectional task specific feature aggregation, taking important features from both decoders as cues to refine segmentation details in the nuclei and contours. In the decoders, bilinear interpolation is used to upsample the feature maps and add to the feature maps from the lateral connections of the encoder. The IAM smoothens these maps and eliminates the grid effect. These features are then passed on to the classifier to determine score maps. The complementary task specific features are concatenated for refinement in the subsequent iteration. Noisy and inaccurate labels can lead to an overfitted model and prevent the learning of essential features. Generally, such outliers tend to have a low prediction probability and result in large errors. A Smooth Truncated Loss is proposed, which reduces the effect of outliers with greater impact for a lower prediction probability. This helps alleviate oversegmentation, helping the network focus on areas with high confidence scores, thus learning more informative features. A Soft Dice Loss is also included in the total loss function along with the Smooth Truncated Loss and a weight decay term to incorporate shape similarity among the nuclei regions. Exploiting the high relevance between the nuclei and contours improves the generalizing ability of the model to unseen data. However, CIA-Net suffers from false negatives in cases of low contrast between the nuclei and background.  

Chen {\em et al.}~\cite{chen2020boundary} proposed a two-stage boundary-assisted region proposal network (BRP-Net), with the first stage proposing possible instances based on boundary detection and the second stage performing proposal-wise segmentation. The first stage consists of the Task Aware Feature Encoder (TAFE), which extracts high-quality features for semantic segmentation and instance contour detection. In this stage, a backbone encoder extracts feature maps of four different sizes from the original input image. These maps are split into segmentation features and boundary features and input to two task-specific encoders that are deeply supervised, similar to CIA-Net \cite{zhou2019cia}. Feature Fusion Models (FFM) based on the IAMs from CIA-Net are devised to aggregate the features from both the task specific encoders. The outputs of the FFM are fed into decoders to perform semantic segmentation and instance boundary detection, respectively. Since TAFE requires postprocessing based on handcrafted hyperparameters, a second stage was introduced to make BRP-Net more robust. A square patch containing the region proposals of different sizes is cropped and classified into two groups using a length threshold. The images in these two groups are resized to a specific size and trained separately using similar networks with dense blocks, which take the image patch and the two probability maps from the previous stage as input. IoU scores are used to train the networks by comparing the image patch with its corresponding ground truth. Patches with an IoU score lower than a threshold are considered to be false positives and inferred as background. 

In a similar approach leveraging boundary and nuclei features, Qin {\em et al.}~\cite{qin2022reu} proposed a region enhanced segmentation network by combining three U-Nets in serial and parallel to form a multi-task architecture (REU-Net). The model uses the attention U-Net as the baseline, with three U-Net like branches to perform the auxiliary tasks of contour extraction and rough segmentation and the main task of fine nuclei segmentation. The predicted results of the auxiliary branches are integrated and multiplied by elements enhancing the saliency of nuclei along with their contours. This region enhances the image, and the original image is fed into the encoder of the fine segmentation branch. The encoded features from all three branches are concatenated and input to the fine segmentation decoder through attention gates to aggregate the spatial and textural features of nuclei and contours. The attention gates diminish the semantic gap between the three branches, removing background elements and providing essential target features to the network. An atrous spatial pooling pyramid (ASPP) structure is used to retrieve a rich set of spatial features by capturing multiscale information to prevent the loss of vital spatial information while encoding. The loss for each branch is computed as a combination of the dice and cross entropy losses. The loss function for the entire network is calculated to be the weighted sum of the losses from each branch, with the fine segmentation branch having double the weight of the auxiliary branches. The results of this approach serve as evidence for the mutual contribution of contour and nuclei information toward segmentation performance. 

\paragraph{Attention Gated}

Attention gates are used in segmentation architectures to focus on the important features and suppress irrelevant features like the background being included in the training. 

\begin{figure*}[t]
\begin{center}
\includegraphics[width=1.0\linewidth]{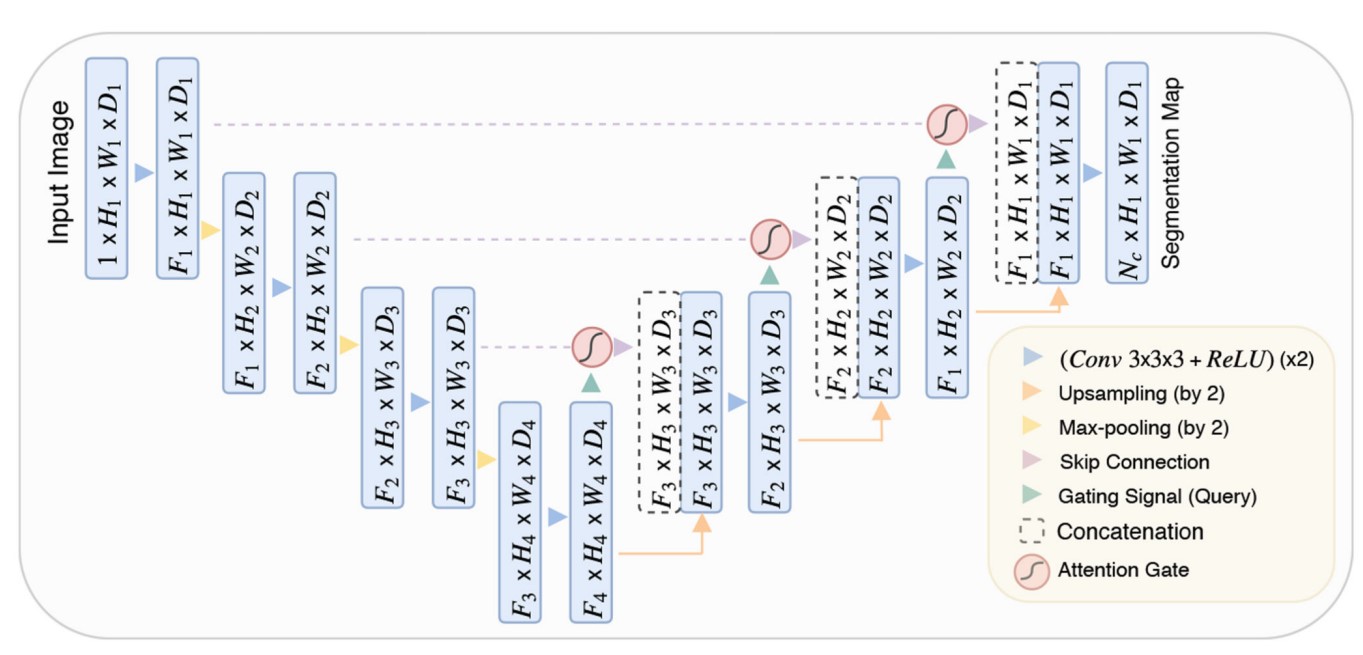}
\caption{ An example of architecture using gated attention for U-Net from \cite{schlemper2019attention} } \label{fig:attention}
\end{center}
\end{figure*}

Schlemper {\em et al.}~\cite{schlemper2019attention} proposed an attention gated U-Net model (see Fig.~\ref{fig:attention}). False positive removal has always required a post processing or multi-stage segmentation approach in many methods. Attention gating attempts to avoid the need for separate removal methods by eliminating the sources of such errors while training. Since additive attention has been found to perform well on high dimensional inputs, the gating coefficient is obtained using this approach. Each global feature vector and activation map are considered at each level to identify the most relevant features for the specific task. Though the attention mechanism doesn't require a separate loss function for optimization, deep supervision appears to allow the feature maps to be discriminative at different scales. In a U-Net model, gating is implemented before concatenating the features between the encoder and decoder paths as a part of the skip connections to combine only the important activations. The attention gates filter activations in forward and backward passes and suppress the irrelevant background information in the backward pass. They use the sigmoid activation function to normalize the attention gated features, resulting in improved training convergence. To account for class imbalance, this method uses the Sorensen Dice loss. They also use a loss term for each scale to ensure the model attends at each scale. 

A three block NucleiSegNet was proposed by Lal {\em et al.}~\cite{lal2021nucleisegnet}, containing a robust residual encoder, an attention based decoder, and a bottleneck block. High level semantic feature extraction is performed by the robust residual block with depth wise and point wise convolution in separable blocks. Four such blocks constitute the encoder. The bottleneck block, with three convolution layers, followed the encoder and contributed toward achieving the best training loss. This block ensured the compressed encoding of the global information from all relevant regions, simplifying the job of the decoder. Features from the encoder and the compressed features from the bottleneck block were merged and input into attention gates in the decoder. In the attention gate, the gating signal was upsampled, as opposed to downsampling the skip connections. They applied a multiplicative attention gate, owing to its memory efficiency and faster computations. The decoder has four upsampling stages with transpose convolutions within each attention block to decrease the model parameters without affecting accuracy. Such an attention mechanism helps in reorganizing fine features, and removing background information. A combined loss function of dice loss and Jaccard loss was implemented to train the model. Overall, this method performs on par with SOTA methods, with a fewer network parameters. 

Yang {\em et al.}~\cite{yang2022gcp} proposed a U-Net based gating context aware pooling network (GCP-Net). GCP-Net uses an ImageNet pretrained ResNet-34 encoder blocker. This is followed by the GCP module that functions as a context extractor, generating high-level semantic features. The GCP module consists of Multi scale Context Gating Residual (MCGR) block, Global Context Attention (GCA) block, and the Multikernel Maxpooling Residual (MMR) block. Context gating (CG) transforms the input feature representation into a new representation with a powerful discriminant capability. To improve on the limited receptive field of CG, they propose the MCGR block with parallelly connected three branches of depth wise convolution, producing weighted feature maps of different resolutions. The input feature map and the weighted feature maps are merged to retrieve multiscale information. Contextual information is proven to improve segmentation results by increasing the size of the receptive field and using attention blocks. The GCA block reweights features to enhance the network's sensitivity to essential information, thus improving performance. In contrast to a single pooling kernel in maxpooling, the MMR block incorporates pooling kernels with four different sizes to capture features with a range of receptive fields, as it can influence the amount of context information used. The decoder module includes four decoder blocks to retrieve the features extracted in the previous stages. Each decoder block consists of two GCA residual blocks, after which the features are concatenated with the information from the skip connection. To overcome the vanishing gradient problem in deep networks, the GCA residual block uses a shortcut connection between layers, forcing the network to learn essential features in each feature map and suppress the irrelevant ones. The performance of this network is comparable to SOTA deep learning frameworks. 

Thi Le {\em et al.}~\cite{thi2022convolutional} proposed the Convolutional Blur Attention (CBA) network, a SOTA approach, and the best performing so far. CBA net is pretrained first, followed by finetuning on the training set. The network consists of the blur attention module and blur pooling operation to retain important features and prevent the addition of noise in the encoding or downsampling process. Initially, the RGB images are converted to grayscale and scaled down. Traditional downsampling algorithms use max pooling, which can cause a loss of features in the early stages. Max pooling is replaced by the blur attention module and blur pooling to improve the segmentation output. This network uses blur convolutional layers with stride 1. The blur pooling operation was an attempt to enforce shift invariance, where any shifts in the input have minimum impact on the output. The blur attention module consists of channel and spatial blur attention. Channel blur attention accumulates spatial information by average and blur pooling to learn spatial statistics and object features. In contrast, spatial blur attention gets channel information from average and blur pooling on the channel axis. Spatially vital regions are identified with the help of a convolution layer that filters the pooling results. Due to the loss of information in the upsampling stage, they propose auxiliary connections between the downsampling and upsampling stages to obtain input features by performing convolutions of different strides while upsampling. The original size features are concatenated with these convolved features to provide generous features to the decoder. In addition, they employ a pyramid blur pooling module to extract multiscale information and identify highly correlated neighboring features. This model is computationally efficient, with parameters multiple times fewer than several SOTA models. 

\subsubsection{Weak Supervision}

Deep learning methods require the availability of large annotated datasets for rich performance. However, with biomedical datasets, it is highly painstaking to manually label such large amounts of data, in addition to inter-observer variability, leading to inaccuracies in certain areas. To alleviate such a tedious task, certain weakly supervised methods have been proposed that require only a subset of the training data to train the model. This subset can refer to only point annotations, i.e., one label per nuclei or a minimum percent of the training set. A brief summary of the weakly supervised methods is presented in Table \ref{table:weak_super}.


\begin{table}
\small
\centering
\caption{A summary of weakly supervised nuclei segmentation methods.} \label{table:weak_super}

\begin{tabular}{p{0.02\textwidth}p{0.2\textwidth}p{0.25\textwidth}p{0.252\textwidth}p{0.1\textwidth}}
\toprule
    \textbf {Ref.} &  \textbf {Dataset} &  \textbf {Methods} &  \textbf {Pre-Processing} &  \textbf {Post-Processing} \\ 
\toprule

\cite{yoo2019pseudoedgenet} & MoNuSeg, TNBC & ResNet-50 backbone segmentation net supervised by auxiliary PseudoEdgeNet & Label assignment - Voronoi and distance transform & Threshold-ing
\\
\midrule
\cite{qu2020weakly}& Lung Cancer Dataset, Kumar Dataset & Semi-supervised nuclei detection followed by weakly supervised segmentation (using ResNet backbone U-Net) & Color normalization, patch extraction, data augmentation, ResNet-34 encoder pretrained, Voronoi and K-means cluster labeling & -
   \\ 
\midrule
\cite{qu2020nuclei} & Lung Cancer Dataset, Kumar Dataset & Uncertainty prediction from Bayesian CNN followed by normal CNN trained with partial points and mask labels & Color normalization, patch extraction, data augmentation & -
\\ 
\midrule
\cite{tian2020weakly}& MoNuSeg, TNBC & Coarse segmentation using self supervision followed by  fine segmentation with contour sensitive constraint & Point distance map and Voronoi edge distance map generation & -
\\ 
\midrule
\cite{lin2022label} & MoNuSeg, CPM-17 & Co-trained U-Net based Segmentation-Colorization Network & Patch extraction, data augmentation, H-component extraction, Voronoi and K-means cluster labeling & -
\\ 
\midrule
\cite{hu2004automated}& MoNuSeg & GAN based nuclei centroid detection followed by peak region backpropagation & Stain normalization, patch extraction & Graph cuts
\\ 
\midrule
\cite{lou2022pixel}  & Kumar, TNBC, MoNuSeg & Conditional SinGAN based training data augmentation from selected patches followed by Mask RCNN for semi-supervised segementation & Patch extraction, data augmentation & -
\\

\bottomrule

\end{tabular}
\end{table}



\paragraph{Point-wise Label Propagation}

This category of weakly supervised methods begins with point annotations. These annotations are extended to generate coarse pixel wise labels that are used to train a nuclei segmentation network. 

\begin{figure*}[htb]
\begin{center}
\includegraphics[width=1.0\linewidth]{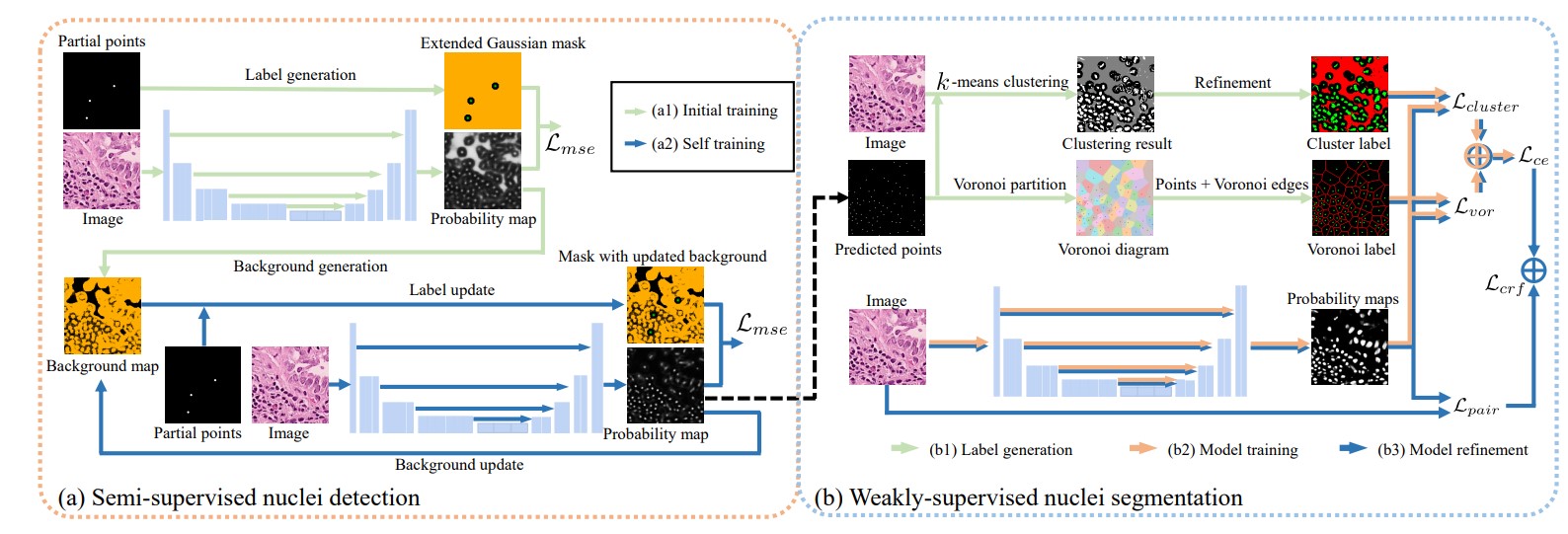}
\caption{ An example of a two stage point wise label propagation from \cite{qu2019weakly}. } \label{fig:pointwise}
\end{center}
\end{figure*}

Yoo {\em et al.}~\cite{yoo2019pseudoedgenet} proposed one of the initial weakly supervised segmentation networks for fine nuclei segmentation, with an auxiliary network for edge detection. To train the segmentation network with point labels, a label assignment scheme assigns positive values to the point annotations and negative values to pixels on the Voronoi boundaries. The binary cross entropy loss is utilized to train the network. Though this network generates nuclei blobs, it lacks information about the boundaries. This led to the design of the auxiliary network for edge detection called the PseudoEdgeNet, a shallow CNN extracting edge information. PseudoEdgeNet is trained with the original image and the point annotations generated by the segmentation network and acts as its supervisory signal. In addition, a large attention module in the PseudoEdgeNet guides it on where to extract edges, thus improving the quality of edge maps generated. The Sobel-filtered result of the segmentation net is used as a reference in calculating the edge loss. Both the networks are trained jointly with a cross entropy based segmentation loss and the edge loss. Though bounded by the performance of fully supervised networks, this approach was a good start to the weakly supervised approach with a promising future. 

Qu {\em et al.}~\cite{qu2020weakly} proposed a two-stage weakly supervised technique from partial point annotations as shown in Fig. \ref{fig:pointwise}. The first stage performs semi supervised nuclei detection. An extended Gaussian mask is generated from the available labeled points based on the distance of a pixel from its nearest labeled point. The background is defined at regions greater than a certain distance, nuclei are identified with an exponential function of the distance, and the remaining pixels are unlabeled and ignored while training. A regression based detection model is trained using the obtained extended Gaussian masks, using the mean square error. The U-Net like model uses ResNet-34 as its encoder. A background map is generated from the first step by thresholding. The second part of the first stage uses an iterative self training method to improve the detection performance by combining information from the initial points and the generated background maps. The background map is updated in each iteration based on an intensity and area threshold. At the end of this stage, the background regions grow, and potential nuclei locations are identified. The second stage requires coarse pixel-level labels to train a CNN. Using the results from the previous stage, Voronoi labeling is used to obtain regions called Voronoi cells that give essential information about the central parts of the nuclei. To understand more about the shapes and boundaries of the nuclei, K-means clustering based on color and spatial information is implemented, obtaining coarse pixel wise labels. A network similar to the one used in the first stage is trained using the generated labels and a weighted loss function based on the cluster labels and Voronoi labels. A dense CRF loss is used to further improve the nuclei boundaries. However, this approach faces difficulties with non-uniform staining. 

In their follow-up paper \cite{qu2020nuclei}, Qu {\em et al.} proposed using a combination of points and masks to enhance the performance of the weakly supervised approach. The first stage comprises an uncertainty prediction task that finds representative complex nuclei to be supervised by annotation masks. Uncertainty maps are generated by a Bayesian CNN built by adding a Gaussian distribution prior to the softmax layer. A probability map is generated on the application of the softmax function to the output of the network. To supervise the training of this model, in addition to the point annotations, proxy labels are generated pixel-wise similar to \cite{qu2020weakly} from cluster and Voronoi labels. This network is trained using the cross entropy loss derived from the two proxy labels. From the generated uncertainty maps, area wise average uncertainty is computed to identify the top 5\% of highly uncertain nuclei predictions that will require mask annotations. In the second stage, the masks for the selected representative nuclei are integrated with the cluster and Voronoi labels, and add some background pixels to aid training with the combined masks. These updated labels train a normal CNN model with the same loss function as in the first stage. This modified approach achieves a slightly improved performance compared to their previous method. 

Tian {\em et al.}~\cite{tian2020weakly} proposed a coarse to fine weakly supervised learning strategy using point annotations. The first stage generates coarse segmentation maps through distance mapping and self supervised learning. To train an FCN, the point annotation map must be transformed into a supervision map. They propose two maps for supervision, the point distance map focusing on highly reliable positive points obtained by dilating the point annotations, and the Voronoi edge distance map focusing on highly reliable negative points indicating non-nuclei pixels. With these generated supervision maps and the original image, an FCN is trained end to end using polarization loss. In addition, sparsely calculated loss concerning the point labels and Voronoi labels are also included as to focus only on the partial labels with high confidence. This coarse segmentation network is trained for about three iterations to obtain reliable masks. For each iteration, while the Voronoi edge map remains the same, the point distance map is updated to result from the previous segmentation round. The second stage focuses on contour refinement. Edge maps are generated from the original image and the coarse masks by applying Sobel filtering. A sparse contour map is obtained by pixel-wise and operation between the edge maps. This auxiliary boundary supervision is implemented using the sparse contour map for supervision and a contour sensitive loss to refine the contours. The second stage, fine tuning, significantly improves the model's performance. 

Lin {\em et al.}~\cite{lin2022label} proposed an alternate approach to learning contour information in weakly supervised methods by means of a sequential Segmentation and Colorization Net called the SC-Net. The initial step generates coarse pixel labels from the point annotations using Voronoi labeling and K-means clustering. Voronoi labeling generates convex polygons with point annotations as the center. To perform K-means clustering, the original image and a distance transformed image are used to cluster the pixels into three categories, nuclei, background, and ignored pixels. They extract the H component from the H\&E stained images, enhancing the contrast between nuclei and non-nuclei regions. A ResU-Net based Segmentation Network is trained with the generated coarse labels and cross entropy losses from the cluster and Voronoi labels. The Voronoi labels help split overlapping nuclei, while the cluster labels provide contour and shape information. To minimize the effect of incorrect cluster labels, they propose a co-training framework with a pair of segmentation networks trained by two non-overlapping sets of data. The training of one network will be supervised by pseudolabels generated by the other network along with the coarse labels. Accurate cross supervision is achieved by using EMA to average the pseudolabels periodically. They also proposed an auxiliary colorization task to obtain precise nuclei contours. The combined SC Net consists of a sequence of two U-Nets, with the first generating probability maps from the H-component and the second reconstructing the H\&E image from the probability map. With the help of the colorization network, the segmentation network gathers more low level features and captures the nuclei-cytoplasm relationship as well. As the training progresses, the segmentation task is given more importance than the colorization task. However, this framework fails if not all nuclei are completely labeled, as it may generate erroneous coarse labels. 

\paragraph{GAN}

With a limited number of annotated training samples, researchers use these masks to generate more samples to be used for training, or to predict the certainty of the annotations. This section describes the use of GANs in a weakly supervised nuclei segmentation framework. 

\begin{figure*}[t]
\begin{center}
\includegraphics[width=1.0\linewidth]{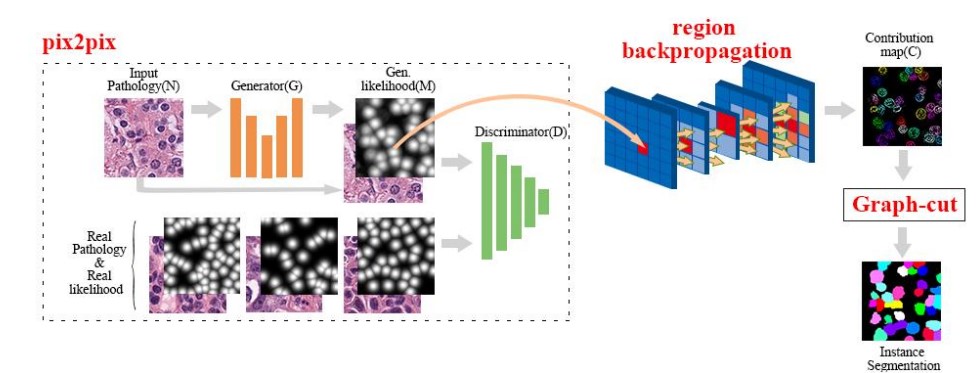}
\caption{ An example framework using GAN to generate a nuclei centroid likelihood map from \cite{hu2020generative}. } \label{fig:GAN}
\end{center}
\end{figure*}

Hu {\em et al.}~ \cite{hu2020generative} proposed a GAN based weakly supervised nuclei segmentation approach depicted in Fig. \ref{fig:GAN}. They first perform stain normalization on the images by decomposing them into stain density maps and obtaining the component distribution. Since not all the point annotations are the exact centroid of the nuclei, they propose the using the nuclei centroid likelihood map as the training set. For this purpose, they use the conditional GAN based pix2pix network for centroid detection. Here, a detection network detects nuclei centroids and generates the likelihood map as the output. A discriminator network distinguished between true and generated centroids from the paired input training data consisting of the original image and the likelihood map. The GAN is trained with a combined loss function of the confrontation loss and an L1 loss. An area threshold is used to detect the central areas in the nuclei from the generated likelihood maps. In addition to these areas, regions surrounding these central areas also contribute to nuclei detection. They use a region guided backpropagation from the central regions to visualize pixels contributing to the centroid detection to obtain a contribution graph for each nucleus. Finally, the graph cuts algorithm is implemented to get the segmented nucleus by considering the contribution graph of each nucleus as the foreground and the remaining as the background. All such fragments are merged to obtain a complete segmentation map. This approach helps in identifying rough nuclei boundaries. 

Lou {\em et al.}~\cite{lou2022pixel} proposed a selection based approach to weakly supervised nuclei segmentation. The annotated image patches to be used for training are determined by two features called representativeness(inter-patch attribute) and consistency (intra-patch attribute). When a patch has the smallest distance from other patches in a cluster, it is considered representative of that cluster. Patches with high similarity within themselves are considered consistent. Such patches with high representativeness and consistency are chosen as the training set. Several similar-sized patches are sampled from each training image. The inter and intra-patch attributes are computed by performing a dual level clustering, which first groups the image patches into several clusters, and each image is split into four equal sub regions. Based on the coarse and fine level representativeness and intra-patch consistency, one image patch from each cluster is chosen to be annotated. Augmentation is performed on each of these masks by random cropping, flipping, and rotation. For each of these image-mask pairs, a mask synthesis algorithm is employed to generate masks. With the original image-mask pairs and these synthetic masks, a conditional SinGAN is trained to generate the nuclei images corresponding to the masks. This network comprises a multi scale conditional generator and a component wise discriminator. For each scale, the training loss combines an adversarial loss and reconstruction loss. To perform segmentation, a Mask RCNN is trained on the real and synthetic image pairs. The trained model then predicts masks for the original training set and these masks act as pseudo labels. The model is finetuned for another 2-3 iterations, including the original training set and the pseudo labels. The model after the final iteration is considered to be the final model. 

\section{Evaluation and Performance Benchmarking}\label{sec:evaluation}

In this section, we start by briefly reviewing the available datasets for training the nuclei segmentation models in \ref{subsec:datasets}. The currently proposed metrics are presented in \ref{subsec:metrics}, since the nuclei segmentation task has different ways that can be evaluated. In \ref{subsec:performance_comp}, we conduct an extensive performance analysis and comparisons of recent methods, both quantitatively and qualitatively. Finally, in \ref{subsec:discussion} we summarize findings drawn from previous comparisons, stressing more the weakly supervised results. 

\subsection{Datasets} \label{subsec:datasets}

Several fully annotated nuclei segmentation datasets have been made publicly available. A summary of a few of the widely used datasets is presented in Table \ref{table:datasets}. These datasets contain exhaustive annotations of all nuclei in the images. 



\begin{table}
\small
\centering
\caption{A summary of publicly available nuclei segmentation datasets.} \label{table:datasets}

\begin{tabular}{p{0.1\textwidth}p{0.1\textwidth}p{0.1\textwidth}p{0.1\textwidth}p{0.25\textwidth}p{0.1\textwidth}}
\toprule
     \textbf {Dataset} &  \textbf {Training Images} &  \textbf {Testing Images} &  \textbf  {Nuclei Count} &  \textbf {Organs Included} &  \textbf {Magnification Level} \\  
\toprule

    MoNuSeg \cite{8880654} & 30 & 14 & 21,623 & 7 (Breast, Liver, Kidney, Prostate, Bladder, Colon, Stomach) & 40x \\
\midrule
Kumar \cite{kumar2017dataset}
& 16
& 14
& 21,623
& 7 (Breast, Liver, Kidney, Prostate, Bladder, Colon, Stomach)
& 40x
    \\ 
\midrule
CPM-15 &
 & 15 &
2,905
& 2
& 40x, 20x
 \\
\midrule
CPM-17 \cite{vu2019methods}
& 32
& 32
& 7,570
& 4
& 40x, 20x
 \\ 
\midrule
CoNSeP \cite{graham2019hover}
& 27
& 14
& 24,319
& 1 (Colorectal Adenocarcinoma)
& 40x
 \\ 
\midrule
TNBC \cite{naylor2018segmentation}
 & & 50 
& 4,056
& 1( Breast)
& 40x
 \\
 \midrule
 CRCHisto \cite{7399414}
& 50
& 50
& 29,756
& 1 (Colon)
& 20x
 \\ 
\midrule
Data Science Bowl 2018~\cite{caicedo2019nucleus}
& 536
& 134
& 37,333
& Combination of tissue from humans, mice and flies 
& Mixed
 \\ 

\midrule
Lung Cancer Dataset \cite{qu2020weakly}
& 24
& 16
& 24,401
& Lung Adenocarcinoma
& 20x
\\
\bottomrule
\end{tabular}
\end{table}


The Multi-Organ Nuclei Segmentation (MoNuSeg) Dataset was originally released for MICCAI 2018 challenge. This dataset comprises 30 training images of size 1000x1000 and magnification 40x and 14 testing images with similar specifications. These images cover samples from Breast, Liver, Kidney, Prostate, Bladder, Colon, and Stomach cells collected from The Cancer Genome Atlas(TCGA). The Kumar dataset is a subset of the MoNuSeg dataset, with its 30 training images split into 16 training and 14 testing samples. With diverse H\&E stained histology images from 7 different organs, good performance on this dataset indicates a high generalization ability.  

The Triple Negative Breast Cancer(TNBC) dataset has 50 H\&E stained images from 11 TNBC patients, annotated by an expert pathologist and research fellows. It has 512x512-sized images of breast cancer tissues at a magnification of 40x. These images include a variety of nuclei annotations, including normal epithelial cells, inflammatory cells, fibroblasts, macrophages, adipocytes, invasive carcinomic cells, and myoepithelial cells. 

Graham {\em et al.} introduced the Colorectal Nuclei Segmentation and Phenotype (CoNSeP) dataset with 41 H\&E stained images obtained from 16 colorectal adenocarcinoma patients. These images are of size 1000x1000 at a 40x magnification and contain a diverse set of tissue components and nuclei types. Two expert pathologists exhaustively annotated each nucleus within every tile with consensus. This dataset displays wide variations within the colorectal adenocarcinoma images, improving performance on unseen images. 

Another colon cancer dataset, called CRCHisto, contains 100 H\&E stained images of size 500x500 at a 20x magnification. The nuclei, however, are not exhaustively annotated with class labels. These images were cropped from 10 whole slide images of 9 patients and annotated by an expert pathologist and a graduate student. 

The Data Science Bowl 2018 (DSB 2018) dataset consists of 670 images with different tissue types, staining modalities, magnification, etc. The nuclei masks have been annotated by a team of experts. This diversity helps in nuclei detection from a wide variety of images and helps in generalization. 

CPM-17 dataset was made publicly available during the MICCAI 2017 Digital Pathology Challenge. The 64 images were extracted from TCGA and contained 16 tiles from four different cancer types each, at magnifications of 20x and 40x. The nuclei were annotated by students and reviewed by pathologists. CPM-15, on the other hand, contains 15 images from two different cancer types. Both these datasets contain images of different sizes. 

The Lung Cancer dataset was generated by Qu {\em et al.} in \cite{qu2020weakly}. It contains 40 H\&E stained lung adenocarcinoma and lung squamous cell cancer images. These images are extracted at a magnification of 20x at a size 900 x 900. Each image was annotated by an expert pathologist with bounding boxes, points, and full masks for the experiment. 

\subsection{Evaluation Metrics} \label{subsec:metrics}

For the task of nuclei segmentation, various evaluation metrics have been used over time. Earlier methods calculated the accuracy of the algorithm based on the number of pixels detected as the nucleus within a region of interest (ROI). However, in medical images, there exists a large class imbalance, often with more background information compared to the relevant object of interest in a single image tile. This imbalance can create a bias in metrics like accuracy, leading to an illusion of excellent performance. 

Segmentation models require measures that can assess localization correctness in addition to classification accuracy. Modified metrics measuring the similarity between the ground truth and the predicted values appear to be a better approach to evaluating performance. The most commonly used evaluation metrics are the F1-score, Dice Similarity Coefficient(DSC), the Jaccard Index(JI) or Intersection over Union score (IoU), Aggregated Jaccard Index (AJI) and Panoptic Quality (PQ). The first two metrics are pixel-level metrics, while the AJI is an object level metric. These metrics are often defined by four values: True positives -(TP - predicted true, actual true), True negatives (TN - predicted false, actual false), False positives (FP-predicted true, actual false), and False negatives (FN- predicted false, actual true). The equations in this section denote X as the set of ground truths and Y as the set of predicted instances corresponding to the ground truth. Table \ref{table:eval_metrics} presents a comparison of the available evaluation metrics. 


\begin{table}
\small
\centering
\caption{A summary of the currently proposed evaluation metrics.} \label{table:eval_metrics}

\begin{tabular}{p{0.2\textwidth}p{0.3\textwidth}p{0.3\textwidth}}
\toprule
     \textbf {Metric} &  \textbf {Advantage} &  \textbf {Disadvantage}  \\ 
\toprule
F1-score & Focuses on evaluating the presence of a predicted object corresponding to the ground truth object & Does not account for pixel-level errors 
\\
\midrule
IoU (Jaccard Index) & Measures the conformance of shape between ground truth and prediction & Does not account for object level errors 
\\ 
\midrule
Dice Similarity Coefficient (DSC)  & Measure of pixel wise agreement between ground truth and prediction & Does not penalize detection errors 
\\ 
\midrule
Aggregated Jaccard Index (AJI) & Penalizes both object level and pixel level errors & Over penalization owing to failed detections
\\ 
\midrule
Panoptic Quality (PQ) & Unified scoring of detection and segmentation & Dependent on IoU with a strict threshold and hence may result in a lower score
\\ 
\bottomrule
\end{tabular}
\end{table}
\subsubsection{F1-score}

The F1-score is defined as the harmonic mean of the precision and recall, calculated from the values of TP, TN, FP, and FN. Precision is defined as the percentage of correct positive predictions of all the positive predictions, while recall is defined as the percentage of actual positives that were correctly predicted. This metric is very commonly used in the field of medical image segmentation and considers each instance as an object, giving a per-object evaluation metric. 
\[Precison = \frac{TP}{TP+FP}\]
\[Recall = \frac{TP}{TP+FN}\]

\[F1-score=\frac{2 \times Precision \times Recall}{Precision + Recall} = \frac{2TP}{2TP + FP + FN}\]

\subsubsection{Intersection over Union score or Jaccard Index}

The IoU score or Jaccard Index (JI) is also defined in terms of TP, FP, FN, and TN. One important difference between this index and the DSC is that, JI penalizes undersegmentation and oversegmentation more than DSC. Higher rates of oversegmentation and undersegmentation lead to a lower JI. This measure also accounts for the level of shape concordance between the ground truth and the predicted map. In general, the IoU score or JI is the ratio of the common elements between the ground truth and predicted map to the union of elements in the ground truth and the predicted map. 

\[IoU score=\frac{TP}{TP + FP + FN}\]
\[IoU score=\frac{| X \cap Y |}{| X \cup Y |}\]

Both the Jaccard Index and the Dice Similarity Coefficient account only for the pixel level errors, and don't account for any object level errors. A suitable metric for a segmentation algorithm must penalize the model for any missed objects and false detections in addition to oversegmentation and undersegmentation errors. The Aggregated Jaccard Index (see Sec.~\ref{subsubsec: AJI}) was proposed in \cite{kumar2017dataset} to account for pixel level and object level errors. 

\subsubsection{Dice Similarity Coefficient}

The Dice Similarity Coefficient (DSC) is computed as twice the set of common elements between the ground truth and predictions divided by the total number of elements in each set. This measure provides an overall score for the quality of instance level segmentation. It gives the level of similarity between the ground truth and the predictions. 

\[DSC=\frac{2 (X \cap Y)}{|X| + |Y|}\]

\subsubsection{Aggregated Jaccard Index}\label{subsubsec: AJI}

The Aggregated Jaccard Index (AJI) is an extension of the Jaccard Index that computes an aggregated intersection cardinality in its numerator and an aggregated union cardinality in the denominator for the predicted nuclei and its ground truth. For each nucleus in a ground truth, the AJI is calculated by adding the pixel count of the intersection between the ground truth and predicted segments to the numerator and adding the pixel count of their union to the denominator. This process aggregates the false positives and false negatives in the denominator. Hence, the AJI ensures that all missed detections, false detections, oversegmentation and undersegmentation are accounted for. In the equation below, N refers to the set of false positives from the prediction set.

\[AJI=\frac{\Sigma_{i=1}^{n} X \cap Y}{\Sigma_{i=1}^{n} X \cup Y + \Sigma_{k\in N} Y_k}\]

\subsubsection{Panoptic Quality}

The Panoptic Quality proposed by \cite{kirillov2019panoptic}, is a metric that assesses the combined detection and segmentation quality. The F1 score measures the detection quality (DQ), and the segmentation quality (SQ) is a measure of similarity between the predicted instance and its ground truth. This metric provides a good evaluation of the detection of individual nuclei instances and their segmentation, and overcomes some limitations of the AJI and DSC. In the equation below, x is the ground truth segment, and y is the predicted segment. Each pair (x,y) is unique if its IoU is greater than 0.5, and this matching splits the segments into TP, FP, and FN. 

\[PQ={DQ  \times  SQ}\]
\[PQ=\frac{|TP|}{|TP| + \frac{1}{2}|FP| + \frac{1}{2}|FN| } \times \frac{\Sigma_{(x,y) \in TP}IoU (x,y)}{|TP|}\]
\subsection{Performance Comparison} \label{subsec:performance_comp}

\subsubsection{Quantitative Comparison}

In this subsection we carry out a performance comparison among different models to deduce how different methods and techniques can enhance the segmentation quality.
Table \ref{table:performance_comparison} shows a quantitative performance comparison of a few state-of-the-art segmentation approaches on the MoNuSeg dataset with AJI, which is the commonly used metric for comparisons and more suitable for instance segmentation problem. In the table, Test 1 and Test 2 refer to the data splitting proposed in \cite{kumar2017dataset}, with Test 1 consisting of 16 images only from breast, liver, kidney and prostate, and Test 2 consisting of 14 images from all the seven organs (see Table \ref{table:datasets}). Combined Test Sets refer to the 30 images, including Test Set 1 and 2, while the MoNuSeg Test set refers to the 14 images used for the challenge. 


\begin{table}
\small
\centering
\caption{Performance comparison on the MoNuSeg dataset with the AJI metric.} \label{table:performance_comparison}

\begin{tabular}{p{0.2\textwidth}p{0.1\textwidth}p{0.1\textwidth}p{0.1\textwidth}p{0.1\textwidth}}
\toprule
   \textbf {Method} &  \textbf {Test~1}&  \textbf {Test~2}&  \textbf {Test~1 \& 2 (comb.)} &  \textbf {MoNuSeg Test Set}\\ 
\toprule

\multicolumn{5}{c}{\em \textbf{Unsupervised}} \\
\midrule
Cell Profiler \cite{carpenter2006cellprofiler, zhou2019cia}& 0.1549  & 0.0809 & - & -
\\ 
\midrule
Fiji \cite{schindelin2012fiji, zhou2019cia}& 0.2508  & 0.3030 & - & -
\\ 
\midrule
DDMRL \cite{kim2019diversify} & - &- & - &0.4860
\\ 
\midrule
Scale-Supervised Attention Net \cite{sahasrabudhe2020self} & - & - & - & 0.5354
\\ 
\midrule
CyC-PDAM \cite{liu2020unsupervised} & 0.5432 & 0.5848 & \textbf{0.5610} & -
\\
\midrule
CBM \cite{magoulianitis2022unsupervised} & - & 0.5808 & - & 0.6142
\\ 
\midrule
HUNIS \cite{magoulianitis2022hunis}& - & \textbf{0.6548} & - & \textbf{0.6387}
\\ 

\midrule
\multicolumn{5}{c}{\em \textbf{Supervised}} \\
\midrule
CNN2 \\ \cite{xing2015automatic, kumar2017dataset} & 0.3558 & 0.3354 & - & -
\\ 
\midrule
U-Net (ResNet-50) \\
\cite{he2016deep, lagree2021review}& - & - & - & 0.4882
\\
\midrule
U-Net (VGG-16) \cite{simonyan2014very, lagree2021review} & - & - & - & 0.4925
\\
\midrule
U-Net (DenseNet-201) \\ \cite{huang2017densely, lagree2021review} & - & - & - & 0.5083
\\
\midrule
CNN3 \cite{kumar2017dataset}& 0.5154  & 0.4989 & 0.5083 \cite{chen2020boundary} & -
\\ 
\midrule
Mask \\
R-CNN \cite{he2017mask} & 0.5978 & 0.5531 & 0.5786 \cite{chen2020boundary} & 0.5282 \cite{lagree2021review}
\\
\midrule
DCAN \cite{chen2017dcan} & 0.6082  & 0.5449\cite{zhou2019cia}  & - & 0.557
\\ 
\midrule
PA-Net \\\cite{liu2018path,zhou2019cia} & 0.6011 & 0.5608  & - & -
\\ 
\midrule
BES-Net \\
\cite{oda2018besnet, zhou2019cia}& 0.5906  & 0.5823 & - & -
\\ 
\midrule
HoVerNet \cite{graham2019hover}& -  & -   & 0.618 & -
\\ \midrule
CIA-Net \cite{zhou2019cia} & 0.6129  & 0.6306  & 0.6205 \cite{chen2020boundary} & -
\\ 
\midrule
REU-Net \cite{qin2022reu}& - & - & 0.636 & -
\\ 
\midrule
BRP-Net  \cite{chen2020boundary}& 0.6196 & 0.6384 & 0.6422 & -
\\ 
\midrule
GCP-Net \cite{yang2022gcp} & - & - & 0.651 & -
\\ 
\midrule
Enhanced lightweight U-Net \cite{hancer2023imbalance}& - & - & - & 0.6895
\\ 
\midrule
SSL \cite{xie2020instance} & - & - & - & 0.7063
\\ 
\midrule
Region Based CNN \cite{liang2022region}& - & - & - & 0.73
\\ 
\midrule
DenseResU-Net \cite{kiran2022denseres} & \textbf{0.7998} & \textbf{0.7684} & \textbf{0.7861} & -
\\ 
\midrule
CBA-Net~\cite{thi2022convolutional}   & - & - & - & \textbf{0.7985}
\\ 
\bottomrule
\end{tabular}
\end{table}

Cell Profiler and Fiji are conventional approaches developed for biomedical image analysis. Cell Profiler applies an intensity threshold, while Fiji performs a watershed based nuclear segmentation. We see their results on the MoNuSeg dataset to be very poor. Domain Diversification and Multi-Domain Invariant Representation Learning (DDMRL) is one of the initial deep learning based unsupervised approaches that set the performance standard for domain adaptive methods. It is seen to perform much better on the MoNuSeg dataset, with an increase of almost 0.18 AJI compared to the conventional methods. The CyC-PDAM, with its nuclei inpainting mechanism and panoptic level adaptation, achieves an AJI of 0.5610 overall on the MoNuSeg dataset. On the same lines as self supervision, the attention-based scale prediction network with segmentation as an auxiliary task \cite{sahasrabudhe2020self} performs even better than the supervised CNN based algorithms, with an AJI of 0.5354. This class of self supervision works on the relevant histology dataset, and doesn't require any labeled data, unlike domain adaptive methods. With the trend of deep learning based unsupervised and self supervised methods producing comparable results to supervised methods, the CBM \cite{magoulianitis2022unsupervised} achieves a high performance with a well-designed unsupervised method and a very small computational complexity using simple image processing techniques. The large performance gap between the deep learning based methods and this method can be owed to the significant domain gap in biomedical images and inherent intensity variations, challenged more by the relatively small amount of training data. The HUNIS paper \cite{magoulianitis2022hunis} further improves upon the performance of CBM by introducing a second stage self supervised refinement on the adaptively thresholded result and obtains a substantial improvement in performance, especially in Test 1. The selective self supervision in the second stage based on the confidence scores of the predicted pixels proposes a condition based supervision that can be further explored as an alternative approach to fully supervised segmentation which suffers from a large memory requirement. Among the unsupervised methods, the HUNIS approach achieves the highest AJI of 0.6548 on Test 2 and 0.6387 on the MoNuSeg Test Set. It outperforms all unsupervised and self supervised deep learning methods and requires only a negligible number of parameters compared to the millions of parameters of deep learning networks. In addition, this method performs on par with U-Net based REU-Net, attention gated GCP-Net, and the enhanced lightweight U-Net. 

CNNs and FCNs (Fully Connected Networks) formed the beginning of deep learning for nuclei segmentation. CNN2, \cite{xing2015automatic} classifying pixels as nuclei or background, performed better than the conventional methods, but is particularly challenged in segmenting dense nuclei clusters. As seen in Table \ref{table:performance_comparison}, CNN 2 obtains an AJI of 0.3558 on Test Set 1 and 0.3354 on Test 2. CNN3 \cite{kumar2017dataset} included a boundary class that showed an improvement in the segmentation of nuclei with diffused chromatin and forming dense clusters. This helped increase the AJI by about 0.16, reaching an AJI of 0.5083. The multi-tasking FCN used in DCAN provided encouraging results for deep learning based methods, with an AJI close to 0.60 in some test sets. 

The concept of upsampling to obtain a pixel-wise segmentation prediction gave rise to the U-Net, an encoder-decoder architecture that performs downsampling to obtain features and builds on these features through upsampling to obtain a segmentation map with similar dimensions as the input. Different configurations in the encoder architecture were developed to extract efficient and representative features. Among the U-Net implementations, the deep DenseNet-201 outperforms networks with other backbones like VGG-16 or the ResNet, by achieving an AJI of 0.5083. This promising approach led to advanced U-Net based architectures like BES-Net~\cite{oda2018besnet}, CIA-Net~\cite{zhou2019cia}, REU-Net~\cite{qin2022reu}, and BRP-Net~\cite{chen2020boundary} that incorporated boundary information to refine the segmentation masks in an end-to-end manner. From the table, we find that boundary supervision contributes to a notable improvement of around 0.09 - 0.14 in AJI, and most of these methods yield an AJI of about 0.60-0.64. HoverNet~\cite{graham2019hover} displays similar performance with an AJI of 0.618 with the help of horizontal-vertical distance map to separate clustered nuclei. GCP-Net is an attention gated network that achieves an AJI of about 0.65. The inclusion of the attention gates suppresses irrelevant pixels from further processing, thereby focusing more on the nuclei regions. Also, adding more context from coarser layers is proven to help the segmentation quality. This approach contributes to improved performance even without contour awareness. 

Histology images often have a class imbalance issue between the nuclei and background pixels. Such an imbalance may introduce a bias in the network. The instance aware self supervised network~\cite{xie2020instance} based on contrastive learning achieves an AJI of 0.70. Though SSL uses nuclei size and quantity priors as the self-supervised pretraining, the best performance is achieved by finetuning the network with 100\% labeled data. An imbalance aware network proposed by Hancer {\em et al.}~\cite{hancer2023imbalance} uses an enhanced lightweight U-Net supervised by the generalized Dice Loss, with an AJI of 0.6895. The DenseResU-Net shows a leap in performance among the U-net methods with AJIs greater than 0.76 on different test sets. Its wise use of atrous blocks in a dense network with residual connections between the encoder and decoder helps reduce the semantic gap. The CBA network~\cite{thi2022convolutional} obtains the highest AJI of 0.7985 among all the supervised deep learning methods. The integration of the attention mechanism and the blur pooling operations overcomes the challenges of variations in staining, while the low pass filtering allows the extraction of enhanced features, thus contributing to SOTA performance. 

In addition to the U-Net, region based CNNs like the Mask RCNN have also shown favorable results. PA-Net and the region based CNN~\cite{liang2022region} employ improved Mask RCNN architectures. While PA-Net achieves an AJI of 0.60, the region based CNN with a guided anchor RPN and Fusioned Box Score hikes the performance by another 0.10 giving a 0.73 AJI. However, it should be noted that Mask RCNN suffers from slow speed, especially with large images, and requires an enormous number of parameters for training. 

\begin{figure*}[t]
\begin{center}
\includegraphics[width=.5\linewidth]{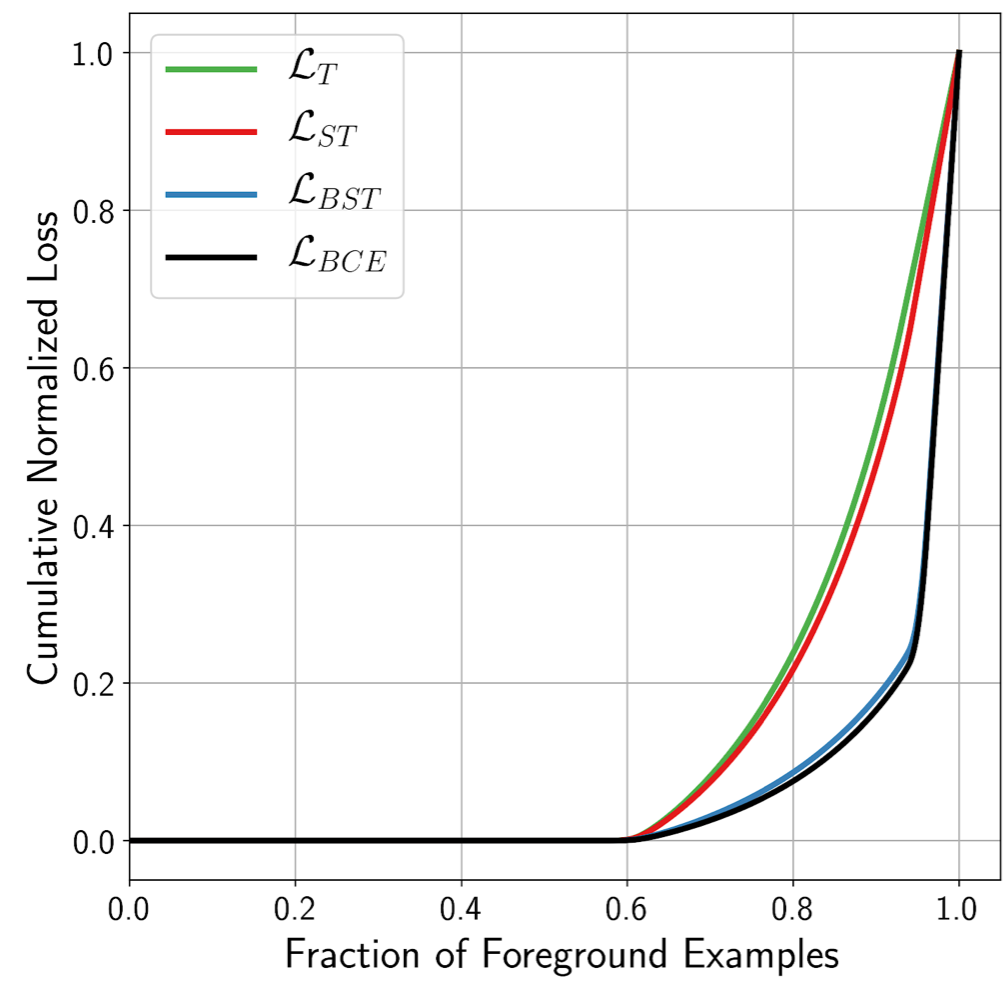}
\caption{ Cumulative distribution of the loss value over the ratio of foreground examples seen from the model. Given the very skewed distribution, a small percent of examples actually contribute in the optimization of the model (results from \cite{zhou2019cia}).} \label{fig:CIA_Weakly}
\end{center}
\end{figure*}

After comparing results from different SOTA methods, we compile some findings with more emphasis on the weakly supervised aspect of the problem. Zhou {\em et. al.}~\cite{zhou2019cia} observed that noise in the staining and digitization process gives rise to ambiguous instances, thus leading to more noise in labels from pathologists' subjective annotations. In Fig.~\ref{fig:CIA_Weakly}, we can see that only the top 10\% of samples have an influence on the 80 \% of the overall cross-entropy loss value, and the really informative regions are more scarce. 

From the weak supervision and point-wise annotations standpoint, it is interesting to analyze how performance is affected by using different fractions of annotated points. Looking into the study of Qu {\em et al.}  \cite{qu2020weakly}, we can realize the more supervision is added the better the results are. As an observation, pixel level metrics, such as the reported accuracy and F1 score (at pixel level) improve marginally by adding more supervision (see Table \ref{tab:Weakly_Comp}). Yet, the performance improvement is more accentuated when using the Dice and AJI metrics object-wise. Another observation from weak supervision, on the Lung Cancer dataset, the performance drops significantly if we reduce the number of training points at half. On the other hand, for the MoNuSeg, the performance is very close in terms of AJI and Dice. Even using 10\% of the points, the AJI performance gap is still small. 

\begin{table}[h]
\centering
\caption{Weak supervision comparison between full supervision and point-wise at different training point ratios (results from \cite{qu2020weakly}).}\label{tab:Weakly_Comp}
\begin{tabular}{|c| c| c| c| c| c|} \hline
 Dataset   & Method  & $Acc_{pixel}$ & $F1_{pixel}$ & $Dice_{obj}$ & $AJI_{obj}$ \\ \hline
\multirow{4}{*}{Lung Cancer (LC)} & Fully-sup & 0.9615 & 0.8771 & 0.8521 & 0.6979 \\ 
    & GT points & 0.9427 & 0.8143 & 0.8021 & 0.6497 \\ 
    \cline{2-6}
    & 5 \% & 0.9262 & 0.7612 & 0.7470 & 0.5742  \\ 
    & 10 \% & 0.9312 & 0.7700 & 0.7574 & 0.5754 \\
    & 25 \% & 0.9331 & 0.7768 & 0.7653 & 0.6003 \\ 
    & 50 \% & 0.9332 & 0.7819 & 0.7704 & 0.6120 \\ \hline
\multirow{4}{*}{MoNuSeg (MO)} & Fully-sup & 0.9194 & 0.8100 & 0.6763 & 0.3919 \\ 
    & GT points & 0.9097 & 0.7716 & 0.7242 & 0.5174 \\ 
    \cline{2-6}
    & 5 \% & 0.8951 & 0.7540 & 0.7015 & 0.4941  \\ 
    & 10 \% & 0.8997 & 0.7490 & 0.7033 & 0.5031 \\
    & 25 \% & 0.8966 & 0.7511 & 0.7087 & 0.5120 \\ 
    & 50 \% & 0.8999 & 0.7566 & 0.7157 & 0.5160 \\ \hline
\end{tabular}
\end{table}

Another question that may arise is how supervision helps the generalization in other datasets. In Table \ref{tab:General_Weakly_Comp}, one can see that the AJI score is not affected when using much less training data (even 5\%)  while training using MoNuSeg and testing on the Lung Cancer dataset. Conversely, there is a small performance improvement when adding more training points when we use the Lung Cancer for training and MoNuSeg for testing. This may be attributed to the smaller size of the LC dataset that challenges the model when a very small fraction of points is included for training. In general, performance drops in both datasets, where LC testing is affected more by the domain shift, while the performance difference because of the training domain shift is less in MoNuSeg.

\begin{table}[h]
\centering
\caption{Weak supervision comparison between full supervision and point-wise at different training point ratios (results from \cite{qu2020weakly}).}\label{tab:General_Weakly_Comp}
\begin{tabular}{|c| c| c| c| c| c|} \hline
 $Train \rightarrow Test$   & Ratio  & $Acc_{pixel}$ & $F1_{pixel}$ & $Dice_{obj}$ & $AJI_{obj}$ \\ \hline
\multirow{4}{*}{$MO \rightarrow LC$} & 5 \% & 0.9271 & 0.7589 & 0.7418 & 0.5609  \\ 
    & 10 \% & 0.9213 & 0.7518 & 0.7297 & 0.5555 \\
    & 25 \% & 0.9222 & 0.7551 & 0.7320 & 0.5588 \\ 
    & 50 \% & 0.9226 & 0.7579 & 0.7336 & 0.5608 \\ \hline
\multirow{4}{*}{$LC \rightarrow MO$} & 5 \% & 0.9004 & 0.7419 & 0.7028 & 0.4884  \\
    & 10 \% & 0.8964 & 0.7338 & 0.6913 & 0.4971 \\
    & 25 \% & 0.8974 & 0.7234 & 0.6886 & 0.4870 \\ 
    & 50 \% & 0.8970 & 0.7232 & 0.6986 & 0.5030 \\ \hline
\end{tabular}
\end{table}

\begin{figure*}[t]
\begin{center}
\includegraphics[width=1.0\linewidth]{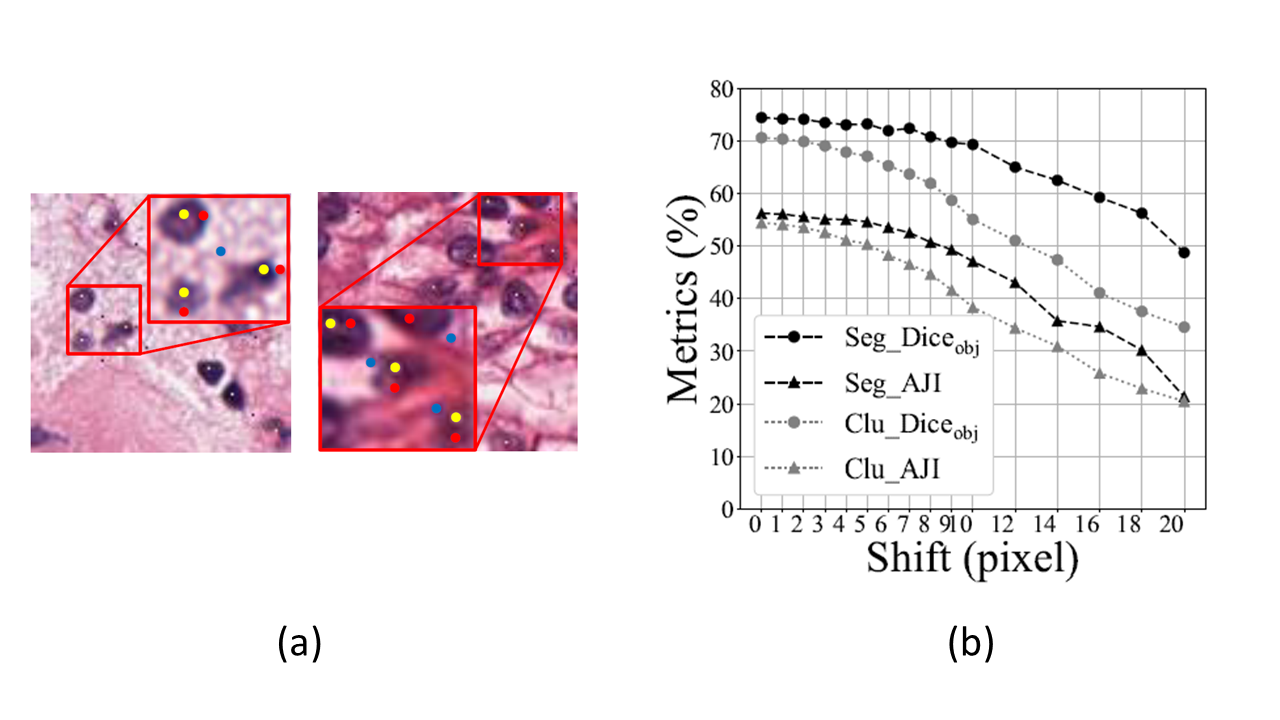}
\caption{(a) Perturbations in point-wise annotations. Yellow points represent nuclei center, while red and blue are points offset by four and eight pixels, respectively. (b) Object-wise metrics for nuclei segmentation over different amounts of perturbations measured in pixel distance from nuclei center. (results from \cite{lin2022label}).} \label{fig:PointPert_Quant}
\end{center}
\end{figure*}

Moving along the point-wise annotations, annotation errors from pathologists can be simulated as perturbation noise and GT points can be some pixels further from the actual center. In Fig.~\ref{fig:PointPert_Quant}, we can see how the noise during annotation can affect the segmentation performance, especially when the points have a distance larger than 8 pixels from nuclei center, reaching closer to nuclei boundaries or even falls out from nuclei.

\subsubsection{Qualitative Comparison}


As has been stressed from many papers, the main source of errors lies in the contours of nuclei that compromises the segmentation performance. Overlapping or touching nuclei in dense clusters and blurred boundaries can cause over or under segmentation problems. Contour aware attention mechanisms in a Deep Neural Network (DNN) turn out to help the network to delineate the nuclei boundaries more accurately. 

\begin{figure*}[t]
\begin{center}
\includegraphics[width=1.0\linewidth]{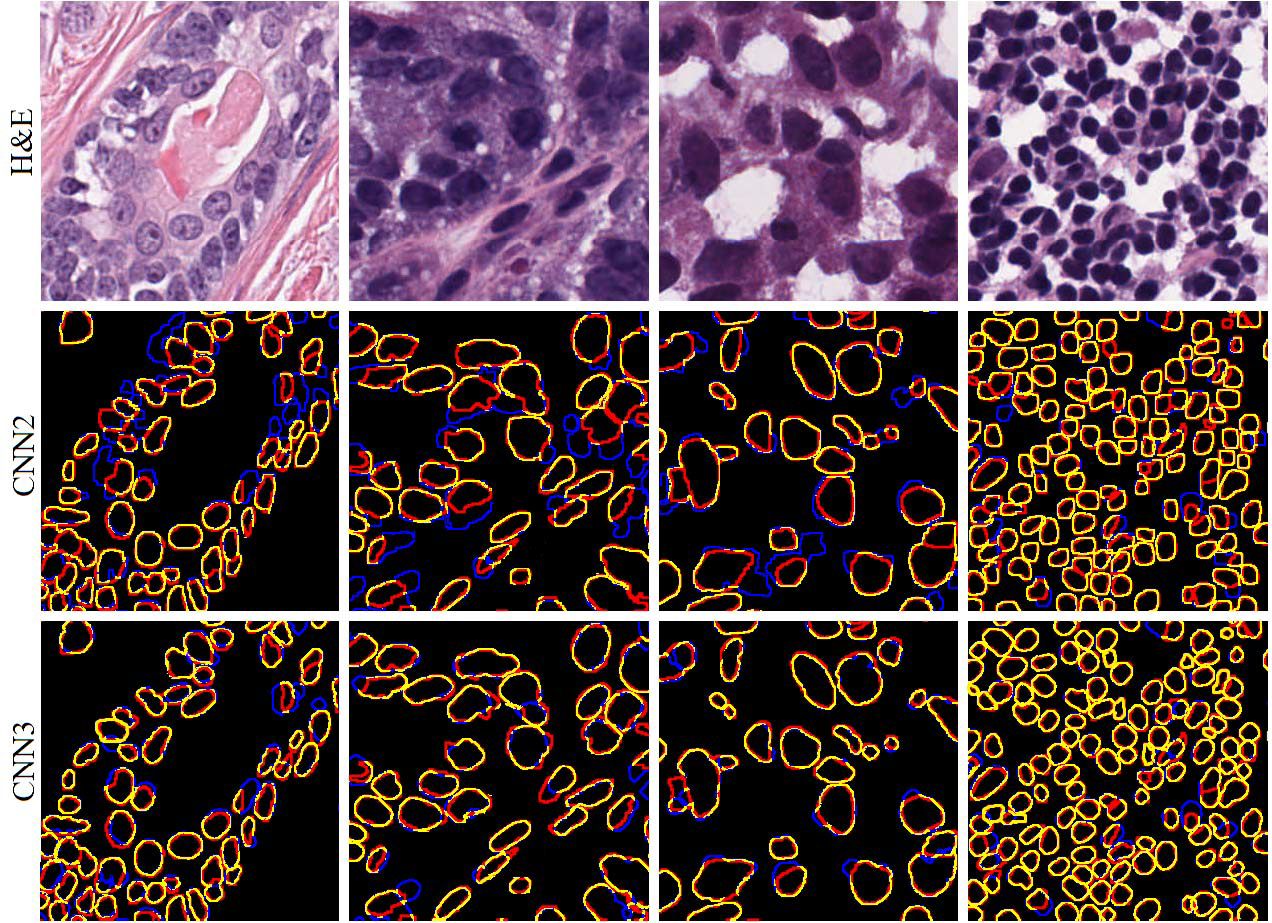}
\caption{In the middle and bottom rows, ground
truth (annotated) boundaries are red, detected are blue, and the overlap between the two is yellow. Segmentation comparisons are shown with the CNN2 baseline model. In bottom row yellow is more prevalent, thus indicating more precise boundaries detection with respect to ground truth. (figure from \cite{kumar2017dataset}).} \label{fig:CNN3_Qual}
\end{center}
\end{figure*}

CNN3 \cite{kumar2017dataset} introduces a third class for detecting the boundaries of cells. The main motivation of that is that in a post-processing step, touching nuclei can be accurately segmented. This is possible by trying to grow the nuclei area in an iterative way, maximizing the boundary probability, without decreasing the nuclei one of other neighboring instances (constrain invasions). They let the nuclei anisotropically grow until boundary-class reaches a local minima, while constraining the growth with the inside (nuclei) and outside (background) classes of surrounding areas. In comparison with CNN2 \cite{xing2015automatic}, it helps to identify boundaries and touching nuclei more precisely (see Fig. \ref{fig:CNN3_Qual}).

\begin{figure*}[t]
\begin{center}
\includegraphics[width=1.0\linewidth]{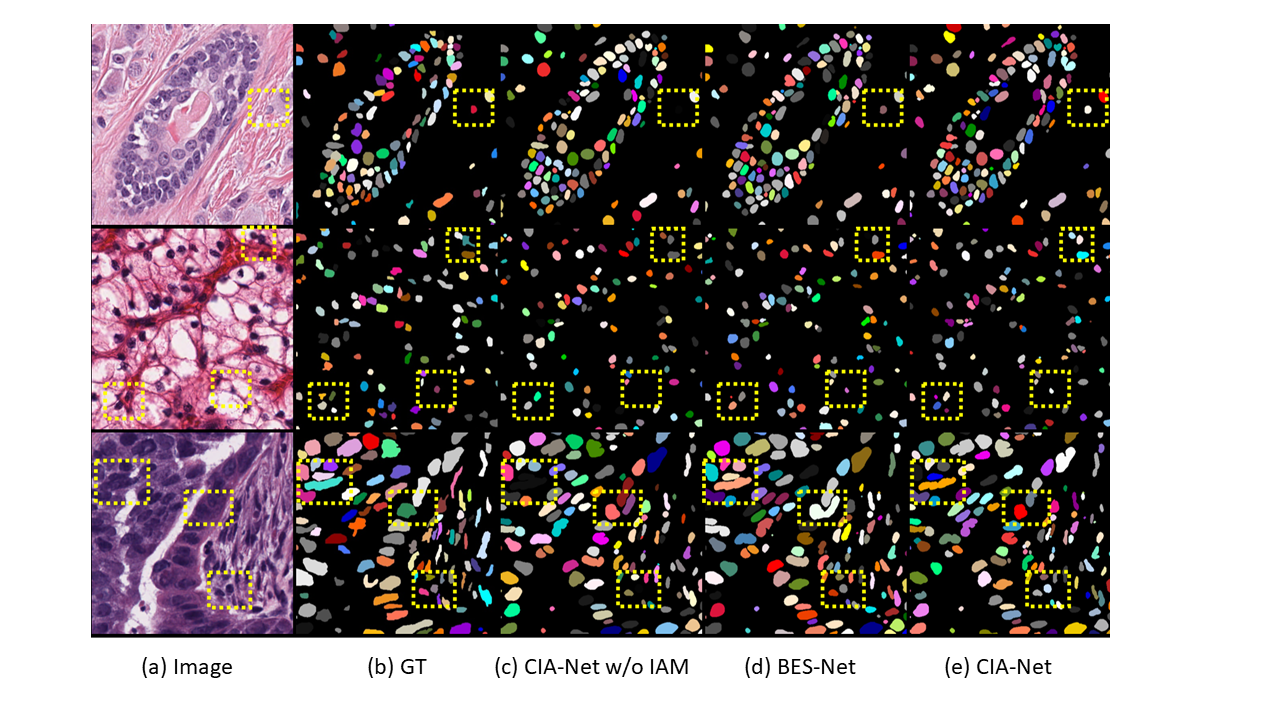}
\caption{ Comparison of CIA-Net without the Information Aggregation Module (IAM) and BES-Net. CIA-Net can identify more accurately connected nuclei that should be slit, even when ground truth is noisy or mislabelled
(figure from \cite{zhou2019cia}).} \label{fig:CIA_Qual}
\end{center}
\end{figure*}

Subjective annotations results in mislabelled instances and inaccurate boundary delineations. The inter-annotator variance is even more evident due to blurred edges and staining artifacts. Based on the observation that noise in labeling has the tendency to statistically dominate the gradients and hence loss calculations. CIA-Net \cite{zhou2019cia} shows an improved performance over earlier methods by adding the ``truncated loss" to diminish the influence in the learning of the outlier regions with high confidence. Focusing on the more informative regions in the training process helps mitigate the over-segmentation problem. In Fig. \ref{fig:CIA_Qual}, one can see the effectiveness of adding the IAM unit that focuses on the texture and spatial dependence between nuclei and their boundaries, as well as that of truncated loss function for more accurate boundary detection, despite the noisy labels.  

\begin{figure*}[t]
\begin{center}
\includegraphics[width=1.0\linewidth]{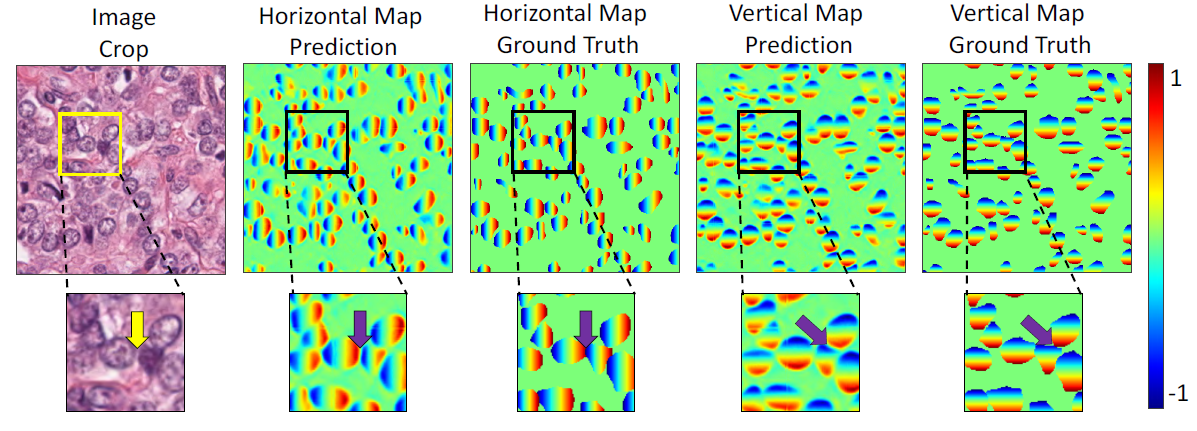}
\caption{Horizontal and vertical prediction maps are shown on areas prone to overlapping nuclei, along with their corresponding ground truth. Distance information alleviates over-segmentation and nuclei splitting phenomena (adjusted figure from \cite{graham2019hover}).} \label{fig:Hover_Qual}
\end{center}
\end{figure*}

As mentioned, another way to mitigate over-segmentation and resolve overlapping instances is proposed in Hover-Net \cite{graham2019hover}. Instance-wise horizontal and vertical distances from their respective center of mass provide rich information to the encoder branch, on top of textural features. In Fig. \ref{fig:Hover_Qual}, we can see how the distance information helps in splitting the nuclei apart.

Cropped image regions show horizontal and vertical map predictions, with corresponding ground truth. Arrows
highlight the strong instance information encoded within these maps, where there is a significant difference in the pixel values.

\begin{figure*}[t]
\begin{center}
\includegraphics[width=1.0\linewidth]{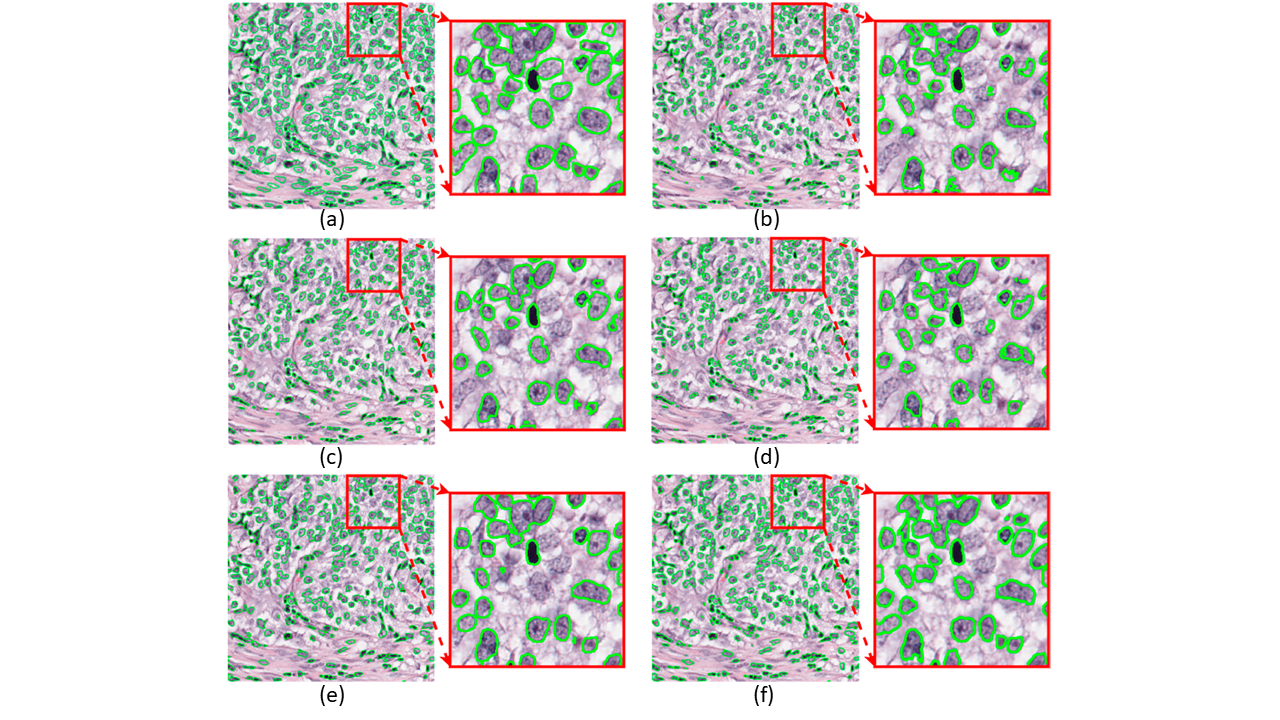}
\caption{A challenging nuclei segmentation comparison among several models, demonstrating the gated attention mechanism. (a) GT, (b) Baseline U-Net \cite{ronneberger2015u}, (c) CNN2 \cite{xing2015automatic}, (d) CNN3 \cite{kumar2017dataset}, (e) Hover-Net \cite{graham2019hover}, (f) NucleiSegNet \cite{lal2021nucleisegnet} (adjusted figure from  \cite{lal2021nucleisegnet}).} \label{fig:SegNet_Qual}
\end{center}
\end{figure*}

Besides overlapping nuclei, another major challenge in this area is nuclei size variability among different organs, datasets, and scanning protocols. Attention mechanisms among the encoder and decoder have been proven efficient because they bring more contextual information from the coarse feature maps with larger receptive fields. In \cite{lal2021nucleisegnet}, the coarser features are used as a gating signal within the gated attention on the skip connections between the encoder and decoder. In Fig. \ref{fig:SegNet_Qual}, we can see the effect of such a gated mechanism over other SOTA models. It can identify and segment more accurately both small and large size nuclei with different textures, thus decreasing the false negative rate.

\begin{figure*}[t]
\begin{center}
\includegraphics[width=1.0\linewidth]{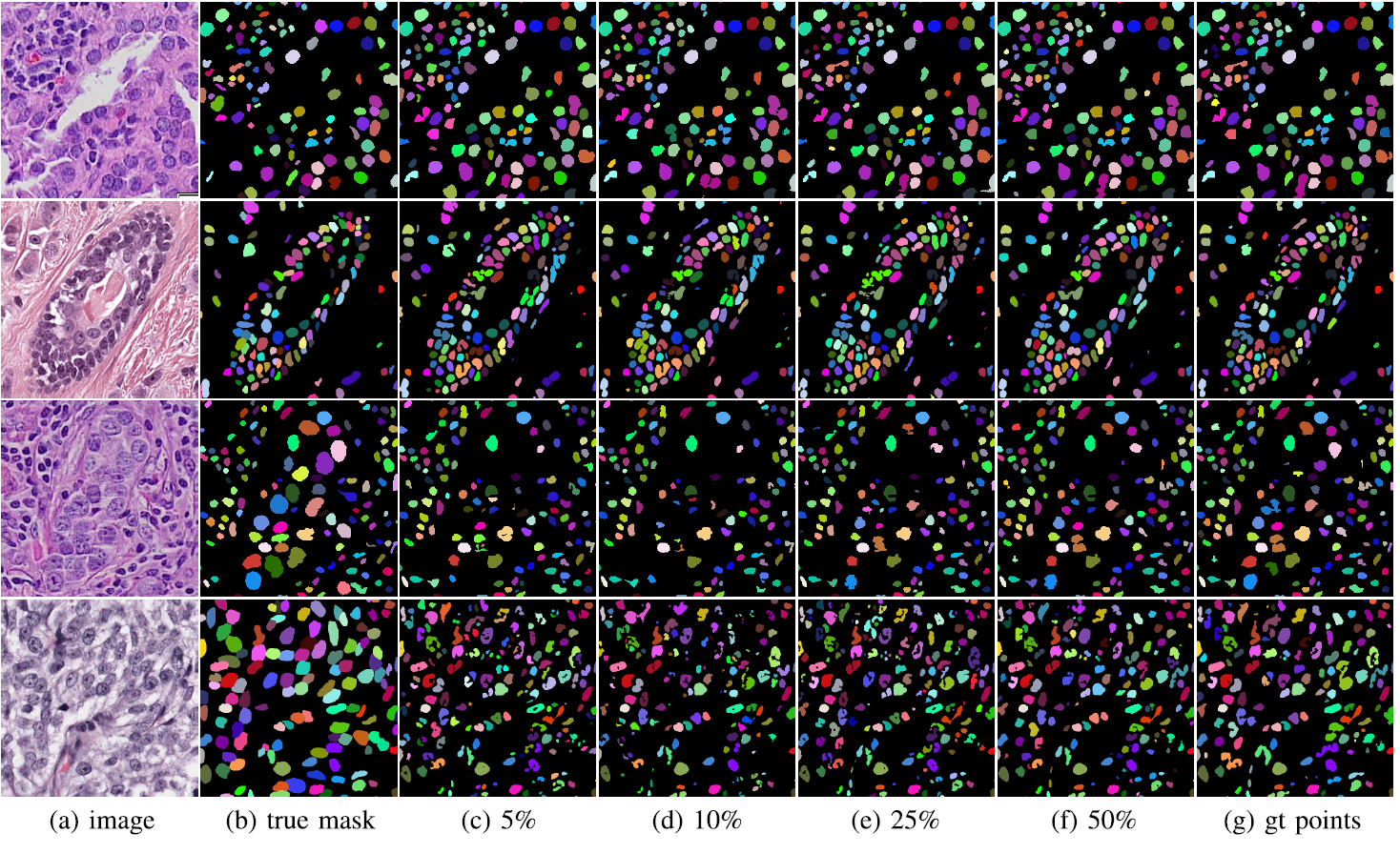}
\caption{Demonstration of weak supervision from point labels, using different ratio of points for training. Starting from top row, odd rows display results from LC dataset, while even rows from the MO dataset. Upper examples show typical segmentation cases, while bottom examples more challenging ones Gt points refer to the full set of available annotations. (combined figures from \cite{qu2020weakly})} \label{fig:Point_Qual}
\end{center}
\end{figure*}

Qu {\em et al.} ~\cite{qu2020weakly} carry out a visualization comparison using certain fractions of point-wise annotations to show how reducing the training points can affect the nuclei segmentation performance. In Fig. \ref{fig:Point_Qual}, we can observe that for easier images where there are not many nuclei texture/color variations, even when less than 50\% of point annotations are used, the segmentation quality is good. On the other hand, for other more challenging images where nuclei texture varies significantly, one can see that nuclei are more under-segmented and only using 50\% or the full set of point annotations, the segmentation result is more accurate.

\newpage

\subsection{Discussion And Conclusions} \label{subsec:discussion}

Having reviewed a large body of papers pertinent to nuclei segmentation from different categories and analyzed their quantitative and qualitative performance on public datasets, it is time to discuss our observations and draw some conclusions. 

\subsubsection{General Remarks}

As a first point, early generic molecular segmentation tools used for nuclei segmentation perform very poorly compared to dedicated models that target the very task. Unsupervised methods using domain adaptation or predictive learning seem to achieve a better performance than methods that use contrastive learning, and they are dominant. Yet, all DL-based methods with no supervision have a much inferior performance compared to the fully or weakly supervised ones. Turns out that transferring meaningful features between two different domains, either from natural or medical images, is very challenging, especially when presented with very few training images, which is the case for nuclei segmentation. On the other hand, the two recent unsupervised methods of CBM and HUNIS, based on more traditional segmentation ways and prior knowledge of the problem, achieve a competitive performance even among supervised DL-based solutions. This is to say that still, traditional techniques, when effectively applied, can provide a high performance solution with a small number of parameters and in a more transparent way. 

Early CNN architectures for binary pixel classification, before U-Net becomes the mainstream baseline for nuclei segmentation, definitely improved the performance over earlier traditional methods. Although, they are challenged by class imbalance problems and segmenting dense nuclei clusters. FCNs further improved the segmentation performance, especially when coupled with multi-tasking branches that focus on the contours of nuclei. 

In more recent approaches, the dominant baseline DNN architecture for nuclei segmentation is the U-Net. The best performance is yielded when DenseNet is used as a backbone model. Also, additional extensions on top of the main model, such as attention mechanisms that bring more contextual information at different scales and exploit the relevance between nuclei and contour, help the classifier to detect and segment nuclei of different sizes. Drawing a piece of evidence from CIA-Net and BRP-Net, this not only increases the segmentation performance but also improves the generalization ability on an unseen organ since those two models perform better on Test-2 (unseen organ) than Test-1 set in MoNuSeg. Moreover, boundary awareness during training definitely helps a model boost its segmentation performance. That can be achieved either using a separate class for boundaries, combined with a post-processing method, or via gated mechanisms that learn more from the informative regions (i.e. boundaries and nuclei), hence suppressing other background information that is more noisy. Mask RCNN has been proven to be effective in other computer vision tasks, albeit in nuclei segmentation, its performance is not convincing compared to other baseline models. It requires many annotated data for training and, thereby is not very practical for this task. Besides, its large model size and slow inference time make it less efficient for deployment. 

\subsubsection{Weak Supervision Standpoint}

One commonly faced challenge in this problem is the noisy labels because that is a laborious and sensitive task, prone to errors and subjectivity. One question that arises is how models can identify the erroneous labels and prevent their influence on the learning model (in terms of gradient propagation). In Fig. \ref{fig:CIA_Weakly}, we can see that only the top 10\% of samples have an influence on the 80 \% of the distribution of overall cross-entropy loss. That is, the really informative regions are scarce, and most foreground examples have a minuscule contribution to the learning process and wrong labels largely affect training. As a remedy, some efficient techniques applied are gated attention for boosting contextual multi-scale information, contour-specific task learning to complement nuclei appearance, and loss functions that can account for class imbalance and suppress noisy regions. The utter goal is to steer the learning process toward more informative regions and rely less on labels that convey noise and may be misleading to gradient descent process. Models are better off trained by putting more emphasis on the nuclei and their corresponding contours, digging out the labels that conform with the model's predictions (more informative about nuclei shape, texture, and boundaries).    

Trying to ease the annotation process (can save almost 88\% \cite{qu2019weakly} of the full pixel annotation time) and enable access to larger annotated datasets, partial (or point-wise) annotations is a recent line of research in nuclei segmentation that has attracted a lot of interest. Observing results that include certain fractions of the overall points, the more data points we include, the higher the performance is. Yet, a remarkable note is that even using a very small fraction of points --5 or 10\%-- the performance does not change much compared to the full set of annotated points. Notably, the initial choice of a set of points seems not to play a significant role in the model's performance. Certainly, models trained on the full pixel-wise annotations seem to have a better performance, nevertheless, the gap is relatively small (about 7 \%, when using point-wise annotations. 

From these findings, it is evident that exhaustively annotated datasets with pixel-wise labels are not much favored. The tremendous savings in annotation time from the point-wise labels --given the small performance gap with the full mask based methods-- provides a better trade off for future research, as it can enable the acquisition of larger datasets at much less cost. Moreover, weak supervision using a reduced training set of points does not affect much the generalization performance across different datasets, although this needs to be validated with more experiments in the future using larger testing datasets. Finally, perturbations in point-wise annotations resulting from human error are tolerated by a certain amount of pixels. After some point, the performance drops significantly since nuclei positions are very misleading, and Voronoi labels become too noisy. It is important then to identify the amount of labels, their type, and the areas that needed annotations for minimizing the labels requirement but still achieving a high performance, satisfying pathologists so that they can use future CAD tools in their everyday cancer diagnosis pipeline.

\section{Future Work}\label{sec:future work}

All in all, supervision seems advantageous to the nuclei segmentation task, but the amount and type of supervision is the key for foreseeing the future line of research. The impractical and costly nature of pixel-wise annotations inhibits the rapid growth of the field, taking into account that DL-models require by their nature a large amount of data samples. Trying to extrapolate from current methods, fully unsupervised methods are hard to achieve a high performance. The nuclei color and texture variations from the staining and digitization process, make domain adaptation very challenging for unsupervised methods, and hence their generalization ability is poor. However, still strong prior knowledge and simple image processing techniques (e.g. HUNIS method) can provide a competitive performance. Although, there is still a gap compared to state-of-the-art DL-based fully supervised methods. 
 
Future models in nuclei segmentation will be mainly trained using the weak supervision paradigm. Point-wise annotations reduce dramatically the annotation time and costs. Taking into account that the performance gap with fully supervised methods is relatively small, as well as this research trend is more recent and has not been fully explored, one can argue that point-wise based methods pave the way for future trends.

Taking it one step further, from our earlier observations, achieving a high performance may not require all the nuclei points to be annotated. This shows a direction, where future algorithms will be trained on points from a few nuclei that are representative of the overall image distribution. This is expected to minimize pathologists' labor and yield datasets with more annotations. Therefore, AI algorithms can be part also of the annotation tool, thus making the nuclei annotation process more to the point. That is, we envision a system that suggests what nuclei need supervision and guides the segmentation process by indicating to pathologists which nuclei they need to pinpoint on their centers. This will reduce human fatigue and hence shifting errors in point-wise annotations that as shown earlier can impact the performance. 

\begin{figure*}[t]
\begin{center}
\includegraphics[width=1.0\linewidth]{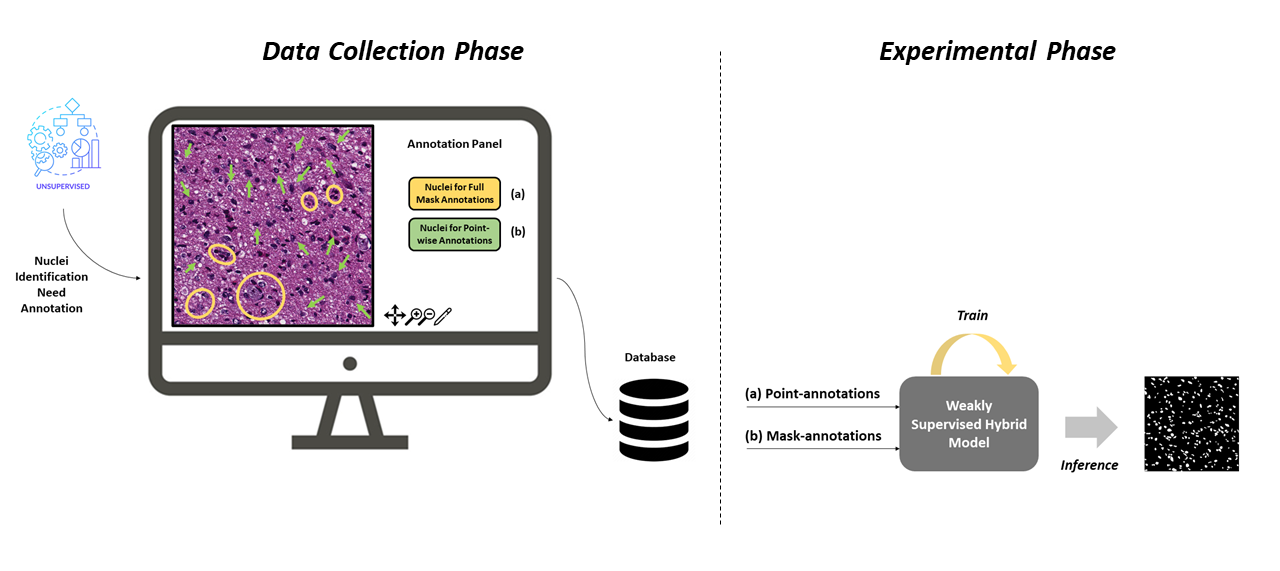}
\vspace{-15mm}
\vspace{5mm}
\caption{The envisioned workflow for CAD-assisted annotations in nuclei segmentation from an unsupervised model to identify areas need annotation. Two modes of annotations may be required from future models to achieve a high performance: (a) point-wise and (b) mask annotations. A hybrid model that can best leverage this information will benefit from a minimum set of annotations to achieve high performance. Certainly, the quality of training depends on the unsupervised model that guides the annotation process.} \label{fig:future_cad}
\end{center}
\end{figure*}

The question arises is for which nuclei we need human supervision. Furthermore, what is the smallest and most representative subset of nuclei points needs to be annotated, so the model under training has a representative distribution about the nuclei of a histopathology image and can potentially maximize the generalization ability to others nuclei as well. For, easy to segment nuclei may not need supervision and unsupervised methods can perform well already. Also, outlier nuclei in appearance should not be included, trying to filter out noisy annotations.
For pointing to the annotator, which pixels need supervision, an unsupervised model would be preferably deployed to identify the group of nuclei whose appearance is more challenging. A supervised model pre-trained on another dataset could also be an option, nevertheless, the bias that would be carried over may give a distorted idea of the areas that need annotation. Hence, we argue that an unsupervised model is more intuitive for identifying areas and nuclei that need supervision. The key to that is to find a representation about nuclei where different appearance aspects can be encoded. Therefore, it would be easy to identify inlier and outlier nuclei from a distribution based on Gaussian distance criteria. 

As another path, in some challenging images, for certain groups of nuclei (e.g. high inner nuclei variation appearance or overlapping nuclei) may also need their full segmentation masks to enhance the annotated data and help in the model training. Thus, it would be interesting in the future to develop hybrid models that could be trained from an annotated dataset that comprises point-wise and fully annotated masks together in areas that models need pixel-level supervision to achieve a very high performance. Fig.~\ref{fig:future_cad} illustrates the workflow pipeline we imagine for future data annotation in nuclei segmentation.

With regards to transparent solutions for medical applications, interpretable models will be favored in the future, so pathologists can understand the segmentation output that comes out of an AI-assisted CAD tool and what factors (i.e., visual features) led to the result. Therefore, apart from a good segmentation performance, explainability and transparency in the nuclei segmentation pipeline are also of high importance, as it will make the future CAD tools more trustworthy to pathologists and hence they can more easily be integrated into their everyday clinical diagnosis pipeline. 

\bibliographystyle{ieeetr}
\bibliography{main}

\begin{thebibliography}{100}

\bibitem{murphy2021mortality}
S.~L. Murphy, K.~D. Kochanek, J.~Xu, and E.~Arias, ``Mortality in the united
  states, 2020,'' 2021.

\bibitem{gurcan2009histopathological}
M.~N. Gurcan, L.~E. Boucheron, A.~Can, A.~Madabhushi, N.~M. Rajpoot, and
  B.~Yener, ``Histopathological image analysis: A review,'' {\em IEEE reviews
  in biomedical engineering}, vol.~2, pp.~147--171, 2009.

\bibitem{guerrero2022software}
R.~E.~D. Guerrero, L.~Carvalho, T.~Bocklitz, J.~Popp, and J.~L. Oliveira,
  ``Software tools and platforms in digital pathology: a review for clinicians
  and computer scientists,'' {\em Journal of Pathology Informatics}, vol.~13,
  p.~100103, 2022.

\bibitem{elmore2016variability}
J.~G. Elmore, H.~D. Nelson, M.~S. Pepe, G.~M. Longton, A.~N. Tosteson,
  B.~Geller, T.~Onega, P.~A. Carney, S.~L. Jackson, K.~H. Allison, {\em
  et~al.}, ``Variability in pathologists' interpretations of individual breast
  biopsy slides: a population perspective,'' {\em Annals of internal \\
  medicine}, vol.~164, no.~10, pp.~649--655, 2016.

\bibitem{stolk2019false}
T.~T. Stolk, I.~J. de~Jong, T.~C. Kwee, H.~B. Luiting, S.~V. Mahesh, B.~H.
  Doornweerd, P.-P.~M. Willemse, and D.~Yakar, ``False positives in pirads (v2)
  3, 4, and 5 lesions: relationship with reader experience and zonal
  location,'' {\em Abdominal Radiology}, vol.~44, pp.~1044--1051, 2019.

\bibitem{ho2022cumulative}
T.-Q.~H. Ho, M.~C. Bissell, K.~Kerlikowske, R.~A. Hubbard, B.~L. Sprague, C.~I.
  Lee, J.~A. Tice, A.~N. Tosteson, and D.~L. Miglioretti, ``Cumulative
  probability of false-positive results after 10 years of screening with
  digital breast tomosynthesis vs digital mammography,'' {\em JAMA Network
  Open}, vol.~5, no.~3, pp.~e222440--e222440, 2022.

\bibitem{kanan2020independent}
C.~Kanan, J.~Sue, L.~Grady, T.~J. Fuchs, S.~Chandarlapaty, J.~S. Reis-Filho,
  P.~G. Salles, L.~M. da~Silva, C.~G. Ferreira, and E.~M. Pereira,
  ``Independent validation of paige prostate: Assessing clinical benefit of an
  artificial intelligence tool within a digital diagnostic pathology laboratory
  workflow.,'' 2020.

\bibitem{lagree2021review}
A.~Lagree, M.~Mohebpour, N.~Meti, K.~Saednia, F.-I. Lu, E.~Slodkowska,
  S.~Gandhi, E.~Rakovitch, A.~Shenfield, A.~Sadeghi-Naini, {\em et~al.}, ``A
  review and comparison of breast tumor cell nuclei segmentation performances
  using deep convolutional neural networks,'' {\em Scientific Reports},
  vol.~11, no.~1, p.~8025, 2021.

\bibitem{zhou2019cia}
Y.~Zhou, O.~F. Onder, Q.~Dou, E.~Tsougenis, H.~Chen, and P.-A. Heng, ``Cia-net:
  Robust nuclei instance segmentation with contour-aware information
  aggregation,'' in {\em Information Processing in Medical Imaging: 26th
  International Conference, IPMI 2019, Hong Kong, China, June 2--7, 2019,
  Proceedings 26}, pp.~682--693, Springer, 2019.

\bibitem{nasir2022nuclei}
E.~S. Nasir, A.~Parvaiz, and M.~M. Fraz, ``Nuclei and glands instance
  segmentation in histology images: a narrative review,'' {\em Artificial
  Intelligence Review}, pp.~1--56, 2022.

\bibitem{hayakawa2021computational}
T.~Hayakawa, V.~S. Prasath, H.~Kawanaka, B.~J. Aronow, and S.~Tsuruoka,
  ``Computational nuclei segmentation methods in digital pathology: a survey,''
  {\em Archives of Computational Methods in Engineering}, vol.~28, pp.~1--13,
  2021.

\bibitem{irshad2013methods}
H.~Irshad, A.~Veillard, L.~Roux, and D.~Racoceanu, ``Methods for nuclei
  detection, segmentation, and classification in digital histopathology: a
  review—current status and future potential,'' {\em IEEE reviews in
  biomedical engineering}, vol.~7, pp.~97--114, 2013.

\bibitem{zhou2020comprehensive}
X.~Zhou, C.~Li, M.~M. Rahaman, Y.~Yao, S.~Ai, C.~Sun, Q.~Wang, Y.~Zhang, M.~Li,
  X.~Li, {\em et~al.}, ``A comprehensive review for breast histopathology image
  analysis using classical and deep neural networks,'' {\em IEEE Access},
  vol.~8, pp.~90931--90956, 2020.

\bibitem{salvi2023impact}
M.~Salvi, A.~Caputo, D.~Balmativola, M.~Scotto, O.~Pennisi, N.~Michielli,
  A.~Mogetta, F.~Molinari, and F.~Fraggetta, ``Impact of stain normalization on
  pathologist assessment of prostate cancer: A comparative study,'' {\em
  Cancers}, vol.~15, no.~5, p.~1503, 2023.

\bibitem{peidirect}
Z.~Pei, S.~Cao, L.~Lu, and W.~Chen, ``Direct cellularity estimation on breast
  cancer histopathology images using transfer learning (2019).''

\bibitem{chan2014wonderful}
J.~K. Chan, ``The wonderful colors of the hematoxylin--eosin stain in
  diagnostic surgical pathology,'' {\em International journal of surgical
  pathology}, vol.~22, no.~1, pp.~12--32, 2014.

\bibitem{fischer2008hematoxylin}
A.~H. Fischer, K.~A. Jacobson, J.~Rose, and R.~Zeller, ``Hematoxylin and eosin
  staining of tissue and cell sections,'' {\em Cold spring harbor protocols},
  vol.~2008, no.~5, pp.~pdb--prot4986, 2008.

\bibitem{tosta2019computational}
T.~A.~A. Tosta, P.~R. de~Faria, L.~A. Neves, and M.~Z. do~Nascimento,
  ``Computational normalization of h\&e-stained histological images: Progress,
  challenges and future potential,'' {\em Artificial intelligence in medicine},
  vol.~95, pp.~118--132, 2019.

\bibitem{hiatt2007tratado}
J.~L. Hiatt and P.~Leslie, ``Tratado de histologia em cores,'' 2007.

\bibitem{michail2014detection}
E.~Michail, E.~N. Kornaropoulos, K.~Dimitropoulos, N.~Grammalidis, T.~Koletsa,
  and I.~Kostopoulos, ``Detection of centroblasts in h\&e stained images of
  follicular lymphoma,'' in {\em 2014 22nd Signal Processing and Communications
  Applications Conference (SIU)}, pp.~2319--2322, IEEE, 2014.

\bibitem{janowczyk2017stain}
A.~Janowczyk, A.~Basavanhally, and A.~Madabhushi, ``Stain normalization using
  sparse autoencoders (stanosa): application to digital pathology,'' {\em
  Computerized Medical Imaging and Graphics}, vol.~57, pp.~50--61, 2017.

\bibitem{bertram2019large}
C.~A. Bertram, M.~Aubreville, C.~Marzahl, A.~Maier, and R.~Klopfleisch, ``A
  large-scale dataset for mitotic figure assessment on whole slide images of
  canine cutaneous mast cell tumor,'' {\em Scientific data}, vol.~6, no.~1,
  p.~274, 2019.

\bibitem{aubreville2021quantifying}
M.~Aubreville, C.~Bertram, R.~Veta, M.and~Klopfleisch, N.~Stathonikos,
  K.~Breininger, N.~ter Hoeve, F.~Ciompi, and A.~Maier, ``Quantifying the
  scanner-induced domain gap in mitosis detection,'' {\em arXiv:2103.16515},
  2021.

\bibitem{roy2018study}
S.~Roy, A.~kumar Jain, S.~Lal, and J.~Kini, ``A study about color normalization
  methods for histopathology images,'' {\em Micron}, vol.~114, pp.~42--61,
  2018.

\bibitem{vijh2021new}
S.~Vijh, M.~Saraswat, and S.~Kumar, ``A new complete color normalization method
  for h\&e stained histopatholgical images,'' {\em Applied Intelligence},
  pp.~1--14, 2021.

\bibitem{gurcan2006image}
M.~N. Gurcan, T.~Pan, H.~Shimada, and J.~Saltz, ``Image analysis for
  neuroblastoma classification: segmentation of cell nuclei,'' in {\em 2006
  International Conference of the IEEE Engineering in Medicine and Biology
  Society}, pp.~4844--4847, IEEE, 2006.

\bibitem{lu2012robust}
C.~Lu, M.~Mahmood, N.~Jha, and M.~Mandal, ``A robust automatic nuclei
  segmentation technique for quantitative histopathological image analysis,''
  {\em Analytical and Quantitative Cytology and Histology}, vol.~34,
  pp.~296--308, 2012.

\bibitem{phoulady2016nucleus}
H.~A. Phoulady, D.~B. Goldgof, L.~O. Hall, and P.~R. Mouton, ``Nucleus
  segmentation in histology images with hierarchical multilevel thresholding,''
  in {\em Medical Imaging 2016: Digital Pathology}, vol.~9791, pp.~280--285,
  SPIE, 2016.

\bibitem{gautam2016automatic}
A.~Gautam, P.~Singh, B.~Raman, and H.~Bhadauria, ``Automatic classification of
  leukocytes using morphological features and na{\"\i}ve bayes classifier,'' in
  {\em 2016 IEEE region 10 conference (TENCON)}, pp.~1023--1027, IEEE, 2016.

\bibitem{win2017automated}
K.~Y. Win and S.~Choomchuay, ``Automated segmentation of cell nuclei in
  cytology pleural fluid images using otsu thresholding,'' in {\em 2017
  International Conference on Digital Arts, Media and Technology (ICDAMT)},
  pp.~14--18, IEEE, 2017.

\bibitem{magoulianitis2022unsupervised}
V.~Magoulianitis, P.~Han, Y.~Yang, and C.-C.~J. Kuo, ``An unsupervised
  parameter-free nuclei segmentation method for histology images,'' in {\em
  2022 IEEE International Conference on Image Processing (ICIP)}, pp.~226--230,
  IEEE, 2022.

\bibitem{magoulianitis2022hunis}
V.~Magoulianitis, Y.~Yang, and C.-C.~J. Kuo, ``Hunis: High-performance
  unsupervised nuclei instance segmentation,'' in {\em 2022 IEEE 14th Image,
  Video, and Multidimensional Signal Processing Workshop (IVMSP)}, pp.~1--5,
  IEEE, 2022.

\bibitem{veta2011marker}
M.~Veta, A.~Huisman, M.~A. Viergever, P.~J. van Diest, and J.~P. Pluim,
  ``Marker-controlled watershed segmentation of nuclei in h\&e stained breast
  cancer biopsy images,'' in {\em 2011 IEEE international symposium on
  biomedical imaging: from nano to macro}, pp.~618--621, IEEE, 2011.

\bibitem{vahadane2013towards}
A.~Vahadane and A.~Sethi, ``Towards generalized nuclear segmentation in
  histological images,'' in {\em 13th IEEE International Conference on
  BioInformatics and BioEngineering}, pp.~1--4, IEEE, 2013.

\bibitem{shu2013segmenting}
J.~Shu, H.~Fu, G.~Qiu, P.~Kaye, and M.~Ilyas, ``Segmenting overlapping cell
  nuclei in digital histopathology images,'' in {\em 2013 35th Annual
  International Conference of the IEEE Engineering in Medicine and Biology
  Society (EMBC)}, pp.~5445--5448, IEEE, 2013.

\bibitem{cui2016self}
Y.~Cui and J.~Hu, ``Self-adjusting nuclei segmentation (sans) of
  hematoxylin-eosin stained histopathological breast cancer images,'' in {\em
  2016 IEEE International Conference on Bioinformatics and Biomedicine (BIBM)},
  pp.~956--963, IEEE, 2016.

\bibitem{koyuncu2016iterative}
C.~F. Koyuncu, E.~Akhan, T.~Ersahin, R.~Cetin-Atalay, and C.~Gunduz-Demir,
  ``Iterative h-minima-based marker-controlled watershed for cell nucleus
  segmentation,'' {\em Cytometry Part A}, vol.~89, no.~4, pp.~338--349, 2016.

\bibitem{rajyalakshmi2017supervised}
U.~Rajyalakshmi, S.~K. Rao, and K.~S. Prasad, ``Supervised classification of
  breast cancer malignancy using integrated modified marker controlled
  watershed approach,'' in {\em 2017 IEEE 7th International Advance Computing
  Conference (IACC)}, pp.~584--589, IEEE, 2017.

\bibitem{hu2004automated}
M.~Hu, X.~Ping, and Y.~Ding, ``Automated cell nucleus segmentation using
  improved snake,'' in {\em 2004 International Conference on Image Processing,
  2004. ICIP'04.}, vol.~4, pp.~2737--2740, IEEE, 2004.

\bibitem{fatakdawala2010expectation}
H.~Fatakdawala, J.~Xu, A.~Basavanhally, G.~Bhanot, S.~Ganesan, M.~Feldman,
  J.~E. Tomaszewski, and A.~Madabhushi, ``Expectation--maximization-driven
  geodesic active contour with overlap resolution (emagacor): Application to
  lymphocyte segmentation on breast cancer histopathology,'' {\em IEEE
  Transactions on Biomedical Engineering}, vol.~57, no.~7, pp.~1676--1689,
  2010.

\bibitem{faridi2016automatic}
P.~Faridi, H.~Danyali, M.~S. Helfroush, and M.~A. Jahromi, ``An automatic
  system for cell nuclei pleomorphism segmentation in histopathological images
  of breast cancer,'' in {\em 2016 IEEE Signal Processing in Medicine and
  Biology Symposium (SPMB)}, pp.~1--5, IEEE, 2016.

\bibitem{beevi2016automatic}
S.~Beevi, M.~S. Nair, and G.~Bindu, ``Automatic segmentation of cell nuclei
  using krill herd optimization based multi-thresholding and localized active
  contour model,'' {\em Biocybernetics and Biomedical Engineering}, vol.~36,
  no.~4, pp.~584--596, 2016.

\bibitem{rashmi2021multi}
R.~Rashmi, K.~Prasad, and C.~B.~K. Udupa, ``Multi-channel chan-vese model for
  unsupervised segmentation of nuclei from breast histopathological images,''
  {\em Computers in Biology and Medicine}, vol.~136, p.~104651, 2021.

\bibitem{7797086}
R.~Saha, M.~Bajger, and G.~Lee, ``Spatial shape constrained fuzzy c-means (fcm)
  clustering for nucleus segmentation in pap smear images,'' in {\em 2016
  International Conference on Digital Image Computing: Techniques and
  Applications (DICTA)}, pp.~1--8, 2016.

\bibitem{otsu1979threshold}
N.~Otsu, ``A threshold selection method from gray-level histograms,'' {\em IEEE
  transactions on systems, man, and cybernetics}, vol.~9, no.~1, pp.~62--66,
  1979.

\bibitem{cai2014new}
H.~Cai, Z.~Yang, X.~Cao, W.~Xia, and X.~Xu, ``A new iterative triclass
  thresholding technique in image segmentation,'' {\em IEEE transactions on
  image processing}, vol.~23, no.~3, pp.~1038--1046, 2014.

\bibitem{roerdink2000watershed}
J.~B. Roerdink and A.~Meijster, ``The watershed transform: Definitions,
  algorithms and parallelization strategies,'' {\em Fundamenta informaticae},
  vol.~41, no.~1-2, pp.~187--228, 2000.

\bibitem{chan2001active}
T.~F. Chan and L.~A. Vese, ``Active contours without edges,'' {\em IEEE
  Transactions on image processing}, vol.~10, no.~2, pp.~266--277, 2001.

\bibitem{mumford1989optimal}
D.~B. Mumford and J.~Shah, ``Optimal approximations by piecewise smooth
  functions and associated variational problems,'' {\em Communications on pure
  and applied mathematics}, 1989.

\bibitem{al2009improved}
Y.~Al-Kofahi, W.~Lassoued, W.~Lee, and B.~Roysam, ``Improved automatic
  detection and segmentation of cell nuclei in histopathology images,'' {\em
  IEEE Transactions on Biomedical Engineering}, vol.~57, no.~4, pp.~841--852,
  2009.

\bibitem{danvek2009segmentation}
O.~Dan{\v{e}}k, P.~Matula, C.~Ortiz-de Sol{\'o}rzano, A.~Mu{\~n}oz-Barrutia,
  M.~Ma{\v{s}}ka, and M.~Kozubek, ``Segmentation of touching cell nuclei using
  a two-stage graph cut model,'' in {\em Image Analysis: 16th Scandinavian
  Conference, SCIA 2009, Oslo, Norway, June 15-18, 2009. Proceedings 16},
  pp.~410--419, Springer, 2009.

\bibitem{zhang2014segmentation}
L.~Zhang, H.~Kong, C.~T. Chin, S.~Liu, Z.~Chen, T.~Wang, and S.~Chen,
  ``Segmentation of cytoplasm and nuclei of abnormal cells in cervical cytology
  using global and local graph cuts,'' {\em Computerized Medical Imaging and
  Graphics}, vol.~38, no.~5, pp.~369--380, 2014.

\bibitem{sauvola2000adaptive}
J.~Sauvola and M.~Pietik{\"a}inen, ``Adaptive document image binarization,''
  {\em Pattern recognition}, vol.~33, no.~2, pp.~225--236, 2000.

\bibitem{sarrafzadeh2015nucleus}
O.~Sarrafzadeh and A.~M. Dehnavi, ``Nucleus and cytoplasm segmentation in
  microscopic images using k-means clustering and region growing,'' {\em
  Advanced biomedical research}, vol.~4, 2015.

\bibitem{win2017k}
K.~Y. Win, S.~Choomchuay, and K.~Hamamoto, ``K mean clustering based automated
  segmentation of overlapping cell nuclei in pleural effusion cytology
  images,'' in {\em 2017 International Conference on Advanced Technologies for
  Communications (ATC)}, pp.~265--269, IEEE, 2017.

\bibitem{chang2016quantitative}
Y.~H. Chang, G.~Thibault, V.~Azimi, B.~Johnson, D.~Jorgens, J.~Link,
  A.~Margolin, and J.~W. Gray, ``Quantitative analysis of histological tissue
  image based on cytological profiles and spatial statistics,'' in {\em 2016
  38th Annual International Conference of the IEEE Engineering in Medicine and
  Biology Society (EMBC)}, pp.~1175--1178, IEEE, 2016.

\bibitem{hsu2021darcnn}
J.~Hsu, W.~Chiu, and S.~Yeung, ``Darcnn: Domain adaptive region-based
  convolutional neural network for unsupervised instance segmentation in
  biomedical images,'' in {\em Proceedings of the IEEE/CVF conference on
  computer vision and pattern recognition}, pp.~1003--1012, 2021.

\bibitem{zheng2018fast}
X.~Zheng, Y.~Wang, G.~Wang, and J.~Liu, ``Fast and robust segmentation of white
  blood cell images by self-supervised learning,'' {\em Micron}, vol.~107,
  pp.~55--71, 2018.

\bibitem{sahasrabudhe2020self}
M.~Sahasrabudhe, S.~Christodoulidis, R.~Salgado, S.~Michiels, S.~Loi,
  F.~Andr{\'e}, N.~Paragios, and M.~Vakalopoulou, ``Self-supervised nuclei
  segmentation in histopathological images using attention,'' in {\em Medical
  Image Computing and Computer Assisted Intervention--MICCAI 2020: 23rd
  International Conference, Lima, Peru, October 4--8, 2020, Proceedings, Part V
  23}, pp.~393--402, Springer, 2020.

\bibitem{ali2022multi}
H.~Ali, M.~Elattar, and S.~Selim, ``A multi-scale self-supervision method for
  improving cell nuclei segmentation in pathological tissues,'' in {\em Annual
  Conference on Medical Image Understanding and Analysis}, pp.~751--763,
  Springer, 2022.

\bibitem{punn2022bt}
N.~S. Punn and S.~Agarwal, ``Bt-unet : A self-supervised learning framework for
  biomedical image segmentation using barlow twins with u-net models,'' {\em
  Machine Learning}, vol.~111, no.~12, pp.~4585--4600, 2022.

\bibitem{liu2020unsupervised}
D.~Liu, D.~Zhang, Y.~Song, F.~Zhang, L.~O'Donnell, H.~Huang, M.~Chen, and
  W.~Cai, ``Unsupervised instance segmentation in microscopy images via
  panoptic domain adaptation and task re-weighting,'' in {\em Proceedings of
  the IEEE/CVF conference on computer vision and pattern recognition},
  pp.~4243--4252, 2020.

\bibitem{xie2020instance}
X.~Xie, J.~Chen, Y.~Li, L.~Shen, K.~Ma, and Y.~Zheng, ``Instance-aware
  self-supervised learning for nuclei segmentation,'' in {\em Medical Image
  Computing and Computer Assisted Intervention--MICCAI 2020: 23rd International
  Conference, Lima, Peru, October 4--8, 2020, Proceedings, Part V 23},
  pp.~341--350, Springer, 2020.

\bibitem{boserup2022efficient}
N.~Boserup and R.~Selvan, ``Efficient self- \\ supervision using patch-based
  contrastive learning for \\ histopathology image segmentation,'' {\em
  arXiv:2208.10779}, 2022.

\bibitem{liang2022region}
H.~Liang, Z.~Cheng, H.~Zhong, A.~Qu, and L.~Chen, ``A region-based
  convolutional network for nuclei detection and segmentation in microscopy
  images,'' {\em Biomedical Signal Processing and Control}, vol.~71, p.~103276,
  2022.

\bibitem{roy2023nuclei}
K.~Roy, S.~Saha, D.~Banik, and D.~Bhattacharjee, ``Nuclei-net: A multi-stage
  fusion model for nuclei segmentation in microscopy images,'' 2023.

\bibitem{graham2019hover}
S.~Graham, Q.~D. Vu, S.~E.~A. Raza, A.~Azam, Y.~W. Tsang, J.~T. Kwak, and
  N.~Rajpoot, ``Hover-net: Simultaneous segmentation and classification of
  nuclei in multi-tissue histology images,'' {\em Medical image analysis},
  vol.~58, p.~101563, 2019.

\bibitem{chanchal2021high}
A.~K. C., S.~Lal, and J.~Kini, ``High-resolution deep transferred asppu-net for
  nuclei segmentation of histopathology images,'' {\em International journal of
  computer assisted radiology and surgery}, vol.~16, pp.~2159--2175, 2021.

\bibitem{kiran2022denseres}
I.~Kiran, B.~Raza, A.~Ijaz, and M.~A. Khan, ``Denseres-unet: Segmentation of
  overlapped/clustered nuclei from multi organ histopathology images,'' {\em
  Computers in Biology and Medicine}, vol.~143, p.~105267, 2022.

\bibitem{saednia2022cascaded}
K.~Saednia, W.~T. Tran, and A.~Sadeghi-Naini, ``A cascaded deep learning
  framework for segmentation of nuclei in digital histology images,'' in {\em
  2022 44th Annual International Conference of the IEEE Engineering in Medicine
  \& Biology Society (EMBC)}, pp.~4764--4767, IEEE, 2022.

\bibitem{hancer2023imbalance}
E.~Hancer, M.~Traor{\'e}, R.~Samet, Z.~Y{\i}ld{\i}r{\i}m, and N.~Nemati, ``An
  imbalance-aware nuclei segmentation methodology for h\&e stained
  histopathology images,'' {\em Biomedical Signal Processing and Control},
  vol.~83, p.~104720, 2023.

\bibitem{chen2023cpp}
S.~Chen, C.~Ding, M.~Liu, J.~Cheng, and D.~Tao, ``Cpp-net: Context-aware
  polygon proposal network for nucleus segmentation,'' {\em IEEE Transactions
  on Image Processing}, vol.~32, pp.~980--994, 2023.

\bibitem{kumar2017dataset}
N.~Kumar, R.~Verma, S.~Sharma, S.~Bhargava, A.~Vahadane, and A.~Sethi, ``A
  dataset and a technique for generalized nuclear segmentation for
  computational pathology,'' {\em IEEE transactions on medical imaging},
  vol.~36, no.~7, pp.~1550--1560, 2017.

\bibitem{chen2020boundary}
S.~Chen, C.~Ding, and D.~Tao, ``Boundary-assisted region proposal networks for
  nucleus segmentation,'' in {\em Medical Image Computing and Computer Assisted
  Intervention--MICCAI 2020: 23rd International Conference, Lima, Peru, October
  4--8, 2020, Proceedings, Part V 23}, pp.~279--288, Springer, 2020.

\bibitem{qin2022reu}
J.~Qin, Y.~He, Y.~Zhou, J.~Zhao, and B.~Ding, ``Reu-net: Region-enhanced nuclei
  segmentation network,'' {\em Computers in Biology and Medicine}, vol.~146,
  p.~105546, 2022.

\bibitem{lal2021nucleisegnet}
S.~Lal, D.~Das, K.~Alabhya, A.~Kanfade, A.~Kumar, and J.~Kini, ``Nucleisegnet:
  Robust deep learning architecture for the nuclei segmentation of liver cancer
  histopathology images,'' {\em Computers in Biology and Medicine}, vol.~128,
  p.~104075, 2021.

\bibitem{yang2022gcp}
G.~Yang, J.~Huang, Y.~He, Y.~Chen, T.~Wang, C.~Jin, P.~Sengphachanh, {\em
  et~al.}, ``Gcp-net: A gating context-aware pooling network for cervical cell
  nuclei segmentation,'' {\em Mobile Information Systems}, vol.~2022, 2022.

\bibitem{thi2022convolutional}
P.~Thi~Le, T.~Pham, Y.-C. Hsu, and J.-C. Wang, ``Convolutional blur attention
  network for cell nuclei segmentation,'' {\em Sensors}, vol.~22, no.~4,
  p.~1586, 2022.

\bibitem{xing2015automatic}
F.~Xing, Y.~Xie, and L.~Yang, ``An automatic learning-based framework for
  robust nucleus segmentation,'' {\em IEEE transactions on medical imaging},
  vol.~35, no.~2, pp.~550--566, 2015.

\bibitem{ronneberger2015u}
O.~Ronneberger, P.~Fischer, and T.~Brox, ``U-net: Convolutional networks for
  biomedical image segmentation,'' in {\em Medical Image Computing and
  Computer-Assisted Intervention--MICCAI 2015: 18th International Conference,
  Munich, Germany, October 5-9, 2015, Proceedings, Part III 18}, pp.~234--241,
  Springer, 2015.

\bibitem{oda2018besnet}
H.~Oda, H.~R. Roth, K.~Chiba, J.~Sokoli{\'c}, T.~Kitasaka, M.~Oda, A.~Hinoki,
  H.~Uchida, J.~A. Schnabel, and K.~Mori, ``Besnet: boundary-enhanced
  segmentation of cells in histopathological images,'' in {\em Medical Image
  Computing and Computer Assisted Intervention--MICCAI 2018: 21st International
  Conference, \\ Granada, Spain, September 16-20, 2018, Proceedings, Part II
  11}, pp.~228--236, Springer, 2018.

\bibitem{schlemper2019attention}
J.~Schlemper, O.~Oktay, M.~Schaap, M.~Heinrich, B.~Kainz, B.~Glocker, and
  D.~Rueckert, ``Attention gated networks: Learning to leverage salient regions
  in medical images,'' {\em Medical image analysis}, vol.~53, pp.~197--207,
  2019.

\bibitem{yoo2019pseudoedgenet}
I.~Yoo, D.~Yoo, and K.~Paeng, ``Pseudoedgenet: Nuclei segmentation only with
  point annotations,'' in {\em Medical Image Computing and Computer Assisted
  Intervention-- \\ MICCAI 2019: 22nd International Conference, Shenzhen,
  China, October 13--17, 2019, Proceedings, Part I 22}, pp.~731--739, Springer,
  2019.

\bibitem{qu2020weakly}
H.~Qu, P.~Wu, Q.~Huang, J.~Yi, Z.~Yan, K.~Li, G.~M. Riedlinger, S.~De,
  S.~Zhang, and D.~N. Metaxas, ``Weakly supervised deep nuclei segmentation
  using partial points annotation in histopathology images,'' {\em IEEE
  transactions on medical imaging}, vol.~39, no.~11, pp.~3655--3666, 2020.

\bibitem{qu2020nuclei}
H.~Qu, J.~Yi, Q.~Huang, P.~Wu, and D.~Metaxas, ``Nuclei segmentation using
  mixed points and masks selected from uncertainty,'' in {\em 2020 IEEE 17th
  International Symposium on Biomedical Imaging (ISBI)}, pp.~973--976, IEEE,
  2020.

\bibitem{tian2020weakly}
K.~Tian, J.~Zhang, H.~Shen, K.~Yan, P.~Dong, J.~Yao, S.~Che, P.~Luo, and
  X.~Han, ``Weakly-supervised nucleus segmentation based on point annotations:
  A coarse-to-fine self-stimulated learning strategy,'' in {\em Medical Image
  Computing and Computer Assisted Intervention--MICCAI 2020: 23rd International
  Conference, Lima, Peru, October 4--8, 2020, Proceedings, Part V 23},
  pp.~299--308, Springer, 2020.

\bibitem{lin2022label}
Y.~Lin, Z.~Qu, H.~Chen, Z.~Gao, Y.~Li, L.~Xia, K.~Ma, Y.~Zheng, and K.-T.
  Cheng, ``Label propagation for annotation-efficient nuclei segmentation from
  pathology images,'' {\em arXiv preprint arXiv:2202.08195}, 2022.

\bibitem{lou2022pixel}
W.~Lou, H.~Li, G.~Li, X.~Han, and X.~Wan, ``Which pixel to annotate: a
  label-efficient nuclei segmentation framework,'' {\em IEEE Transactions on
  Medical Imaging}, vol.~42, no.~4, pp.~947--958, 2022.

\bibitem{qu2019weakly}
H.~Qu, P.~Wu, Q.~Huang, J.~Yi, G.~M. Riedlinger, S.~De, and D.~N. Metaxas,
  ``Weakly supervised deep nuclei segmentation using points annotation in
  histopathology images,'' in {\em International Conference on Medical Imaging
  with Deep Learning}, pp.~390--400, PMLR, 2019.

\bibitem{hu2020generative}
W.~Hu, H.~Sheng, J.~Wu, Y.~Li, T.~Liu, Y.~Wang, and Y.~Wen, ``Generative
  adversarial training for weakly supervised nuclei instance segmentation,'' in
  {\em 2020 IEEE International Conference on Systems, Man, and Cybernetics
  (SMC)}, pp.~3649--3654, IEEE, 2020.

\bibitem{8880654}
N.~Kumar, R.~Verma, D.~Anand, Y.~Zhou, O.~F. Onder, E.~Tsougenis, H.~Chen,
  P.-A. Heng, J.~Li, Z.~Hu, Y.~Wang, N.~A. Koohbanani, M.~Jahanifar, N.~Z.
  Tajeddin, A.~Gooya, N.~Rajpoot, X.~Ren, S.~Zhou, Q.~Wang, D.~Shen, C.-K.
  Yang, C.-H. Weng, W.-H. Yu, C.-Y. Yeh, S.~Yang, S.~Xu, P.~H. Yeung, P.~Sun,
  A.~Mahbod, G.~Schaefer, I.~Ellinger, R.~Ecker, O.~Smedby, C.~Wang,
  B.~Chidester, T.-V. Ton, M.-T. Tran, J.~Ma, M.~N. Do, S.~Graham, Q.~D. Vu,
  J.~T. Kwak, A.~Gunda, R.~Chunduri, C.~Hu, X.~Zhou, D.~Lotfi, R.~Safdari,
  A.~Kascenas, A.~O’Neil, D.~Eschweiler, J.~Stegmaier, Y.~Cui, B.~Yin,
  K.~Chen, X.~Tian, P.~Gruening, E.~Barth, E.~Arbel, I.~Remer, A.~Ben-Dor,
  E.~Sirazitdinova, M.~Kohl, S.~Braunewell, Y.~Li, X.~Xie, L.~Shen, J.~Ma,
  K.~D. Baksi, M.~A. Khan, J.~Choo, A.~Colomer, V.~Naranjo, L.~Pei, K.~M.
  Iftekharuddin, K.~Roy, D.~Bhattacharjee, A.~Pedraza, M.~G. Bueno,
  S.~Devanathan, S.~Radhakrishnan, P.~Koduganty, Z.~Wu, G.~Cai, X.~Liu,
  Y.~Wang, and A.~Sethi, ``A multi-organ nucleus segmentation challenge,'' {\em
  IEEE Transactions on Medical Imaging}, vol.~39, no.~5, pp.~1380--1391, 2020.

\bibitem{vu2019methods}
Q.~D. Vu, S.~Graham, T.~Kurc, M.~N.~N. To, M.~Shaban, T.~Qaiser, N.~A.
  Koohbanani, S.~A. Khurram, J.~Kalpathy-Cramer, T.~Zhao, {\em et~al.},
  ``Methods for segmentation and classification of digital microscopy tissue
  images,'' {\em Frontiers in bioengineering and biotechnology}, p.~53, 2019.

\bibitem{naylor2018segmentation}
P.~Naylor, M.~La{\'e}, F.~Reyal, and T.~Walter, ``Segmentation of nuclei in
  histopathology images by deep regression of the distance map,'' {\em IEEE
  transactions on medical imaging}, vol.~38, no.~2, pp.~448--459, 2018.

\bibitem{7399414}
K.~Sirinukunwattana, S.~E.~A. Raza, Y.-W. Tsang, D.~R.~J. Snead, I.~A. Cree,
  and N.~M. Rajpoot, ``Locality sensitive deep learning for detection and
  classification of nuclei in routine colon cancer histology images,'' {\em
  IEEE Transactions on Medical Imaging}, vol.~35, no.~5, pp.~1196--1206, 2016.

\bibitem{caicedo2019nucleus}
J.~C. Caicedo, A.~Goodman, K.~W. Karhohs, B.~A. Cimini, J.~Ackerman,
  M.~Haghighi, C.~Heng, T.~Becker, M.~Doan, C.~McQuin, {\em et~al.}, ``Nucleus
  segmentation across imaging experiments: the 2018 data science bowl,'' {\em
  Nature methods}, vol.~16, no.~12, pp.~1247--1253, 2019.

\bibitem{kirillov2019panoptic}
A.~Kirillov, K.~He, R.~Girshick, C.~Rother, and P.~Doll{\'a}r, ``Panoptic
  segmentation,'' in {\em Proceedings of the IEEE/CVF conference on computer
  vision and pattern recognition}, pp.~9404--9413, 2019.

\bibitem{carpenter2006cellprofiler}
A.~E. Carpenter, T.~R. Jones, M.~R. Lamprecht, C.~Clarke, I.~H. Kang,
  O.~Friman, D.~A. Guertin, J.~H. Chang, R.~A. Lindquist, J.~Moffat, {\em
  et~al.}, ``Cellprofiler: image analysis software for identifying and
  quantifying cell phenotypes,'' {\em Genome biology}, vol.~7, pp.~1--11, 2006.

\bibitem{schindelin2012fiji}
J.~Schindelin, I.~Arganda-Carreras, E.~Frise, V.~Kaynig, M.~Longair,
  T.~Pietzsch, S.~Preibisch, C.~Rueden, S.~Saalfeld, B.~Schmid, {\em et~al.},
  ``Fiji: an open-source platform for biological-image analysis,'' {\em Nature
  methods}, vol.~9, no.~7, pp.~676--682, 2012.

\bibitem{kim2019diversify}
T.~Kim, M.~Jeong, S.~Kim, S.~Choi, and C.~Kim, ``Diversify and match: A domain
  adaptive representation learning paradigm for object detection,'' in {\em
  Proceedings of the IEEE/CVF Conference on Computer Vision and Pattern
  Recognition}, pp.~12456--12465, 2019.

\bibitem{he2016deep}
K.~He, X.~Zhang, S.~Ren, and J.~Sun, ``Deep residual learning for image
  recognition,'' in {\em Proceedings of the IEEE conference on computer vision
  and pattern recognition}, pp.~770--778, 2016.

\bibitem{simonyan2014very}
K.~Simonyan and A.~Zisserman, ``Very deep convolutional networks for
  large-scale image recognition,'' {\em arXiv preprint arXiv:1409.1556}, 2014.

\bibitem{huang2017densely}
G.~Huang, Z.~Liu, L.~Van Der~Maaten, and K.~Q. Weinberger, ``Densely connected
  convolutional networks,'' in {\em Proceedings of the IEEE conference on
  computer vision and pattern recognition}, pp.~4700--4708, 2017.

\bibitem{he2017mask}
K.~He, G.~Gkioxari, P.~Doll{\'a}r, and R.~Girshick, ``Mask r-cnn,'' in {\em
  Proceedings of the IEEE international conference on computer vision},
  pp.~2961--2969, 2017.

\bibitem{chen2017dcan}
H.~Chen, X.~Qi, L.~Yu, Q.~Dou, J.~Qin, and P.-A. Heng, ``Dcan: Deep
  contour-aware networks for object instance segmentation from histology
  images,'' {\em Medical image analysis}, vol.~36, pp.~135--146, 2017.

\bibitem{liu2018path}
S.~Liu, L.~Qi, H.~Qin, J.~Shi, and J.~Jia, ``Path aggregation network for
  instance segmentation,'' in {\em Proceedings of the IEEE conference on
  computer vision and pattern recognition}, pp.~8759--8768, 2018.

\end{thebibliography}

\end{document}